\pdfoutput=1

\documentclass[11pt,a4paper,twoside,openright]{report}

\usepackage[mieic,final,backrefs]{feupteses}

\usepackage{graphicx}
\usepackage{float}
\usepackage[utf8]{inputenc}
\DeclareUnicodeCharacter{2212}{-}
\usepackage{amssymb}
\usepackage{multirow}
\usepackage{multicol}
\usepackage{caption}
\usepackage{fancyvrb}
\usepackage{subcaption}
\usepackage{url}
\usepackage{array}
\usepackage{breakcites}
\usepackage{enumerate}

\usepackage{color}
\definecolor{cloudwhite}{cmyk}{0, 0, 0, 0.025}

\usepackage{siunitx}
\graphicspath{{figures/}}

\usepackage{listings}
\usepackage[acronym,toc]{glossaries}

\makeglossaries

\newacronym{UGC}{UGC}{User Generated Content}

\newacronym{SMC}{SMC}{Social Media Content}

\newacronym{NLP}{NLP}{Natural Language Processing}

\newacronym{UI}{UI}{User Interface}

\newacronym{ICT}{ICT}{Information and Communications Technology}

\newacronym{ITS}{ITS}{Intelligent Transportation Systems}

\newacronym{SM}{SM}{Smart Mobility}

\newacronym{NLTK}{NLTK}{Natural Language Toolkit}

\newacronym{BoW}{BoW}{Bag-of-words}

\newacronym{BoE}{BoE}{Bag-of-embeddings}

\newacronym{IQR}{IQR}{Interquartile-range}

\newacronym{UTC}{UTC}{Coordinated Universal Timezone}

\newacronym{API}{API}{Application Programming Interface}

\newacronym{LDA}{LDA}{Latent Dirichlet Allocation}

\newacronym{REST}{REST}{Representational State Transfer}

\newacronym{GPS}{GPS}{Global Positioning System}

\newacronym{ROC}{ROC}{Receiver Operating Characteristic}

\newacronym{TPR}{TPR}{True Positive Rate}

\newacronym{FPR}{FPR}{False Positive Rate}

\newacronym{AUC}{AUC}{Area Under the Curve}

\newacronym{HTML}{HTML}{HyperText Markup Language}

\newacronym{JSON}{JSON}{JavaScript Object Notation}

\newacronym{SVM}{SVM}{Suport Vector Machines}

\newacronym{OLS}{OLS}{Ordinary Least Squares}

\newacronym{RF}{RF}{Random Forests}

\newacronym{MLP}{MLP}{Multilayer Perceptron}

\newacronym{NB}{NB}{Naïve Bayes}

\newacronym{DT J48}{DT J48}{Decision Trees J48}

\newacronym{CRF}{CRF}{Conditional Random Fields}

\newacronym{POS}{POS}{Part-of-the-Speech Tagging}

\newacronym{WSD}{WSD}{Word Sense Disambiguation}

\newacronym{SMA}{SMA}{Social Media Analytics}

\newacronym{NED}{NED}{Name Entity Disambiguation}

\newglossaryentry{MicroBlog}{
    name = MicroBlog,
    description = {It is a tool that allows quick abd short status updates, and if possible, through multiple different platforms}
}

\newglossaryentry{Crowdsensing}{
	name = Crowdsensing or mobile crowdsensing,
	description = {Technique used to collectively share and extract information from large groups of individuals in order to analyse, infer or even measure processes of common interest}
}

\newglossaryentry{Twitter_Firehose}{
	name = Twitter Firehose,
	description = {It is a paid Twitter service that guarantees the delivery of 100\% of the tweets matched with certain criteria}
}

\newglossaryentry{Influenza_A}{
	name={Influenza A},
	description={Influenza A is a type  of virus capable of infecting animals, although it is more common for people to suffer the ailments associated with this type of flu}
}

\newglossaryentry{k-fold-cross-validation}{
	name={k-fold cross-validation},
	description={It is a technique where the original dataset is randomly partitioned into \emph{k} equal sized sub-datasets. Of the \emph{k} sub-datasets, only one is retained as the validation data for testing the model, and the remaining \emph{k − 1} sub-datasets are used as training data}
}

\newglossaryentry{bounding_box}{
	name={bounding-box},
	description={A bounding-box is a rectangle obtained by two coordinate pairs (latitude and longitude, for the South-West point and the North-East point)}
}

\usepackage[tight]{minitoc}

\dominitoc

\begin{document}

\title{Social Media Text Processing and Semantic Analysis for Smart Cities}
\author{João Filipe Figueiredo Pereira}

\supervisor{Supervisor}{Rosaldo José Fernandes Rossetti}
\supervisor{Co-supervisor}{Pedro dos Santos Saleiro da Cruz}

\thesisdate{June 27, 2017}



\logo{uporto-feup.pdf}


\begin{Prolog}
  \chapter*{Abstract}

With the rise of Social Media, people obtain and share information almost instantly on a 24/7 basis. Many research areas have tried to gain valuable insights from these large volumes of freely available user generated content. The research areas of intelligent transportation systems and smart cities are no exception. However, extracting meaningful and actionable knowledge from user generated content is a complex endeavor. First, each social media service has its own data collection specificities and constraints, second the volume of messages/posts produced can be overwhelming for automatic processing and mining, and last but not the least, social media texts are usually short, informal, with a lot of abbreviations, jargon, slang and idioms.
 
In this thesis, we try to tackle some of the aforementioned challenges with the goal of extracting knowledge from social media streams that might be useful in the context of intelligent transportation systems and smart cities. We designed and developed a framework for collection, processing and mining of geo-located Tweets. More specifically, it provides functionalities for parallel collection of geo-located tweets from multiple pre-defined bounding boxes (cities or regions), including filtering of non complying tweets, text pre-processing for Portuguese and English language, topic modeling, and transportation-specific text classifiers, as well as, aggregation and data visualization.

We performed an extensive exploratory data analysis of geo-located tweets in 5 different cities: Rio de Janeiro, São Paulo, New York City, London and Melbourne, comprising a total of more than 43 millions tweets in a period of 3 months. Furthermore, we performed a large scale topic modelling comparison between Rio de Janeiro and São Paulo. As far as we know this is the largest scale content analysis of geo-located tweets from Brazil. Interestingly, most of the topics are shared between both cities which despite being in the same country are considered very different regarding population, economy and lifestyle.

We take advantage of recent developments in word embeddings and train such representations from the collections of geo-located tweets. We then use a combination of bag-of-embeddings and traditional bag-of-words to train travel-related classifiers in both Portuguese and English to filter travel-related content from non-related. We created specific gold-standard data to perform empirical evaluation of the resulting classifiers. Results are in line with research work in other application areas by showing the robustness of using word embeddings to learn word similarities that bag-of-words is not able to capture. The source code and resources developed in this dissertation will be publicly available to foster further developments by the research community in smart cities and intelligent transportation systems.

\chapter*{Resumo}

Devido à ascensão das Redes Sociais, as pessoas obtêm e partilham informação quase que instantaneamente 24/7. Muitas áreas de investigação tentaram extrair informações importantes destes grandes volumes de conteúdo, gerado por utilizadores, e livremente disponíveis. As áreas de investigação de sistemas inteligentes de transportes e de cidades inteligentes (\textit{smart cities}) não são excepção. Contudo, extrair conhecimento acionável e significativo de conteúdo gerado por utilizadores exige um esforço complexo. Primeiro, cada serviço de social media possui as suas próprias especificidades e restrições para o método de recolha dos dados; em segundo lugar, o volume de mensagens produzidas pode ser esmagador para o processamento automático e prospeção; e por último, não menos importante, os textos das redes sociais são, geralmente, curtos, informais, com muitas abreviações, jargões, gírias e expressões idiomáticas. 

Nesta dissertação, tentamos abordar alguns dos desafios acima mencionados com o objectivo de extrair conhecimento de mensagens das redes sociais que possam ser úteis no contexto de sistemas inteligentes de transportes e cidades inteligentes (\textit{smart cities}). Nós idealizamos e desenvolvemos uma \textit{framework} para a recolha de dados, processamento e prospeção de Tweets geo-localizados. Mais especificamente, a \textit{framework} fornece funcionalidades para a recolha paralela de tweets geo-localizados de \textit{bounding-boxes} (cidades ou regiões), incluindo filtragem de tweets não preenchidos, pré-processamento de texto para a língua portuguesa e inglesa, modelagem de tópicos e classificadores de texto específicos para transportes, bem como, agregação e visualização de dados.

Nós realizamos uma análise exploratória extensiva relativamente a tweets geo-referenciados para 5 cidades diferentes: Rio de Janeiro, São Paulo, Nova Iorque, Londres e Melbourne, perfazendo um total de mais de 43 milhões de tweets num período de 3 meses. Posteriormente, nós realizámos modelação de tópicos em grande escala entre as cidades do Rio de janiero e São Paulo. Tanto quanto nós conhecemos, esta é a análise de conteúdo em maior escala para tweets  geo-referenciados no Brasil. Curiosamente, a maioria dos tópicos detectados são partilhados por ambas as cidades, que apesar de pertecerem ao mesmo país, são muito diferentes em termos de população, economia e estilo de vida.

Nós tiramos partido dos desenvolvimentos recentes em \textit{word embeddings} e treinamos tais representações a partir das coleções de tweets geo-referenciados. Nós então usamos a combinação dos \textit{bag-of-embedding} e dos tradicionais \textit{bag-of-words} para treinar os classificadores relacionados com viagens, tanto em Português como em Inglês, para filtrar conteúdo relacionado com transportes de conteúdo não relacionado. Nós criamos dados \textit{gold-standard} específicos para realizar análise empírica dos classificadores resultantes. Os resultados estão coerentes com o trabalho de investigação realizado em outras áreas de aplicação demonstrando a robustez da utilização de word embeddings para aprender similaridades que os \textit{bag-of-words} não são capazes de capturar. O código fonte e os recursos desenvolvidos nesta dissertação estarão publicamente disponíveis a fim de motivar outros desenvolvimentos pela comunidade científica em \textit{smart cities} e sistemas de transportes inteligentes. 
  \chapter*{Acknowledgements}

First of all, my deep gratitude to my friends for being on my side when I was a bit down.

\medskip

To my companions at Lab I120, João Neto, José Pinto, João Pedro Dias and Luís Reis ($\rho$7 Boyz): thank you for the funny moments during the whole dissertation period, specifically, during the tough process of writing up the document.

\medskip

To my colleagues, specially, Henrique Ferrolho: thank you for the friendship, patience and support in these five long years. Now, I am sure that more challenges are coming to us which may imply distance but besides that I truly believe that in the future we still would cross paths at the professional or even academic course.

\medskip

To Professor Rosaldo Rossetti and Pedro Saleiro, thank you very much for all support, dedication, enthusiasm and knowledge passed to me. During each task you defined in the dissertation period, I was able to improve myself in both academic and social levels.

\medskip

To the institution that host me, Faculty of Engineering of University of Porto (FEUP), as well as to all of its docents that guide me during this Master's program, I am thankful for everything I have learn until now.

\medskip

Last and more important, I would like to express my deep gratitude to my mother, Ana Brito, and my father, Júlio Pereira, for all the sacrifice and effort made to assure my future and concede me this opportunity to fulfil a dream: be graduated. I hope this achievement of mine make you very proud and I wish all success for both mine and your's ambitions and goals in the future. Like always, you know that you can count on me for everything you need.

\vspace{10mm}
\flushright{João Pereira}
  \cleardoublepage
\thispagestyle{plain}

\vspace*{8cm}

\begin{flushright}
   \textsl{``Life is too short for long-term grudges.''} \\
\vspace*{1.5cm}
           Elon Musk
\end{flushright}
  \cleardoublepage
  \pdfbookmark[0]{Table of Contents}{contents}
  \tableofcontents
  \cleardoublepage
  \pdfbookmark[0]{List of Figures}{figures}
  \listoffigures
  \cleardoublepage
  \pdfbookmark[0]{List of Tables}{tables}
  \listoftables
\end{Prolog}


\StartBody

\chapter{Introduction} \label{chap:intro}

\minitoc \mtcskip \noindent

\section{Scope and Motivation}\label{sec:scope_motivation}
With the rise of Social Media, people obtain and share information almost instantly on a 24/7 basis. Many research areas have tried to extract valuable insights from these large volumes of user generated content. The research areas of intelligent transportation systems and smart cities are no exception. Transforming this data into valuable information can be meaningful and useful for city governance, support traffic management and control, transportation services or even ordinary citizens wanting to be constantly informed about their cities. For instance, if a social media message is about the service quality of the \texttt{bus} transportation mode - "My bus is delayed." - then the responsible entities should have the necessary tools and information to tackle the existent problem. However, to extract the knowledge to handle the solving task, first it is necessary to filter travel-related messages from the non-related and only then explore the topic, sentiment, or even the target of its content.

\gls{SMC} is still in the process of maturation regarding its use in the \textit{smart cities}~\cite{batty2012smart} and transportation~\cite{gal2014potential} fields; users tend to publicly share events in which they participate, as well as the ones related to the operation of the transportation network, such as accidents and other disruptions. Indeed, this type of content is also targeted by studies about opinion mining, human behavior and respective activity patterns, political issues, social communication (e.g. news websites). Such studies focus their efforts on ways of understanding what people think and talk about and transform this knowledge into actionable and meaningful content such as the identification of terrorist movement~\cite{zhang2016mining}, potential news to be reported or even the prediction of the sentiment polarity regarding politicians, for example.

The exploration of \gls{SMC} brings particular advantages, under virtually no cost, such as real-time data and content authenticity due to its human generated nature~\cite{kuflik2017automating} since users express the reason why they prefer a certain product or a certain transportation mode service to others. The availability of this kind of data may be seen as its main advantage. Social media companies provide tools to the developers community that do not require additional costs regarding its exploration, allowing the local storage of data and the possibility of performing off-line analysis.

Among the existing social networks, Twitter is probably the most adequate for these purposes due to its microblog nature in which users publicly share short messages about their daily lifes. Twitter has already proved its value and potential in domains ranging from news detection~\cite{sankaranarayanan2009twitterstand} to real-time traffic sensing~\cite{carvalho2010real}. Other social networks, such as Facebook are not so accessible as users tend to publish content within a private circle of friends. Twitter is the 11th most visited website\footnote{\url{http://www.alexa.com/topsites}} in the planet. Its community is continuously growing and, nowadays, the number of active users is about 313 million\footnote{\url{https://about.twitter.com/company}}, registering a daily average of 400 million new posts. Although only 1-2\% of all tweets published everyday are geo-located ~\cite{ikeda2013twitter}, these tweets have the advantage of having an explicit geographic relevance to the city where users published those messages.

\section{Problem Statement}\label{sec:problem}

que classificar travel-related tweets será um 1o passo para filtrar ruido
que é preciso agregar o conhecimento
etc
fazer uma ponte entre o scope e os goals
mas este problem statement ta assim um pouco mais scope do que descrição concreta dos problemas que se vai atacar
mas está muito melhor do que dantes
por isso...

Mining Twitter data is a laborious and time-consuming process due to the restrictions and difficulties in its content. The informal language, the existence of slang, abbreviations, jargons and the short length of the message are some of the problems when analyzing this data. Harvesting tweets automatically and, at the same time, extracting valuable information for \textit{smart cities} and transportation domains makes the task even more complex. Geo-located tweets can have a fundamental role since the 
study of its content can serve to characterize topics that may be associate the current subjects, events or even interests in specific locations.
The lack of gold standards datasets is the most disturbing problem since we are not able to benchmark any analysis performed to these aforementioned domains.

Having this considered, the problem on focus in this dissertation is to find a way of demonstrating social media text analysis about cities/regions/countries that can be valuable for entities, governments or even ordinary citizens during decision-making policies, such as, for example, which city presents the most safety level or which mode of transport may an individual choose to travel across a city. However, demonstrate valuable information to these stakeholders is complex since a continuous flow of social media streams is composed of a large number of noisy messages which do not bring additional meaningful and actionable knowledge and due to the previously mentioned restrictions present in a social media message, the text analysis needs to follow a predefined group of operations to erase/minimize these limitations. On the other hand, the analytics process flow may be automatic, i.e. the collection, filtering, text processing, results aggregation and visualization tasks may be associated with each other performing a tool capable of monitoring automatically the current state of a city's transportation domain or even identify differences between cities apart from the demographic, economic and lifestyle factors.

The first point to be tackled here is the continuous collection of geo-located tweets from multiple cities. Each social network, Twitter inclusive, has its own particular specificities regarding the data collection methodology. To solve this problem, it is necessary to know what are the targets (stakeholders) of the resulting information and what are the methods available in the collecting tools provided by social networks, as well as its limitations.

Second, it is necessary to filter noisy messages from the final datasets of each city. We must ensure that every tweet is actually located inside the city, that knowledge can be extract from messages besides the aforementioned restrictions (short text, informality, existence of abbreviations, jargon, slang and idioms) in its content and what are the geo-located tweets that are indeed related to specific domains such as the transportation area. In the transportation domain, we can identify multiple modes of transport such as "bike", "car", "taxi", "train", and for this reason, a important challenge to surpass is the automatic classification of geo-located tweets related to travel, allowing through this an preliminary filtering process of what messages are relevant for the domain in study. Nowadays, new modes of transport emerges unexpectedly and conventional approaches to filter travel-related tweets are limited to a set of terms during the training process of the text classifiers. This challenges to use recent advances in text mining such as the usage of word embeddings to capture syntactic and semantic similarities reducing the previously mentioned problem of limited number of transportation modes a text classifier tackle.

Other issue to be concerned is the ambiguity in the social media text messages where the name of a transportation mode may also have other meaning such as "train" which can be seen as a mode of transport or even a physical or mental activity. For this reason, it is important the application of text mining techniques that allows the correct identification of a certain entity in a text message~\cite{saleiro2013popstar}. Nonetheless, the community does not publicly share datasets for this domain (transportation) and due to this we are obligated to build our own datasets if we choose to use supervised learning methods for the conceptualization of a domain-specific text classifier.

Third, the volume of data retrieved by social media collecting tools is overwhelming and to automatically process and mine these data it is necessary to study what are the most valuable and less time consuming approaches to extract the desired information, useful to entities in the context of Smart Cities and \gls{ITS}. Nowadays, there is already a large group of \gls{NLP} techniques applied to social media streams, such as topic modelling which consists in discovering latent topics within the messages, to monitor what is being talked about in cities. Several authors have already explored this text mining technique to monitor and also characterize regions and cities as reported in the work of Lansley and Longley~\cite{lansley2016geography} where the city on focus was London. Here, the main problem focus on the discovering of latent topics and further aggregation at geographic and temporal level to characterize each city and then compare them. 

Last but not least, the final tool needs to be built having into consideration not only the aforementioned text analysis but also needs to be scalable to implement further and complementary tasks, such as sentiment analysis which is not tackled in this dissertation. Hence, another problem here is the design and implementation of a framework, or at least a prototype, which has as basis components modularity, scalability and flexibility, and might be capable of collecting, processing, aggregate and demonstrate results for multiple cities in parallel during a continuous period of time. Moreover, the final system must be capable of using in a easier way, being only necessary to turn on the tool, defining a specific area and the results are performed automatically.

\section{Aim and Goals}\label{sec:aim_goals}
This thesis aims to design and develop of a research framework for text processing and semantic analysis of geo-located Tweets within a pre-defined geographical area (e.g. cities). More specifically, to practically implement such a framework we shall accomplish the following goals: 

\begin{itemize}
	\item Continuous collection of geo-located tweets from multiple bounding boxes in parallel and in compliance with Twitter API usage limits;
	
	\item Tackling Twitter Geo API inconsistencies and filtering noisy tweets;
	
	\item Implement standard text pre-processing methods for social media texts;
	
	\item Content analysis using topic modeling and comparative characterization among different bounding boxes (e.g. cities);
	
	\item Travel-related classification of tweets using supervised learning;
	
	\item Train word embeddings from geo-located tweets;
	
	\item Study the impact of word embeddings in travel-related classification;
	
	\item Creation of gold-standard data for travel-related supervised learning;
	
	\item Aggregation and visualization of results.
\end{itemize}

\section{Document Structure}\label{sec:doc_structure}
The remainder of this document is organized as follows.
Chapter~\ref{chap:sota} starts with a brief contextualization in the Smart Cities and \gls{ITS} domains, as well as previous related works using social media content as its basis.
The proposed framework is referenced in Chapter~\ref{chap:framework}, being each its composing modules depth described.
Experiments performed to test each module of the framework are reported in Chapter~\ref{chap:experiments}.
We end the document with Chapter~\ref{chap:conclusions} where conclusions, future work and a few final remarks are exposed.
\chapter{Background and Literature Review} \label{chap:sota}

\minitoc \mtcskip \noindent

This section aims the analysis and reflection about some works that has as final goal, similarly to ours, the development of a framework with the purpose of exploring social media data to extract meaningful domain-specific information. Nonetheless, studying works from other authors may help or even find already proposed solutions in order to solve the aforementioned problems.

Hence, this section will contemplate a brief contextualization about how can an intelligent system contribute to the improvement of a city or transportation services. Moreover, technologies and methods that allow extraction of information from a text document or, in this particular case, from tweets will be described. Finally, an exploration through already existent frameworks regarding the information extraction from social media content as well as the identification of its application domain.

\section{Smart Cities}\label{sec:smart_cities}

Smart City is a concept appeared thanks to the continuous growth of a city's population which contributed to an aggressive level of urban and technological developments~\cite{ulloa2016mining}. In the last few years, several definitions for its meaning have emerged but its main idealization is not yet fully known~\cite{komninos2009intelligent}. Angelidou~\cite{angelidou2015smart} defined Smart City as

\emph{"Conceptual urban development model on the basis of the utilization of human, collective, and technological capital for the development of urban agglomerations"},

enhancing \textit{knowledge} and \textit{innovation economy} as the primary factors that support the development of a city. Alongside with the previous factors, the author identifies other three distinct forces that shape the concept of a Smart City:

\begin{enumerate}
	\item \textit{Technology Push}: The need of new products and solutions are introduced into the market due to a fast advance in science and technology.
	\item \textit{Demand Pull}: Current problems are solved originating new possibilities to respond society demands such as the continuous growth of the population.
	\item \textit{Urban Future}: Represents the final goal of a city constituting for that reason an important role in the whole transformation process.
	\item \textit{Knowledge and Innovation Economy}: The creation of new products using the most recent technologies is associated to solution for the efficiency and sustainability of a city.
\end{enumerate}

\begin{figure}
	\centering
	\includegraphics[width=0.6\textwidth]{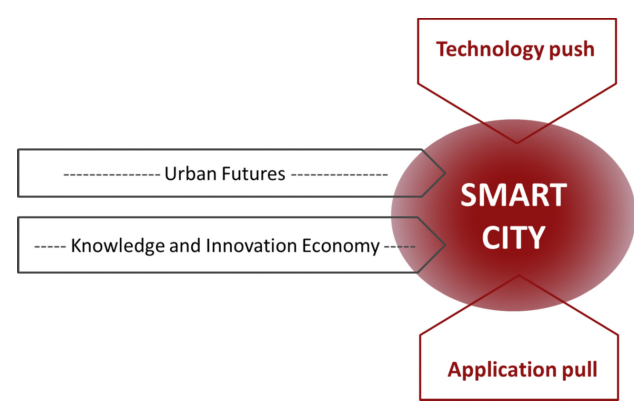}
	\caption{\textit{Smart City} conjecture of four forces. Source:~\cite{angelidou2015smart}}
	\label{fig:four_forces}
\end{figure}

The first two forces previous mentioned are directly dependent of the other ones as it is showed in Figure~\ref{fig:four_forces}. However, the absence of desire to reach a better future having into consideration the city's economy and resources can result in the break of its dynamics and healthy, affecting services of a city due to the population discontentment.

The development environment of a city tagged as \textit{smart} is another key factor to reach the success. Komninos~\cite{komninos2009intelligent} associates collective sources of innovation to the improvement of life quality in cities. The globalization of innovation networks is responsible for the emergency of another types of environments and infrastructures, as so \emph{"global innovation clusters and i-hubs, intelligent agglomerations, intelligent technology districts and intelligent clusters, living labs"} allowing the testing of products or services by the ordinary citizens in order to identify problems or even analyse their behaviour  and reactions regarding what have experimented~\cite{komninos2009intelligent}. Hence, it is possible to affirm that the development of a city has its starting point in the community but also depends on the quality of \glspl{ICT}~\cite{hollands2008will}, an essential requirement in the city's evolution process.

Last but not least, a Smart City may focus its efforts in several sectors, such as the environment, culture and recreation, education, social and economic aspects, demography, and travel and transportation~\cite{caragliu2011smart} in order to have equally advances in all of them.

\section{Intelligent Transportation Systems}\label{sec:intelligent_transportation_systems}

The transportation system is inherently connected to the progress of a city because people uses on a daily-basis transportation modes, i.e. bus, private cars, metropolitan, and others, in order to go to their jobs and make their own life and through that they contribute to the economic progress of it. Although this connection, such system is also influenced by the problem of population growth being relevant and necessary the finding of solutions to minimize or even erase it~\cite{caragliu2015smart}. Hence, "a \textit{smart city} should be focused on its actions to become smart", coming up the concept of innovation~\cite{ulloa2016mining}.

To understand what are \gls{ITS}, it is necessary to introduce the meaning of \gls{SM}. \gls{SM} is a combination of comprehensive and smarter traffic service with smart technology, enabling several intelligent traffic systems which provide control in the signals regarding the traffic volume, information about smooth traffic flows, times of bus, train, subway and flight arrivals, their routes or even the knowledge of what citizens thought about the city's services~\cite{chun2015review}. The majority of \gls{ITS} are expressed through smart applications where transportation and traffic management has became more efficient and practicable, allowing the users to access important information about the transportation systems in order to make correct decisions about what they want to use in their cities~\cite{caragliu2015smart}. \gls{ICT}-based infrastructures are the main support for \textit{smart cities} and due to that, they also serve as support to \gls{ITS}, since the information provide by such infrastructures allows the piloting of activities such as traffic operations, as well as its  management over a long period of time~\cite{ulloa2016mining}.

The \gls{ITS} research community have made multiple advances through the last years in the prediction of traffic flow using GPS-traced mechanisms~\cite{sandim2016using} and bluetooth~\cite{filgueiras2014sensing}. Besides these real-time traffic prediction methods~\cite{barros2015short}, the exploration of social media in this area is still in a maturing process. This kind of data allows new ways to study human behaviour, such as the mood and reactions of people, and smart mobility allowing the usage of new techniques instead of conventional ones~\cite{liu2013understanding}.

Nowadays, cities are exploring some initiatives of sensing to support the development of technological projects. Areas such as utilities management (where, for example, is monitored the consumption level of power, water and gas), traffic management (using vibration sensors to measure the traffic flows on bridges, or even the full capacity of a parking lot), environment awareness (using video cameras to monitor the population behaviour and sensors to measure the level of air pollution) make use of physical sensors, i.e. some devices that can capture information to study and improve the quality of life in a daily basis~\cite{doran2015social}. Szabo et al.~\cite{szabo2013framework} and Doran et al.~\cite{doran2015social} reported the highly economic cost to this kind of sensing, since it is require maintenance and replacement of this devices, as well as a tracking infrastructure to store and process the collected information. Hence, a new form of sensing has emerged - \gls{Crowdsensing} - offering to the cities several ways to improve their services by exploring the participation of the citizens through social networks where there is a publicly sharing of  opinions and thoughts regarding some problems around the city where they live are passing in~\cite{roitman2012harnessing}. This type of sensing consists in human-generated data provided by the population through the usage of mobile devices and social networks web-based platforms. Such data can be further used to extract some analytics regarding specific services in a city, namely the urban transportation system~\cite{roitman2012harnessing}. Having this considered, social media can be seen as a good source of data to extract valuable information aiming the direct use of it into the smartness evolution process of a city~\cite{szabo2013framework}. Recently, it is possible to verify that cities are increasingly opting for technological opportunities based on \textit{crowd sensing}, once this type of exploration brings a considerable reduction of costs and support in the development of news valuable technologies.
witter
In the last few years, several authors have published a widely range of social-media-based contributions focusing this specific domain. Kurkcu et al.~\cite{kurkcu2016evaluating} use geo-located tweets to try and discover human mobility and activity patterns. The subject of transport modes was explored by Maghrebi et al.~\cite{maghrebi2016transportation} in the city of Melbourne, Australia. From a dataset of 300,000 geo-located tweets, authors tried to extract tweets related to several modes of transport using a keyword-based search method. 

Additionally, there were also different efforts focused on the tracking of accidents using Twitter social media data. Mai and Hranac~\cite{mai2013twitter} tried to establish a correlation between the California Highway Patrol incident reports and the increased volume of tweets posted at the time they were reported. On the other hand, Rebelo et al.~\cite{rebelo2015twitterjam} implemented a system capable of extract and analyse events related to road traffic, coined  TwitterJam. In that study, authors also used geo-located tweets that were already confirmed as being related to events on the roads and compared their counts with official sources.

Performing robustness experiments over this domain is challeging since although the large number of recently publications, gold standards are yet not defined or even public being for this reason difficult to prove the methodology chosen or suppositions made. Maghrebi et al.~\cite{maghrebi2016transportation} enhances some terms related to the transportation domain, however they are limited and also very common ones. After a tough investigation work, it is worth noting a list produced by Gal-Tzur et al.~\cite{gal2014potential} containing a large number of terms whose may serve as support for new  scientific contributions using social media in studies of the transportation domain.

\section{Social Media Analytics}

In the last few years social networks have made impact on the business communications since users assumed the role of costumers through the publication of content on these networks, rising high levels of interaction between them, as well as with businesses entities~\cite{ulloa2016mining}. A proof of that is the amount of information produced since 2011 which is equivalent to a number over than 90\% of the available data online~\cite{sintef2013bigdata}. Facebook\footnote{\url{https://www.facebook.com/}}, Twitter\footnote{\url{https://twitter.com/}} and other social networking websites are nowadays used as business tools by companies aiming the efficient use of digital marketing techniques to publicize their products~\cite{royle2014digital}. Besides the business field, the population turn widely into this new communication technologies publicly sharing real-life events, their opinions about certain topics and their on-time feelings in the network through a simple message \cite{dandrea2015realtime}.

\gls{SMA} can be described as a type of digital analytics which focus is the study of interactions between, their opinions/thoughts, their own life, companies as so its products or services through the social media data. Such study provides important information to "analysts, brands, agencies or vendors" facilitating the generation of economic value to many organizations~\cite{judah2012social}. In order to achieve the main goal of the \gls{SMA}, companies focus their effort in the development automatic systems to make possible an easy collection, analysis, summarization and visualization of processed social media data establishing specific points about what is necessary to improved in their products~\cite{zeng2010social}.

However the potential value that \gls{SMA} can provide, Phillips~\cite{judah2012social} enhance some important factors to be considered in the analytics process: (1) Users permissions; (2) Awareness/listening of real-time information; (3) Search mechanisms; (4) Text analysis methodologies and techniques; (5) Data access and integration; (6) System integration, customization and growth.

The previous mentioned factors will help during the identification and comprehension of possible necessary features in a social media analytics tool, as well as to establish potential parameters/metrics to test and evaluate such tool. Without careful conduction in the social media tool elaboration, for instance, use of a wrong technique of \gls{SMA} could have a bad business impact for the company resulting possible bankruptcies and increase the unemployment tax of a city.


\section{Text Mining}

Text mining comprises a richer set of fields such as information retrieval, data mining, machine learning, statistics and computational linguistics which aims the extraction of valuable information from unstructured textual data~\cite{he2013social}. The intensively usage of this analysis methodology is due to the massive amount of information stored in text documents being necessary automatic techniques to identify, extract, manage and integrate the knowledge acquired from these texts exploration in a efficiently and systematically way~\cite{ananiadou2015textmining}. On the other hand, the emergency of social media applications have also contributed to the widely growth of text mining usage because of the "application’s perspective and the associated unique technical and social science challenges and opportunities"~\cite{zeng2010social}.

Text mining shares some of the issues presented by the \gls{NLP} field. Texts are performed by humans and due to that, some problems in its construction can appear, such as spelling mistakes, wrong phrasal construction, slang among other. Before the mining process of a text, it is important to apply some preprocessing steps in order to eliminate or, at least reduce, undesired content (words) in the primary analysis process.
Stavrianou et al.~\cite{stavrianou2007overview} cite these issues very well alongside their work and some of them are observable in Table~\ref{table:textminingissues}.

\begin{table}[htbp]
	\centering
	\caption{Text mining issues by Stavrianou \cite{stavrianou2007overview}}
	\label{table:textminingissues}
	\begin{tabular}{ | l | p{7cm} |}
		\hline \textbf{Issue}            & \textbf{Details}\\
		\hline Stop list                 & Should we take into account stop words?\\ 
		\hline Stemming                  & Should we reduce the words to their stems?\\ 
		\hline Noisy Data                & Should the text be clear of noisy data?\\ 
		\hline Word Sense Disambiguation & Should we clarify the meaning of words in a text?\\ 
		\hline Part-of-speech Tagging    & What about data annotation and/or part of speech characteristics?\\ 
		\hline Collocations              & What about compound or technical terms?\\ 
		\hline Grammar / Syntax          & Should we make a syntatic or grammatical analysis? What about data dependency, anaphoric problems or scope ambiguity?\\ 
		\hline Tokenization              & Should we tokenize by words or phrases and if so, how?\\ 
		\hline Text Representation       & Which terms are important? Words or phrases? Nouns or adjectives? Which text model should we use? What about word order, context, and background knowledge? \\ 
		\hline Automated Learning        & Should we use categorization? Which similarity measures should be applied? \\ \hline
	\end{tabular}
\end{table}

The removal of words from text may sometimes not be desirable because some sentences can lose its information or even leads to a different meaning compared with its original form. The generation of a stop list words should be a supervised task as long as little words could induce distinct results in the text classification~\cite{riloff1995little}.

Stemming is a task that depends mostly from the speaking language of the text than its specific domain~\cite{stavrianou2007overview}. The main goal of this technique is to reduce a word to its root form helping in the calculus of distances between texts, keywords or phrases, or even in the text representation.

The noisy data is derived from spelling mistakes, acronyms and abbreviations in texts and to solve this, a conversion of these terms should be done to maintain the integrity of data. Commonly solution approaches involve text edit distances (Levenshtein Distance\footnote{\url{https://en.wikipedia.org/wiki/Levenshtein_distance}}) and phonetic distances measures between known words and the misspelling ones to achieve good corrections~\cite{bontcheva2013twitie} 

\gls{WSD} is focused on solving the ambiguity in the meaning of a word. Other similar field to \gls{WSD} is \gls{NED} where the disambiguation target are named-entities mentions, while \gls{WSD} focus on common words. WordNet\footnote{\url{https://wordnet.princeton.edu/}} is a commonly used resource to extinguish this ambiguity \cite{chang2016comparison}. There are two types of disambiguation: the unsupervised, where the task is support by a dictionary or a thesaurus \cite{stavrianou2007overview}; and, the supervised one, where different meanings of a word are unknown and normally learning algorithms with training examples are used to achieve good results regarding the performance of the disambiguation task~\cite{yarowsky1995unsupervised}.

Tagging can be describe as the process of labeling each term of the text with a part-of-speech tag, i.e. classify each word as a noun, verb, adjective, and others \cite{hotho2005brief}. Collocations are groups and constitutes a very important step in some text mining approaches. Grouping two or more words to give its correct meaning is sometimes crucial to perform tasks such as sentiment analysis where negations (e.g. "don't like") needed to be composed by two or more words in order to assure the negative value of, for example, a verb. Collocations are usually made before the \gls{WSD} task since some compound technical terms have different meaning from the individual words which composed it \cite{stavrianou2007overview}.

Tokenization serves to pick up all the terms presented in a text document and to achieve this it's necessary splitting its content into a stream of words implying the removal of the punctuation marks and non-text characters \cite{hotho2005brief}. Some authors also see tokenization as a text representation form since one of the most used models to represent texts is \gls{BoW}. This model broke down texts into words and stores it in a term-frequency vector according the occurrence of a word in the text. Hence, each word may represent a feature \cite{sriram2010short}. Another commonly used model to represent texts is Vector Space Models that represent all the documents in a multi-dimensional space where documents are converted to vectors and each vector may be seen as a feature. This model provides some advantages since the documents can be compared with each other by performing some specific vector operations \cite{hotho2005brief}.

Once been introduced some of the most preliminary important steps in text mining, the remainder subsection are focused in two different text analytics approaches: topic modelling and text classification. The majority of \gls{SMA} approaches focus its efforts in modelling and classification tasks in order to understand the large range of data collected and support commonly used techniques to extract information from it, such as sentiment analysis, trend analysis and topic modeling~\cite{fan2013unveiling}.

\subsection{Topic Modelling}
\label{subsec:topic_modelling_sota}

Topic modelling is a text mining unsupervised technique/method aiming the identification of similarities in unlabeled texts. Usually, this technique is applied over texts of large volume since to correctly identify the resulting patterns in its content requires the existence of lots of information.

One of the first studies made using Twitter data was proposed by Kwak et al.~\cite{kwak2010twitter} and consisted in the collection of messages to classify the trends in its content. Results showed that almost 80\% of the trends in Twitter are related to real-time news and the period in which each trend maintains itself in the top is limited. The authors proved that Twitter can be seen as a mirror of real-time occurring events/incidents in the world.

Several works were already proposed to identify social patterns in the population daily-basis life and mapping such patterns geographically by topic modelling techniques to discover latent topics in social media streams. Usually, studies about topic modelling, in particular \gls{LDA} model, to text mining problems follow unsupervised approaches~\cite{lansley2016geography,oliveira2016sentibubbles} - where is not required the creation of a training dataset. Others improved the model and made it an supervised approach~\cite{ramage2010characterizing}, dependent of training data, and compare to the traditional one in order to prove better results.

Using entity-centric aggregations and topic modelling techniques, Oliveira et al.~\cite{oliveira2016sentibubbles} built a system focused in data visualization that allows an user to search for an entity during a specific period and shows which are the main topics identified in the Twitter messages.  Ordinary weekday patterns were identified by Lansley and Longley~\cite{lansley2016geography} in their study regarding the inner region of London. The authors used a \gls{LDA} model to distribute 20 topics over 1.3M tweets. After crossing the results of the experiment with land-uses datasets it was possible to observe interesting patterns in specific zones and places of the British city. Nonetheless, Ramage et al.~\cite{ramage2010characterizing} improved a \gls{LDA} model by adding a supervised layer that automatic label each tweet used in their experiment.

\gls{CRF} are explored by Nikfarjam et. al~\cite{nikfarjam2015pharmacovigilance} which have applied word embeddings in combination to other text features, such as adverse drug reactions lexicons, \gls{POS} and negation collocations in order to train a supervised model. Such model was able to demonstrate high performances on the extraction of concepts/topics from the social media user-generated content. To prove robustness and efficiency in the model, authors have compared the obtained results with DailyStrength corpus and were able to notice that due to the limited size of text in a tweet, the detection of different reactions about drugs is more complex, which could be simplified with access of greater amount of information provided in the training process of the model.

Differently from the majority of works involving topic modelling techniques, Tuarob and Tucker~\cite{tuarob2015quantifying} take support of a \gls{LDA} model to extract the most frequent words for groups of tweets previously collected. The overall work is focused in sentiment analysis approaches and aims the perception of what people fells about a specific product as well as its composing features. Authors used the \gls{LDA} model to find what were the main 2 topics present in each product set of tweets and considered the most frequent 30 words. Moreover, POS tagging, disambiguation and stemming techniques were used in order to filter out and normalized words related to the product. Finally, an unsupervised method to calculate the sentiment polarity was applied to the data being final results coherent to the product feature/aspect extracted.

Topic modeling techniques consisting in supervised learning approaches were explored by Zhang et al.~\cite{zhang2016mining}, where authors have compared the results obtained from a SVM classification for accident-related tweets with a classification using a two-topic generative model SLDA (Supervised \gls{LDA}). Contrarily to the unsupervised method, this one takes into consideration the label assigned to the training examples and can be trained as a genuine classification model. By comparing the final results between both models, it is possible to observe a significative increase of the precision and a decrease of only 0.04 points in the accuracy meaning that supervised topic modelling techniques to binary classification may compete well with conventional classification models, with respect to tweets.

\begin{table}[htbp]
	\centering
	\caption{Brief overview of the related work for topic modelling}
	\label{tab:topic_related_work}
	\resizebox{\textwidth}{!}{\begin{tabular}{c|c|c|c|c}
			\hline
			\textbf{Approach} & \textbf{Features} & \textbf{Methods} & \textbf{Goal} & \textbf{Potential Domain}\\ \hline
			Kwak et al.~\cite{lansley2016geography} & Twitter metadata & \begin{tabular}[c]{@{}c@{}} Aggregation of trending topics using \\external information \end{tabular} & \begin{tabular}[c]{@{}c@{}} Quantitative study in order to reveal Twitter\\ as both social media and news media platform \end{tabular} & Smart City \\ \hline
			
			Oliveira et al.~\cite{oliveira2016sentibubbles} & specific-entity words & \begin{tabular}[c]{@{}c@{}}Unsupervised Latent Dirichlet \\ Allocation\end{tabular} & \begin{tabular}[c]{@{}c@{}} Extract the most relevant entity-related topics \end{tabular} & Smart City \\ \hline
			
			Lansley and Longley~\cite{lansley2016geography} & Bag-of-words & \begin{tabular}[c]{@{}c@{}}Unsupervised Latent Dirichlet \\ Allocation\end{tabular} & \begin{tabular}[c]{@{}c@{}} Study social dynamics of London using Twitter topics\end{tabular} & Smart City \\ \hline
			
			Tuarob and Tucker~\cite{tuarob2015quantifying} & Bag-of-words & \begin{tabular}[c]{@{}c@{}}Unsupervised Latent Dirichlet \\ Allocation\end{tabular} & \begin{tabular}[c]{@{}c@{}} Extraction of people's polarization sentiment about a\\ specific feature of a product (aspect sentiment analysis)\end{tabular} & Smart City - Economy \\ \hline
			
			Ramage et al.~\cite{lansley2016geography} & Labeled bag-of-words & \begin{tabular}[c]{@{}c@{}}Supervised Latent Dirichlet \\ Allocation\end{tabular} & \begin{tabular}[c]{@{}c@{}} Proving the applicability of supervised approaches\\ in conventional \gls{LDA} model \end{tabular} & Smart City \\ \hline
			
			Zhang et al.~\cite{zhang2016mining} & Labeled bag-of-words & \begin{tabular}[c]{@{}c@{}}Supervised Latent Dirichlet \\ Allocation~\cite{mcauliffe2008supervised}\end{tabular} & \begin{tabular}[c]{@{}c@{}} Comparing performances with SVMs models \\ to accident-related tweets \end{tabular} & \begin{tabular}[c]{@{}c@{}}Smart City - Travel and \\Transportation\end{tabular} \\ \hline
			
			Nikfarjam et. al~\cite{nikfarjam2015pharmacovigilance} &  \begin{tabular}[c]{@{}c@{}}ADR Lexicons, POS Tagging \\ Negation, Word Embeddings\end{tabular} & CRF & 
			\begin{tabular}[c]{@{}c@{}} Discrimination of adverse drug reactions in tweets\\ content\end{tabular} & Smart City - Health \\ \hline
		\end{tabular}}
	\end{table}
	
Probabilistic topic models, such as \gls{LDA}, are the most used techniques in topic detection tasks. Although high applicability, authors question themselves regarding the performance of this technique over social media data which present limitations, starting at the size of the message and ending in the bad phrasal construction and informality~\cite{mehrotra2013improving}. In this dissertation we will tackle this question and try to answer it by presenting results obtained in a real-world scenario experiments.

\subsection{Text Classification}

Text classification is a text mining task which main goal is the discrimination or characterization of a piece of text into a specific format value. Such value can vary from number (sentiment analysis tasks), labels (multi-labeling tasks), classes (binary or multi-class tasks). Classification in text analysis is a widely used methodology and had already been reported in several scientic contributions regarding the smart cities and transportation domains.

\gls{SVM}, \gls{OLS}, \gls{RF}, \gls{MLP}, \gls{NB} and \gls{DT J48} are some of the supervised classification models used to analyse social media data over fields such as health~\cite{signorini2011use} and pharmacovigilance, political opinion~\cite{saleiro2016sentiment}, transportation (travel classification~\cite{carvalho2010real, kuflik2017automating}, traffic and incidents detection~\cite{zhang2016mining}), financial sentiment analysis~\cite{saleiro2017feup} and \textit{online} reputation monitoring~\cite{saleiro2017texrep}.

Sifnorini et al.~\cite{signorini2011use} reported a study which main goal was the tracking of the disease \gls{Influenza_A} virus. Tweets collected by the authors using term-based search sum up more than 300 million examples. Their methodology consists in training \gls{SVM} models with sets of frequency features composed by the most used weekly-terms over the whole dataset. Each model was specifically trained according a certain set of keywords and follow an iterative process, i.e. authors firstly have classified all illness-related tweets related and than used the resulting related subset of data to perform new classification regarding specific keywords, such as what was the disease source, countermeasures used and infected people characteristics. Final results allowed the verification of a decrease of Twitter activity while more new cases were appearing meaning less concerning about this epidemic through time.

Accident-related classification for Twitter data was proposed by Zhang et al.~\cite{zhang2016mining}. Authors explored the Twitter Streaming API to collect geo-located tweets from Northern Virginia during a completed year, January to December of 2014, and recurring to auxiliary loop detectors that are, in intervals of 15 minutes, recording the traffic flow. In order to automatize the detection of accidents in that interval of time (were the sensors are not recording the scene), authors have built a binary classification model using Linear \gls{SVM} with a balanced dataset composed by 400 training examples for each of the accident-related and non-related classes composed by a boolean-vectors according the final 3,000 tokens resulted from the token filtering and stemming process. Performance was improved by submitting the model to a 5-fold cross validation which was proved by values of accuracy and precision over than 70\% of success.


Considering the task of discriminate travel-related tweets, Carvalho et al.~\cite{carvalho2010real, kokkinogenis2015mobility} have constructed a bag-of-words dependent classification model and achieved improvements at the model's performance with support of a bootstrapping approach implying a two phases train to the \gls{SVM} model.  By assuming the similarities, i.e. all four works were related to binary text classifications, we can induce an hypotheses that Linear \gls{SVM} models have superior performances relatively to other models for this type of classification tasks.

Multi-class classification models were also applied to the transportation domain through text analysis of social media content. Kufliket et al.~\cite{kuflik2017automating} build multiple classification models using methods such as \gls{NB} and \gls{DT J48} to predict multiple modes of transport during three different sports events. Tweets sum up a total of 3.7M and were submitted to the models classification task in order to prove that an harvesting automatically information from \gls{SMC} is possible and may help transportation entities in the planning and management of their services during social occasions as it is demonstrate in theirs use cases.

On the other hand, Saleiro et al.~\cite{saleiro2016sentiment} tried to predict the 2011 Portuguese bailout results analysing opinion within the tweets about all five political parties candidates. The opinion was measure using a \gls{OLS} model trained with specific sentiment aggregate functions and proved to be capable of correctly predict who would be elected prime minister of Portugal only exploring sentiment analysis in social media data. In SemEval-2017 Task 5, Saleiro et al.~\cite{saleiro2017feup} explored word embeddings techniques to extract the sentiment polarity and intensity in financial-related tweets. Authors have proved good performance of models trained with bag-of-words and bag-of-embeddings features together although the approach been applied to a specific domain. The usage of features representing syntactic and semantic similarities of texts, such as word embeddings, can be seen with great potential namely to the area of travel-related text classification.

\begin{table}[htbp]
	\centering
	\caption{Brief overview of the related work for text classification - Best Experiments}
	\label{my-label}
	\resizebox{\textwidth}{!}{\begin{tabular}{c|c|c|c|c}
			\hline
			\textbf{Approach} & \textbf{Features} & \textbf{Classification Methods} & \textbf{Goal} & \textbf{Potential Domain}\\ \hline
			Sifnorini et al.~\cite{signorini2011use} & Bag-of-words & Linear SVM & \begin{tabular}[c]{@{}c@{}}Tracking the evolution of public sentiment\\ and increasing of social media activity about\\ the H1N1 pandemic\end{tabular} & Smart City - Health \\ \hline
			
			Zhang et al.~\cite{zhang2016mining} & \begin{tabular}[c]{@{}c@{}}Boolean vectors matrix \\ (3,000 different tokens)\end{tabular} & Linear SVM & \begin{tabular}[c]{@{}c@{}}Improve transportation control by automatic \\discriminate accident-related tweets \end{tabular} & \begin{tabular}[c]{@{}c@{}}Smart City - Travel and \\Transportation\end{tabular} \\ \hline
			
			
			Kuflik et al.~\cite{kuflik2017automating} & Bag-of-words &  \begin{tabular}[c]{@{}c@{}} Naïve Bayes, \\ DT J48\end{tabular} & \begin{tabular}[c]{@{}c@{}}Multi-class mode of transport classification and the \\ purpose behind it\end{tabular} & \begin{tabular}[c]{@{}c@{}}Smart City - Travel and \\Transportation\end{tabular} \\ \hline
			
			Carvalho et al.~\cite{carvalho2010real} & Bag-of-words & \begin{tabular}[c]{@{}c@{}}Linear SVM with \\ Bootstrapping\end{tabular} & Discrimination of travel-related tweets & \begin{tabular}[c]{@{}c@{}}Smart City - Travel and \\Transportation\end{tabular}\\ \hline
			
			Saleiro et al.~\cite{saleiro2016sentiment} & \begin{tabular}[c]{@{}c@{}} Sentiment Aggregate\\ Functions \end{tabular} & OLS & \begin{tabular}[c]{@{}c@{}} Predicting Portuguese polls results through \\ opinion mining \end{tabular} & Smart Cities - Government \\ \hline
			
			Saleiro et al.~\cite{saleiro2017feup} & \begin{tabular}[c]{@{}c@{}}Word Embeddings,\\ Bag-of-words,\\ domain-specific lexicons\end{tabular} & RF & \begin{tabular}[c]{@{}c@{}} Extraction of sentiment polarity and intensity from social\\  media content and web news \end{tabular} &  Smart City - Economy \\ \hline
			
		\end{tabular}}
	\end{table}

\medskip

There is a wide diversity in text classification approaches. A worth noting fact in this review at the literature is that word embeddings have been supporting conventional techniques in order to improve performances in text classification tasks. Transportation domain lacks in studies having this particular feature in the training process of its classification models. Hence, it is of major importance perform experiments about this domain aiming conclusions and additional content to support the potential advantages brought by word embeddings.

\subsubsection{Classification Evaluation Metrics}
\label{subsubsec:evaluation_metrics}
In order to measure the performance of a text classification model, there are several types of metrics that can help this process, depending of course the context of the task. Regarding binary classification tasks, the most common evaluation metrics used are precision, recall (sensitivity) and F1-score which is the harmonic mean or the weighted average of the previous two. Therefore, it is described each of these metrics as well the mathematical equation used in its calculation.

\begin{itemize}
	\item \textbf{Precison:} Represents the fraction of correct predictions for the travel-related class (Equation~\ref{eq:precision}).
	
	\item \textbf{Recall:} Represents the fraction of travel-related tweets correctly predicted (Equation~\ref{eq:recall}).
	\begin{multicols}{2}
		\begin{equation}\label{eq:precision}
		Precision = \frac{tp}{tp+fp}
		\end{equation}
		
		\begin{equation}\label{eq:recall}
		Recall = \frac{tp}{tp+fn}
		\end{equation}
		\end{multicols}
		
		where \textbf{$tp$} is related to the true positives classified tweets, \textbf{$fp$} represents the false positives and \textbf{$fn$} are the false negatives.
		
		\item \textbf{F1-score:} Represents the harmonic mean of precision and recall.
		\end{itemize}
		
		\begin{equation}
		{F1}_{score} = 2*\frac{precision*recall}{precision+recall}
		\end{equation}
		
These first three metrics only showed us the performance of the classifier for a discrimination threshold of 0.5. The \gls{ROC} curve gives us the \gls{TPR} and the \gls{FPR} for all possible variations of the discrimination threshold. Through the \gls{ROC} curve, it is possible to compute the \gls{AUC} to see what was the probability of the classifier to rank a random positive higher than a random negative one.

\section{Related Social Media Frameworks}

In the last few years, the number of proposals of frameworks to treat social media content and produce valuable information to the end-users has widely increased. For instance, each framework has it own domain of application and generalization is not the center focus. Event detection, \textit{online} reputation monitoring, socio-semantic analysis to human reactions and traffic sensing are some of the application domains that research community present their contribute through framework proposals.

Liu et al. \cite{liu2012using} have made a study in three different transportation modes (private cars, public transportations and bicyclists) using theirs channels on Twitter to estimate a percentage of the majority gender that uses this services in the city of Toronto. They have extracted all the channel's tweets appealing only to the \textit{non-protected} followers and applied an already developed classification model to label each tweet with its creator gender: male or female. Author decided to implement a system that produce automatically analysis since they have find interesting results in the experiment conducted.

Regarding the field of event/incident detection, Abel et al~\cite{abel2012twitcident} developed Twitcident, a real life accidents-aware web-based framework that is connected to a emergency broadcast system in order to detect incidents across the world. Then, an automatically system starts the collection and filtering of content from social media platforms and extracts information about entities using Named Entity Recognition and Disambiguation techniques. Data temporal distributions are also produced to analyse the time line of the events.

Anastasi et al. \cite{anastasi2013urban} proposed a framework which objective was the promotion of flexible transportation systems usage, i.e. encouraging people to share transport or to opt for the use of bicycles in order to minimize infrastructural and environmental problems. Their tool takes advantages of the crowd sensing techniques by exploring social media streams to predict accidents or traffic congestion and alert the users of their service about this type of events.

Ludwig et al. \cite{ludwig2015crowdmonitor} proposed a tool capable of collect and display social media streams to help the integration and coordination of volunteers in emergency services to prevent engagement in dangerous areas. Their tool present to the end-users map visualization of a city where they could identify public calls of the emergency services to accept or deny them.

Traffic sensing over the city of Rio de Janeiro, Brazil, was studied by Rebelo et al.~\cite{rebelo2015twitterjam} which have implemented a system capable of extract and analyse events related to road traffic, coined  TwitterJam. In that study, authors used geo-located tweets that were already confirmed as being related to events on the roads and compared their counts with official sources. Finally but not least, authors depicted geographic visualizations to the end-users in order to understand what is the current traffic-state of a certain road.

Social Media is used by Ludwig et al.~\cite{ludwig2015crowdmonitor}, in a framework that attempts the creation of voluntary and emergency activities, coined CrowdMonitor. The systems allows through the analyse of human mobality through tweets posted in the platform. Although absence of text analysis methodologies, such system intents to promote more cooperation between citizens and also promotes the applicability of crowd sensing, a crucial factor for the smartness evolution of a city.

Technological companies is the main target of the framework proposed by Lippizzi et al.~\cite{lipizzi2015extracting}. The system analyses social media content having in consideration specific products, such as mobile phones, tablets and others, and tries to extract information of what their customers talk about it. By measuring the sentiment of word clusters produced by the system, companies may take profit and additional insights about what in needed to be improved in their products.

CrowdPulse is a domain-agnostic framework proposed by Musto et al.~\cite{musto2015crowdpulse} which main objective is the presentation of text analytics to the end-users. Such framework is rich regarding implemented text methods, which range from entities disambiguation to sentiment analysis. Authors followed unsupervised approaches to implement all the framework composing methods, and applied the resulting system in two real-world scenarios, the earthquake of L'Aquila city and The Italian Hate Map. Further analysis of the results proved that simple techniques can provide faster insights about people sentiment regarding any type of domain.

POPMine~\cite{saleiro2015popmine} is a framework capable of performing real-time analysis based on entity filtering and sentiment analysis techniques for the political domain. Later on, Saleiro et al.~\cite{saleiro2017texrep} extend this tool to a large group of scenarios and present a full-based text mining framework for \textit{online} reputation monitoring that explores and extracts multiple types of information from a wide range of Web sources. TextRep is divided in several modules in order to perform correctly the different text mining techniques, such as the collection of data, disambiguation and sentiment analysis. The system is adaptable to different domains as well and applications of it to political opinion mining and financial sentiment analysis are two of the use cases presented by the authors.

\section{Summary}

The literature review shows positives and negatives points that are necessary to be reported. First, the conceptualization of a meritorious system capable of bringing value to the smartness evolution of a city is a labourious and time-consuming process. Although iterative steps, it is necessary the stipulation of a detailed work-plan and what are/is indeed the final target/s and objectives of such system. Crowd sensing is a type of sensing that enables the study of what citizens say about a specific topic, and social media platforms can easily be explored in order to take its content to futher analysis and support the construction of a adaptable and profitable tool for the city's entities. Nowadays, text mining techniques allows the extraction of information from social media content, which can be represented, after accurate aggregations on the results, in visualization views facilitating analysis by the end-users of these systems. Last but not least, we could identify two unexplored approaches in this literature. Word embedding is a technique which has not been applied to transportation domain using social media content. Domain-agnostic frameworks using supervised learning methods are an hard task regarding its conception, however, due to the learning phase, models could learn new similarities from the text, and we see potential in this approach since it is not necessary construction of auxiliar dictionaries to perform the desired tasks.
\chapter{Framework}
\label{chap:framework}

\minitoc \mtcskip \noindent

In this chapter we describe the details and specificities of the framework proposed in this dissertation. First, we enunciate the necessary requirements to fulfill and achieve the mentioned development. Moreover, it is present the framework architecture design, as well as it inner pipeline. The modules that constitutes such architecture are described afterwards as so the required methodologies and algorithms incorporated in each of its tasks. Finally but not least, we mention and explain the different data visualizations available in the framework.

\section{Requirements}\label{sec:requirements}

The development of frameworks to the domain of \textit{smart cities} and intelligent transportation systems using human-generated content (e.g. text messages) is a laborious and time-consuming process. The source of the data to fed such system is one of the biggest challenges in this kind of developments, ranging from social media, smart phones and urban sensors. In this dissertation we tackle the problem of exploring social media data since this kind of data have, recently, been seen as a new opportunity and source to mine valuable information to the cities services and corresponding responsible entities~\cite{musto2015crowdpulse}.

Social media data is mostly represented by text messages being necessary the application of \gls{NLP} methodologies in order to extract information from its content. Such methodologies are usually complex and composed by several different steps (e.g. some related to the syntax of the sentences while others are related to the semantics of its content) before the achievement of the desired results. Social media streams are no exception, indeed, the analysis of such texts is even more complex since messages are usually short and present lots of informal characteristics.

A framework for the domain of social media content requires, in the first place, a data collection module. Depending on the social network, the data collection module can have different heuristics with respect to the data retrieving. Here, the choice of such heuristics is important and needs to be made according the final users expectations, or at least, according the framework final use case. Towards the application of \gls{NLP} techniques, a module in charge of preprocessing tasks is required. The main purpose of this module establishes in the performance and robustness of the results obtained by the previously mentioned techniques. \gls{NLP} techniques can provide different types of information, however in this dissertation the focus is on the classification of travel-related tweets and characterization of the topic associated with a tweet. Each technique is represented as an independent module whose belongs to the boundary of text analytics. This framework needs also to be capable of processing information regarding the creation date of a tweet, \textit{metadata} and geographic distribution associated to it. For the fast retrieving of this informations to the data visualization view, some aggregations need to be made. This requirement is due to one of the big data demands, the instantly availability of the results. Such demand is important for the framework end-users since it helps in the entities' decision-making process making easier and faster the improvement of its services.

The construction of this complex system requires careful planning since there are dependency between a task and the one that follows it, at least with respect to the filtering and preprocessing of data. Adaptability to different languages is considered and further addiction of new ones may be possible. For the same reason, but this time regarding new functionalities, the framework needs to follow a modular architecture allowing new text analysis layers as well as other type of data visualizations. The domain of \textit{smart cities} is vast in terms of indicators and fields that constituting it. For this reason, the final architecture may be designed in a way that allows configuration about the user's field of interest, if he do not desire analytics visualization from all fields.

\section{Architecture Overview}\label{sec:architecture}

The framework proposed in this dissertation is divided into four different modules: (1) collection and filtering; (2) text pre-processing and analysis; (3) aggregation and (4) data visualization.

The current collection module is implemented to retrieve geo-location tweets from a specific \gls{bounding_box}, however if the user demands, multiple locations can be explored at the same time. Other collection heuristics are also available, such as the keyword-search and users following. Depending on the target scenario and analytics to be explored, these two heuristics will need to be added in the module. This detail was considered during implementation period and flexibility was assured into the module composition.

\begin{figure}[!ht]
	\centering
	\includegraphics[width=\textwidth]{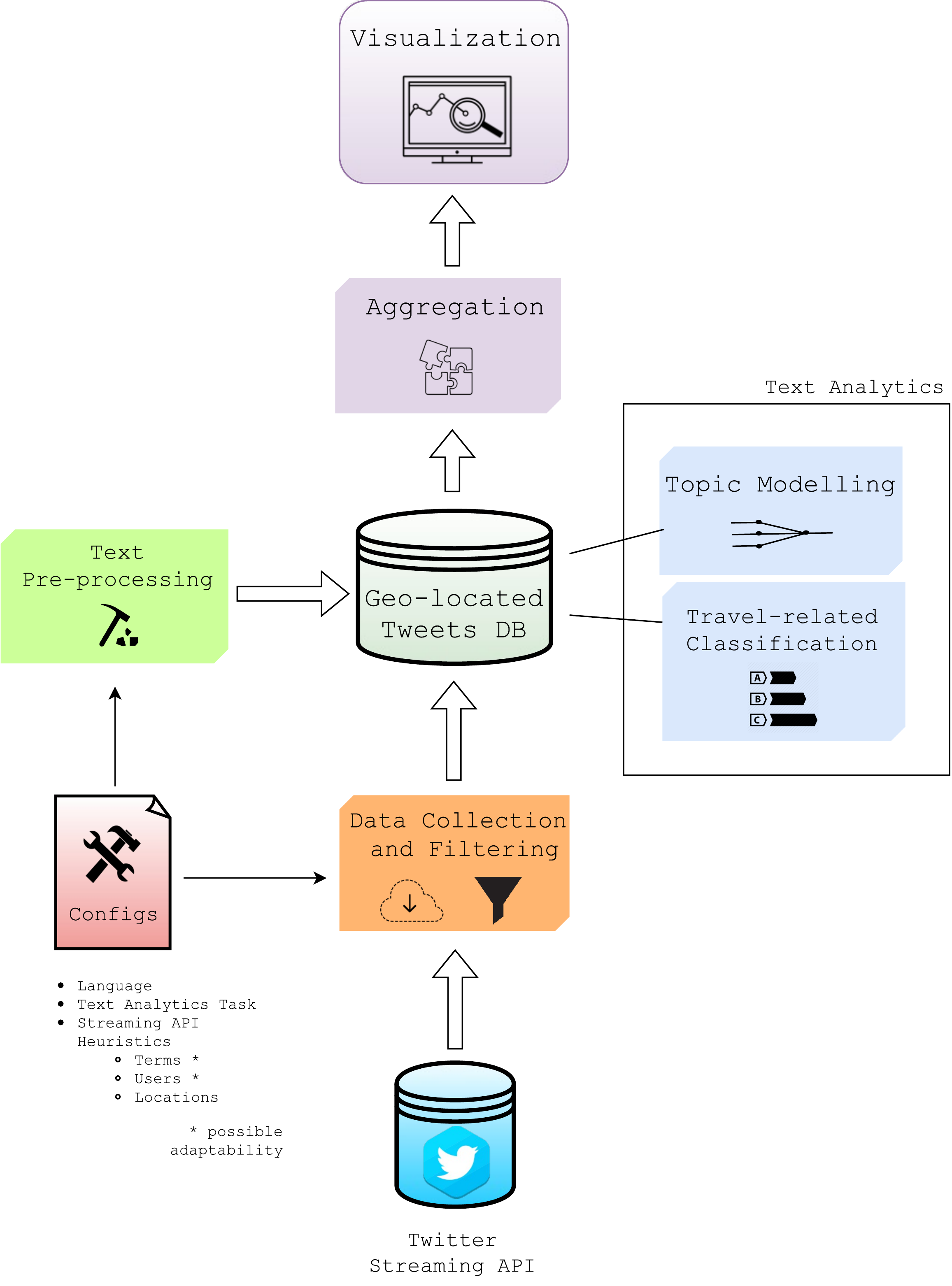}
	\caption[Framework Architecture Overview]{Architecture overview of the implemented framework}
	\label{fig:architecture}
\end{figure}

Filtering tasks are directly related to \texttt{locations} heuristic of the collection module. Since this framework is designed to analyse cities or specific regions/zones of it, it is necessary guarantee if a tweet is actually inside of the searching \gls{bounding_box} to do not induce information in the analysis from places far away of the target location. For instance, a bounding-box is a rectangle obtained by two coordinate pairs (latitude and longitude, for the South-West point and the North-East point). If other heuristics will be implemented, the filtering module can be configured to support other filtering-specific operations.

The text pre-processing module has into consideration the future task in the framework. Having this considered, we implemented a segmented pipeline allowing the user a definition of the desired tasks he wants to analyse in the text messages since different text analysis may have different operation in the preprocessing routine. Methods implemented here are carefully described in Section~\ref{sec:text_preprocessing}.

Text analytics module is composed by two different sub-modules, both of them focusing in a specific text analysis method. Travel-related classification of tweets for two different speaking languages is available since one of the final goals regarding domain-agnostic framework is its adaptability into different scenarios and the language of texts constitutes one of them. Topic Modelling sub-module is available as a text analytics method provided by the framework. We trained a model over a sample of tweets and characterize each topic generated in order instantly characterize future tweets by only being necessary passing it over the transformation process to have their topic identified.
In terms of generalization, the main module, text analytics module, was construct following adaptability and flexibility approaches to, in the future, new analysis be integrated.

By adding new functionalities, new aggregations are required in order to present the specific-task final results to the end-user. The aggregation module is structured into integrative methods facilitating future extensions or updates on it. Last but not least, aggregation results are communicated to the visualization module, where, similar to other modules, it is possible the inclusion of new data visualization charts, according to the new integrated functionalities.

\section{Data Collection}\label{sec:data_collection}

In Section~\ref{sec:requirements}, we explain the importance of the decision made to the data collection's heuristics. Twitter allows the developers' community two different tools to collect data, the Search and the Streaming \glspl{API}. The Search \gls{API} is based on the \gls{REST} protocol and only looks up for tweets published in the last 7 days, while the Streaming \gls{API} creates basic endpoints (independent of the \gls{REST}ful endpoints) and retrieves up to 1\% of the \gls{Twitter_Firehose}. Regarding the proposed and developed framework, we chose the Streaming \gls{API} due to its free-access for the community, smooth integration in the module implementation and due to the availability of real-time information. A positive point about the Streaming \gls{API} is the three available heuristics to the data collection, allowing the retrieval of tweets that match a specific text query (e.g. tweets with the word \texttt{bus} or \texttt{car}), the retrieval of tweets associated to a variable  number of users - being necessary previous knowledge about these users \textit{ids} - or even the retrieval of tweets located inside a bounding-box~\cite{mac2016effects}. There are two negative points regarding the Twitter Streaming \gls{API}: first, Twitter imposes limits in its data exploration, where only 400 words can be tracked, 5,000 users can be followed and 25 different bounding-boxes can be explored\footnote{\url{https://dev.twitter.com/streaming/reference/post/statuses/filter} (Accessed on 18/06/2017)}; second, the previously mentioned heuristics cannot be used together, i.e. we can not track specific tweets from an user that match with certain words. Although the negative points, we remain with the choice made, of using the Twitter Streaming \gls{API} as our source of information and limiting the heuristic to the one that retrieves tweets located inside a pre-defined bounding-box. Our choice is additionally supported by the need of studying cities and exploring the information derived from it. In this way, we know, a priori, that if the data collection method is able to retrieve tweets with precise geo-location then this makes our work easier since the exploration of specific regions of a city is already available taking into consideration the information available in tweets.

After the method selection, as well as the selection of its heuristic, we conduct an experiment regarding the amount of tweets being retrieved by one Twitter client for a city. Twitter has into consideration the number of clients used in the data collection process by tracking the IP address of the machine in the network. This constitutes a restriction to explore several cities with the same client since the Streaming \gls{API} retrieves only 1\% of the total overcome. In the experiment, we tested the capacity of a client to retrieve all the tweets posted in New York City and used four different clients for it: one defined with the city bounding-box, and the other three defined with bounding-boxes of three boroughs in the city: Bronx, Brooklyn and Manhattan. Considering the bounding-boxes creation, we took support of an open-source \textit{online} tool coined BoundingBox~\footnote{\url{http://boundingbox.klokantech.com/} (Accessed on 23/06/2017)}, which is integrated with the Google Maps \gls{API} and allows an user to create a bounding-box for an existing place in that \gls{API}.

Results showed that the client defined with the greatest bounding-box, New York City, was able to retrieve 100\% of the tweets from the three different boroughs. This experiment is consolidated with the work of Morstatter et al.~\cite{morstatter2013sample} where it was compared the Streaming \gls{API}'s capacity, regarding geo-located tweets, against the Twitter Firehose. Authors concluded that the percentage of geo-located tweets corresponds to 1-2\% of total overcome from Twitter and the Streaming \gls{API} is able to retrieve almost 90\% of it. Hence, we do not need to be concerned about how many bounding-boxes are used in the collection process because if we did so, we would need to be aware of 90\% of the world, which is not the case.

\subsection{Data Filtering}
\label{subsec:data_filtering}
In the first attempts to study the data collected geographic distribution, we discover that not all tweets had a precise coordinate attached to it. Nonetheless, there were cases where tweets from other cities were collected by our crawler and this phenomenon is not supposed to happen when the collection method is based in geo-located characteristics. By studying the Twitter mobile application, we found out that a user can tag himself in the tweet by two different ways: (1) a user can activate the \gls{GPS} in the mobile application and associate to the tweet his precisely geo-location; (2) a user can choose a place from a predefined list provide by Twitter and associate the place to the tweet.

The second method of tagging the geo-location to the tweet can arise some conflicts when this kind of tweets is used to perform scientific studies or even development of system to help the cities in the regularization, control and improvement of its services. Having this considered, it was necessary to understand how the Twitter Streaming \gls{API} works and what kind of heuristics follows in order to retrieve such type of tweets. The documentation~\footnote{\url{https://dev.twitter.com/streaming/overview/request-parameters\#locations} (Accessed on 17/06/2017)} enhances two different heuristics:

\begin{enumerate}
	\item If the coordinates field is populated, the values there will be tested against the bounding-box;
	\item If the coordinates field is empty but place is populated, the region defined in place is checked for intersections against the locations bounding-box. Any overlapping areas will yield a positive match.
	\end{enumerate}
	
The first heuristic only happens if a user is able/willing to tag a post with his precise geo-location associated with it; otherwise, the user can tag the post associated with a place and in this case the second heuristic is applied. Each place contained in the previous mentioned list, which is provided by Twitter, is composed by a bounding-box, and if any piece of it overlaps the bounding-box used in the collecting process, then a positive match is yielded and the tweet is retrieved. For instance, if a tweet has a place such as Portugal and our filter bounding-box is defined for Porto, all tweets from place Portugal will be in our dataset, regardless the fact some tweets are posted elsewhere, such as in the city of Lisbon, very far away from Porto.

\begin{figure}[!htbp]
	\centering
	\includegraphics[width=\textwidth]{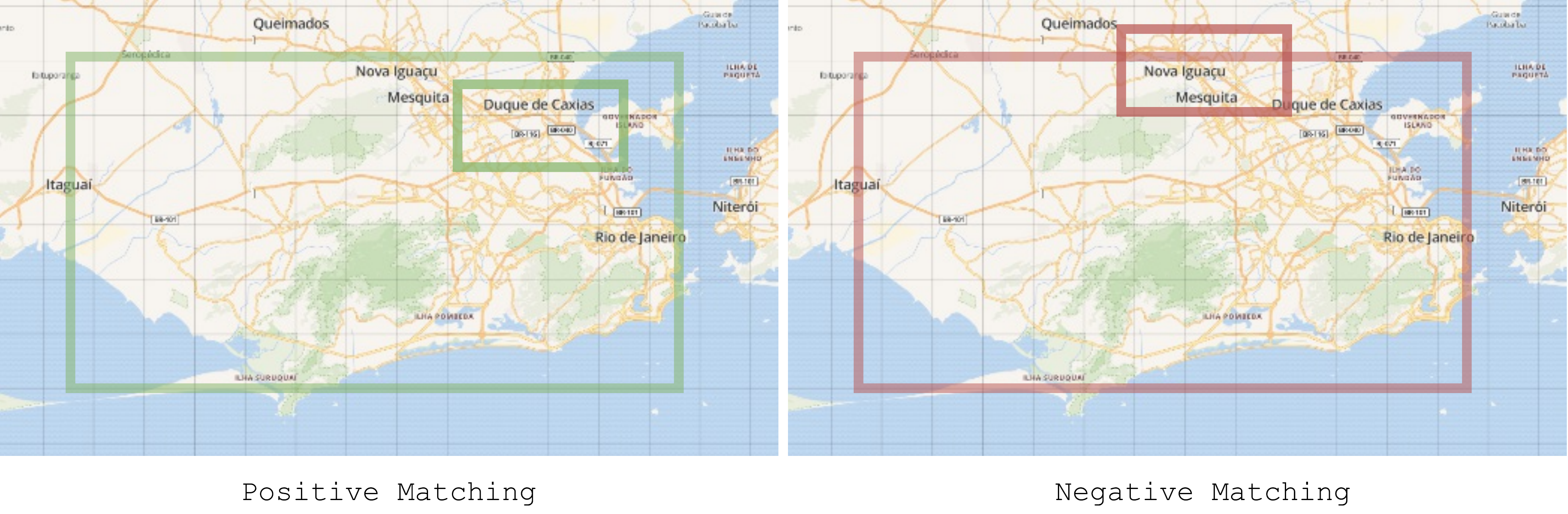}
	\caption[Bounding-boxes filtering process]{Example of our filtering method for geo-located tweets with variable bounding-boxes}
	\label{fig:matching_example}
\end{figure}

This restriction required the development of a external layer which was responsible for the filter of tweets located outside the area of each city. To built this so, it was necessary \textit{a posteriori} information and, thus, we extract the Twitter default bounding-box of each city in study appealing to the tweets \textit{place} field. Such information was then used as the limited area in order to filter out tweets which \textit{coordinates} field was not populated. The methodology behind the filtering process consists in the matching of the Twitter default bounding-box of the city against all places' bounding-boxes in tweets. In Figure~\ref{fig:matching_example}, we illustrate an example of our method in which the green color represents the matching of a tweet attached with place Duque de Caxias yielding a positive result, while the red color represents a tweet with place Nova Iguaçu yielding a negative match result with the Twitter default bounding-box for the city of Rio de Janeiro.

\section{Text Pre-processing}\label{sec:text_preprocessing}
In the requirements section (\ref{sec:requirements}), we mentioned some problems of social media streams as the short length and informality of the text message. The informality problem ranges from the writing style of each person to the existence of lots of abbreviations, slang, jargons, \textit{emoticons} and bad usage of punctuation signs. The preprocessing module presented in this section has as main goal the submission of the text messages under several operations in order to remove, or at least, reduce this type of informality characteristics and make easier the work of future tasks.

Below, we enumerate and described the different preprocessing methods implemented:

\begin{itemize}
	\item \textbf{Lowercasing:} This operation is responsible for the conversion upper case characters to lower representation. The advantages provided by this operation are centered in the analysis of words written in different ways. An representative example is \texttt{london} and \texttt{London} whose meaning is the same but due to the different casing in one letter, its representation/interpretation by text mining techniques may be disparate.
	
	\textit{\textbf{Travel-related Classification}} and \textit{\textbf{Topic Modelling}} modules explore this pre-processing operation.
	
	\item \textbf{Lemmatization:} Only plural words are transformed into singular ones (e.g. cars -> car).
	
	\textit{\textbf{Topic Modelling}} module explores this pre-processing operation.
	
	\item \textbf{Tokenization:} Is the method of dividing each sentence in a list of tokens/words.
	Since we are dealing with social media content, standard tokenizations techniques available in packages, such as the \texttt{tokenize}~\footnote{\url{http://www.nltk.org/api/nltk.tokenize.html}} from Python's \gls{NLTK}, perform poorly and are not capable of dealing with \textit{\#hashtags}, \textit{@mentions}, abbreviations, strings of punctuation (e.g. \texttt{...} or \texttt{\%\&\/\$}), \textit{emoticons} (e.g. \texttt{:)} or \texttt{:-)} or \texttt{=D}) and \texttt{unicode} glyphs which are very common in Twitter. Having considered this, we used a Twitter-based tokenization package, coined Twokenize and firstly presented by O'Connor et al.~\cite{o2010tweetmotif}, which is capable of dealing with these special characteristics of tweets.
	
	\textit{\textbf{Topic Modelling}} module explores this pre-processing operation.
	
	\item \textbf{Transforming repeated characters:} Sequences of characters repeated more than three times were transformed, e.g. "loooool" was converted to "loool".
	
	\textit{\textbf{Travel-related Classification}} and \textit{\textbf{Topic Modelling}} modules explore this pre-processing operation.
	
	\item \textbf{Punctuation removal:} Every punctuation symbols are removed from the text message, including the previous mentioned \textit{emoticons}.
	
	\textit{\textbf{Topic Modelling}} module explores this pre-processing operation.
	
	\item \textbf{Cleaning \textit{Entities} and Numerical Symbols:} Removing \textit{URLs}, user mentions, \textit{hashtags} and digits from the text messages.
	
	\textit{\textbf{Travel-related Classification}} and \textit{\textbf{Topic Modelling}} modules explore this pre-processing operation.
	
	\item \textbf{Stop and short words removal:} This operation consists in the removing of the most common words in the language in analysis. We used the standard words of the \gls{NLTK} Corpus package for the stop words removal task. Other type of words, such as 'kkk' or 'aff' represent short words that do not bring any valuable information from the message analysis. For this reason, we conceive a short dictionary containing these words and removed it from the message.
	
	\textit{\textbf{Travel-related Classification}} and \textit{\textbf{Topic Modelling}} modules explore this pre-processing operation.
\end{itemize}

Regarding other fields in a tweet, this module was also in charge of convert the date of creation of a tweet to the city timezone. The field \textit{created\_at} in a tweet is given in the \gls{UTC} and in order to have knowledge about the most active local hours and days on Twitter, we used the Python timezone package \texttt{pytz}~\footnote{\url{https://pypi.python.org/pypi/pytz}} to convert the world timezone to the one desired.

Although the existence of more text preprocessing techniques, in this dissertation we only used the ones previously described since each of them is associated to, at least, one text analytics module whose are described in the following section.

\section{Text Analytics}
\label{sec:text_analytics}
The extraction of information from texts can vary in several types depending on the task performed to achieve it. In this dissertation, we explored two different types of analysis to the tweets: topic modelling and travel-related classification.

\subsection{Topic Modelling}
\label{sec:topic_modelling_framework}

Social media, more specifically, microblog services are platforms where people publicly share their opinions and due to that they are seen as a rich source of content to explore. In order to mine such information, we implement in our framework a generative module using topic modelling techniques.

Topic modelling is a text mining technique which goal is the identification of latent topics in a collection of documents. During the last decade, the research community had been using this technique in a vast range of works aiming the test of its applicability in different domains. Here, we also used topic modelling to characterize different cities and provide this type of information to the framework's end-users.

Latent Dirichlet Allocation (\gls{LDA}) is a generative statistical model proposed by D. Blei et al.~\cite{blei2003latent} that makes possible the discovering of unknown groups and its similarities over a collection of text documents. The model tries to identify what topics are present in a document by observing all the words that composing it, producing as final result a topic distribution. 

\begin{figure}[htbp]
	\centering
	\includegraphics[scale=0.41, keepaspectratio]{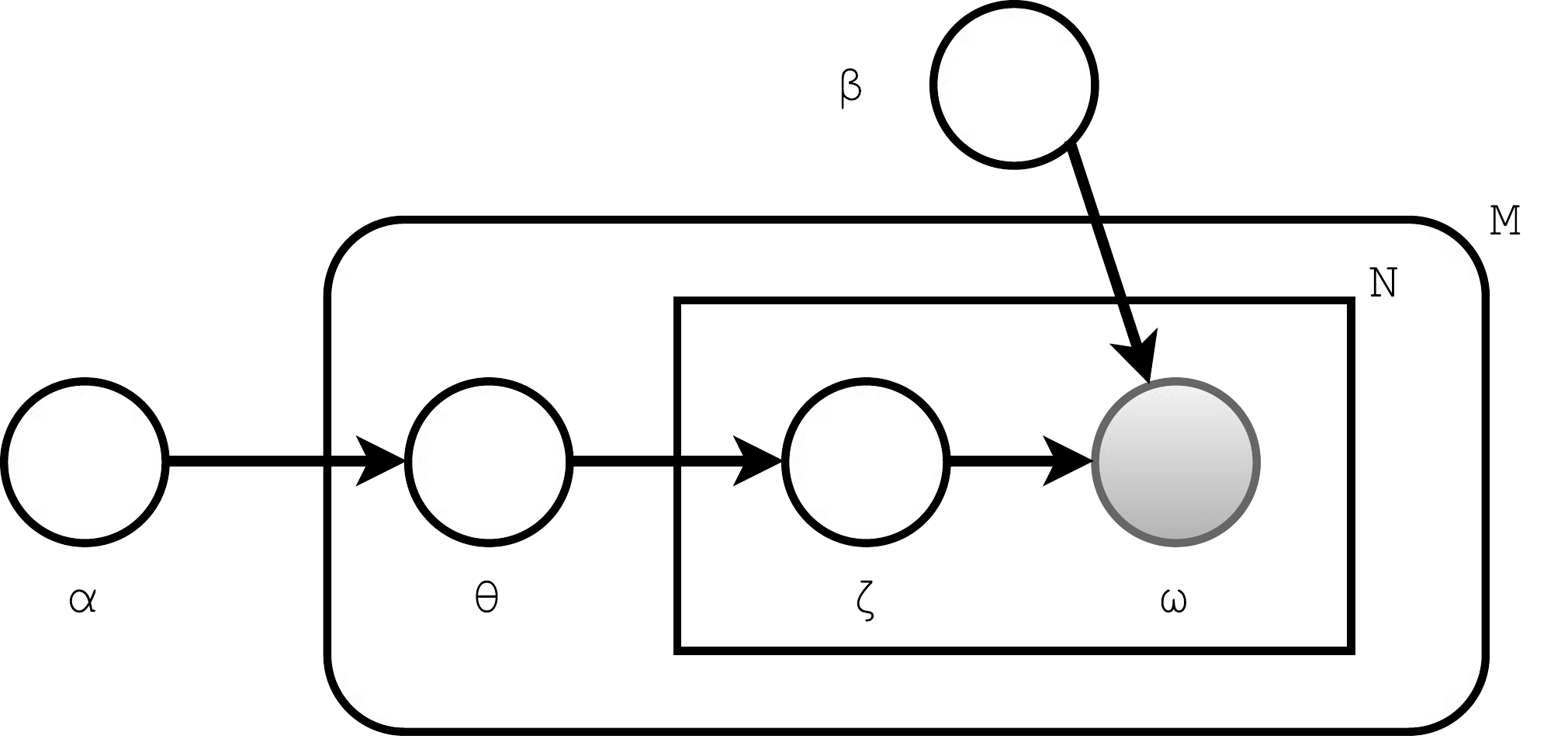}
	\caption[Plate notation of \gls{LDA} by Blei et al. ~\cite{blei2003latent}]{Plate Notation of the graphical model representation of Latent Dirichlet Allocation by Blei et al. Source:~\cite{blei2003latent}}
	\label{fig:lda_graphical_model_representation}
\end{figure}

In Figure~\ref{fig:lda_graphical_model_representation} it is illustrated the plate notation to the graphical model of \gls{LDA}. There, we can observe that for a collection of documents $M$, each one composed by a sequence of $N$ words, the model tries to attribute a per-document topic distribution, using an $\alpha$ dirichlet prior, to a topic-word distribution $\xi$ (associated also with a dirichlet prior $\beta$), inducing that each topic's probability $\theta$ is focused in a small set of words $w$ which characterize that topic.

The most important advantage this model provides is related to the group of features involved in its training process. Conventional application of this model uses only as features a bag-of-words matrix representation, and for this reason the task of topic modelling becomes very simple since only the frequency of words in documents are taken into account. Last but not least, \gls{LDA} model performs two different distributions: (1) distribution of words over topics and (2) distribution of topics over the documents, resulting in the assumption that each document is random mixture of topics, whose in turn are composed by a probabilistic distribution of words.

The cities' characterization provided by our framework centers in the topics being talked about at the time. We conduct an experiment to evaluate if such information could bring added-value for the cities entities and the results although being very promiscuous proved to have potential in certain occasions. The overall experiment is described in Section~\ref{sec:topic_modeling} as well as potential improvements to the generated model.

\subsubsection{Features}
Topic modelling requires, like in other learning model, a group of features to be trained. In this case, we used the \glossary{BoW} representation matrix - which is a representation where each document is converted to a frequency vector according to the number of occurrences of each word in the message. The set of features was limit to a dictionary containing 10,000 words and it only took into account uni-grams in the message content. The dictionary was also limited to words that occur in a maximum percentage of 40$\%$ in the whole dataset, avoiding common words that were not removed because they were not included in the Python's \gls{NLTK} stop words list for the specific language in analysis. The minimal occurrence value for a word being considered was set to 10.

\subsubsection{LDA Model Resulting Topics}
The final model used in the implementation of our framework is defined to characterize a tweet into 50 different topics. Although that, in the experiment made to comprove the added-value brought by the model,  we were obligated to cluster some of the topics due to the similarity presented in words constituting them. The final list of possible topics can be seen in Section~\ref{subsec:lda_results}, more specifically in Table~\ref{tab:topic_labels}.

\subsection{Travel-related Classification}
\label{sec:travel_classification}

\emph{Prima facie}, we tried to extract and characterize travel-related tweets from large datasets in order to study the geographical and temporal distributions of such specific content. The transportation entities may take advantages from this kind of information since human mobility can be study, as well as citizens' opinions regarding the transportation services. The Twitter Streaming \gls{API} provides a massive amount of data and filter out the relevant in a short period of time is a laborious process. In order to be successful in this task, we created an automatic text classifier capable of discriminating travel-related tweets from non-related ones. Due to the absence of gold standard datasets in this domain, there was the need of creating a training and testing set of data in order to proceed the experiment and evaluate the performance of the produced model. Conventional classification tasks in the domain of intelligent transportation systems follow traditional approaches by constructing their group of features using standard bag-of-words techniques. In our experiment, we tried to combine a \gls{BoW} features with \gls{BoE} (word embeddings representation matrices), producing, for the best of our knowledge, the first travel-related classification model with both type of features.

\subsubsection{Features}
\label{sec:travel_features}
\gls{BoW} representation matrix is a list of lists, where each entry of the matrix is associated to a sentence of the document and takes the form of a term-frequency vector. In this group of features, we only considered uni-grams as the basis of text representation form. The final dictionary of this form was produced with the 3,000 most frequent terms across the training set excluding the ones found in more than 60$\%$ of the documents (tweets).

The technique of word embeddings is used by Mikolov et al.~\cite{mikolov2013efficient} in the implementation of a powerful computational method named \emph{word2vec}. This method is capable of learning distributed representations of words, and each word is represented by a distribution of weights across a fixed number of dimensions. Authors have also proved that such representation is robust when encoding syntactic and semantic similarities in the embedding space.

The training objective of the skip-gram model, as defined by Mikolov~et~al.~\cite{mikolov2013linguistic}, is to learn the target word representation, maximizing the prediction of its surrounding words given a predefined context window. For instance, to the word $w_t$, present in a vocabulary, the objective is to maximize the average log probability:

\begin{equation}
\frac{1}{T}  \sum_{t=1}^{T}  \sum_{-c \leq j \leq  c, j \neq 0} \textnormal{log } P(w_{t+j} | w_t)
\end{equation}

where $c$ is the size of the context window, $T$ is the total number of words in the vocabulary and $w_{t+j}$ is a word in the context window of $w_t$. After training, a low dimensionality embedding matrix $\textbf{E}$ encapsulates information about each word in the vocabulary and its use (i.e. the surrounding contexts). For instance, by using the skip-gram model over our datasets we were able to verify that words such as \texttt{ônibus} and \texttt{busão} are used in the similar contexts, as a mode of transport.

Later on, Le and Mikolov~\cite{le2014distributed} developed \emph{paragraph2vec}, an unsupervised learning algorithm operating on pieces of text not necessarily of the same length. The model is similar to \emph{word2vec} but learns distributed representations of sentences, paragraphs or even whole documents instead of words. Hence, we explored \emph{paragraph2vec} to learn the vector representations of each tweet and tried several configurations in the model hyper-parameterization.

Using \textit{paragraph2vec}~\cite{le2014distributed}, we created \gls{BoE} representation matrices for the tweets in order to explore the learning distributed representations of words where each word is represented by a distribution of weights across a fixed number of dimensions. Mikolov et al.~\cite{mikolov2013linguistic} proved that this kind of text representation is robust when encoding syntactic and semantic similarities in the embedding space. The training process of our classification models involved 10 iterations over the datasets using a context window of value 2 and feature vectors of 50, 100 and 200 dimensions. Then, the corresponding embedding matrix yielded the group of features fed into our classification routine.

Both previous described methods are available in the collection of Python scripts we used in this dissertation, coined \texttt{Gensim}~\footnote{\url{https://radimrehurek.com/gensim/about.html} (Accessed on 20/06/2017)}, presented and lately improved by \v{R}eh\r{u}\v{r}ek and Sojka~\cite{rehurek2010software}.

The overall experiments regarding the travel-related classification of tweets are described and detailed in Sections~\ref{subsec:rio_de_janeiro_sao_paulo_experiment} and~\ref{subsec:new_york_city_experiment}. Concluded the experiments, we select the best classifiers for each case and used it in the implementation of the framework's travel-related modules allowing discrimination of potential new tweets related to the transportation domain.

\section{Data Storage and Aggregation}\label{sec:storage_aggregations}

Besides the few percentage of geo-located tweets provided by Twitter (1-2\% of the total Firehose overcome), this data requires, in the first place, large storage capability  and, secondly, a tool that allows the easy manipulation and quick access of data. Having considered this, we opted for the use MongoDB, an open-source cross-platform document-oriented database, as the data warehouse technology for our framework. MongoDB allows storage of \gls{JSON} documents which is the retrieved format of tweets by the Streaming \gls{API}. Since in this dissertation we developed the framework as a prototype of a system capable of extracting information related to \textit{smart cities} and transportation services, the large physical space to storage data was not a priority.

MongoDB presents, alongside the high performance, availability and scaling, an inner framework that allows the aggregation of data according to specific user-generated queries. Here, we took advantage of such a pipeline in order to produce interesting statistics regarding the processed data. Map-reduce is the processing paradigm behind the aggregating operations allowing high performance even when applied to large volumes of data, as in this particular case where it is necessary to process thousands or millions of tweets in a short period of time.

\section{Visualization}\label{sec:visualization}

One of the most laborious and time-consuming tasks in the development of this social-media-based framework is the selection of data visualizations to illustrate the results provided by the previous mentioned modules. Visualization of information retrieved from the social networks needs to be easy of understanding and the majority of systems explores key performance indicators while others use more complex visual representations such as egocentric networks~\cite{saleiro2016timemachine}. Due to the amount of data being processed, the generation of data visualization using an atomic implementation is sometimes poorly in terms of response time. Hence, we needed to adopt a different approach in order to solve this non-efficient procedure.

After a long period of research, we found a solution to this problem by creating a set of routines (bash scripts) that are called periodically (hourly) to execute all type of necessary aggregations and update its corresponding data collections in the database. Then, other routine is invoked to generate all type of data visualizations and store its visual representation in \gls{HTML} files. In the implementation of this module, these files - containing the data visualization - were embedded inside several view pages. \texttt{Plotly}~\footnote{\url{https://plot.ly/python/}} is a Python graphing library that has available the saving of the visualizations produced in files with \gls{HTML} format. Besides that, the library offers an extensive range of graphical representations, such as basic charts (bar charts, scatter plots, etc), scientific charts (heatmaps), financial charts (time series) and maps (choropleth, bubble and line maps), which facilitates the construction and designing of dynamic dashboards. Here, we explore mostly the section of basic charts to build simple representations of the results obtained from the analytics phase and also added top lists about some metadata of the tweets, as so the overall, daily and hourly top \textit{hashtags} and uni-grams.

\section{Summary}
In this Chapter we detail the implementation of our framework, the modules that compose it, as well as the methodologies and methods chosen for each module conceptualization. 

During the construction and implementation phase, we tried to maintain modularity in the whole system in order to make possible future extensions to it or even complementarity of the existing modules. The final framework is composed by four different modules connected between them. However, if the situation demands, each module can be adaptable to other systems.

Concluded the overview of the main characteristics of our framework, we provide the final users information that have been collected, filtered and analysed in a almost real-time scenario. It is worth noting that due to the laborious tasks each tweets passes through, it is impossible to provide instantaneous results because the system should be coherent and not treat each message separately. Information is given to the final user in short periods of time (e.g. 30 minutes), making possible that all text analytics modules conclude its tasks and all type of analysis is actually available.

An interesting future improvement to the framework is the incorporation of an extra module to perform tasks of sentiment analysis. Work together with the two already developed the information provided is of additional value to the services of a smart city, including the transportation domain.
\chapter{Exploratory Data Analysis}
\label{chap:exploratory_data_analysis}

\minitoc \mtcskip \noindent

The main goal of this chapter is the devise of relevant analysis taking into consideration the five different collected datasets. Since this dissertation is supported in experiments using real-world data, such analysis is crucial in order to gain better knowledge of the intrinsic characteristics of it. A tweet provides some fields of interest, such as, the text message, date of creation, language, and the \emph{entities}, which are constantly analysed in several data analytics systems. An \emph{entity} is metadata and additional contextual information contained in the tweet and is composed by the \emph{hashtags}, \emph{user mentions}, \emph{urls} and \emph{media} fields. We count the amount of tweets containing this kind of information for all the cities, London, New York, Melbourne, Rio de Janeiro and São Paulo, and projected some data visualizations for different temporal frequencies. The following subsections are divided into three different categories:  (1) Geographical Distribution, (2) Temporal Frequencies and (3) Metadata Composition. Additionally, we discuss the results of each city, as well as the main observable differences.

\section{Geographic Distributions}\label{sec:geographical_distribution}

\begin{figure}[t]
	\centering
	\begin{subfigure}[t]{0.31\textwidth}
		\centering
		\includegraphics[width=1\linewidth]{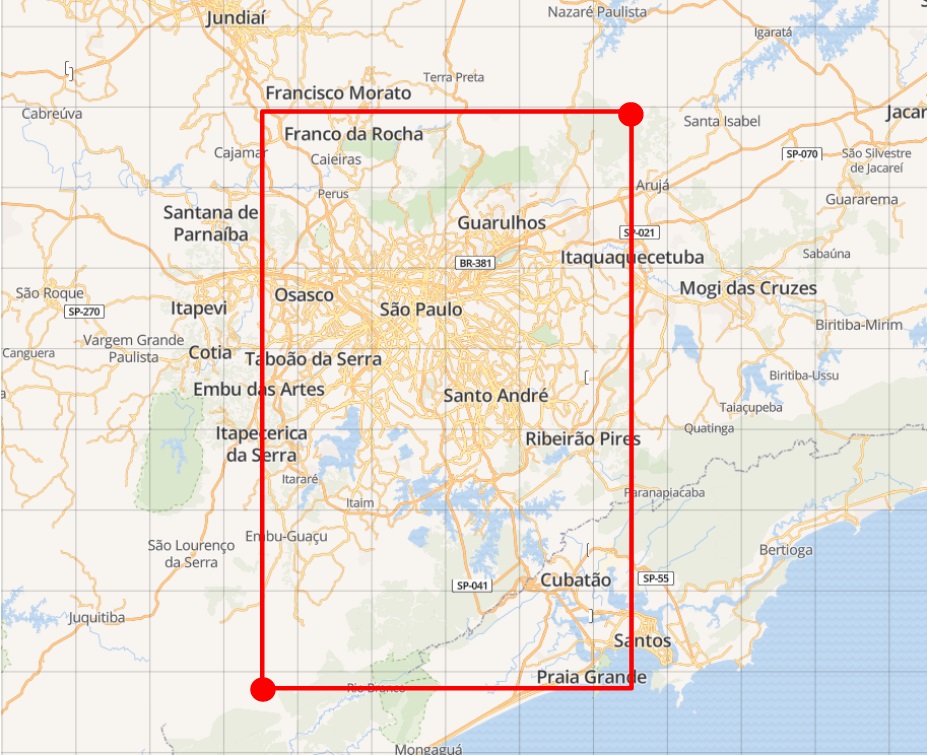}
		\caption{São Paulo}
		\label{fig:saopaulo_bounding_box}
	\end{subfigure}%
	\quad
	\begin{subfigure}[t]{0.3\textwidth}
		\centering
		\includegraphics[width=1\linewidth]{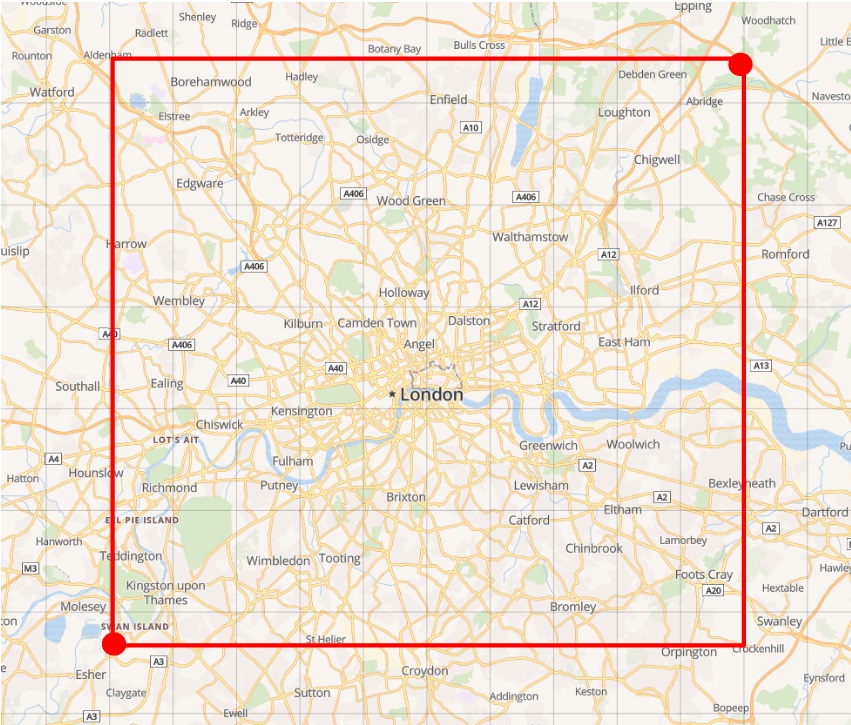}
		\caption{London}
		\label{fig:london_bounding_box}
	\end{subfigure}
	\quad
	\begin{subfigure}[t]{0.3\textwidth}
		\centering
		\includegraphics[width=1\linewidth]{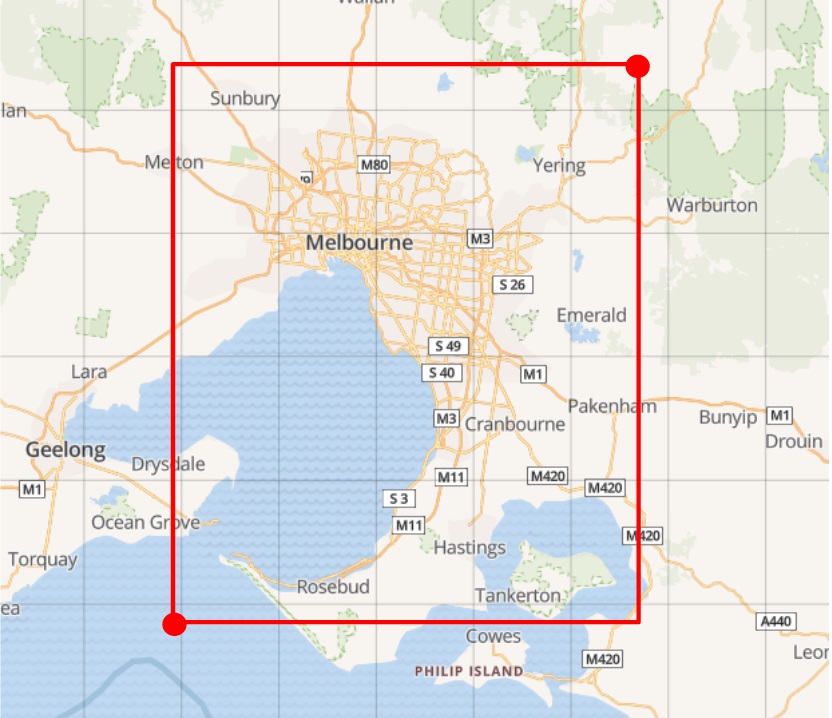}
		\caption{Melbourne}
		\label{fig:melbourne_bounding_box}
	\end{subfigure}
	
	\medskip
	
	\begin{subfigure}[t]{0.42\textwidth}
		\centering
		\includegraphics[width=1\linewidth]{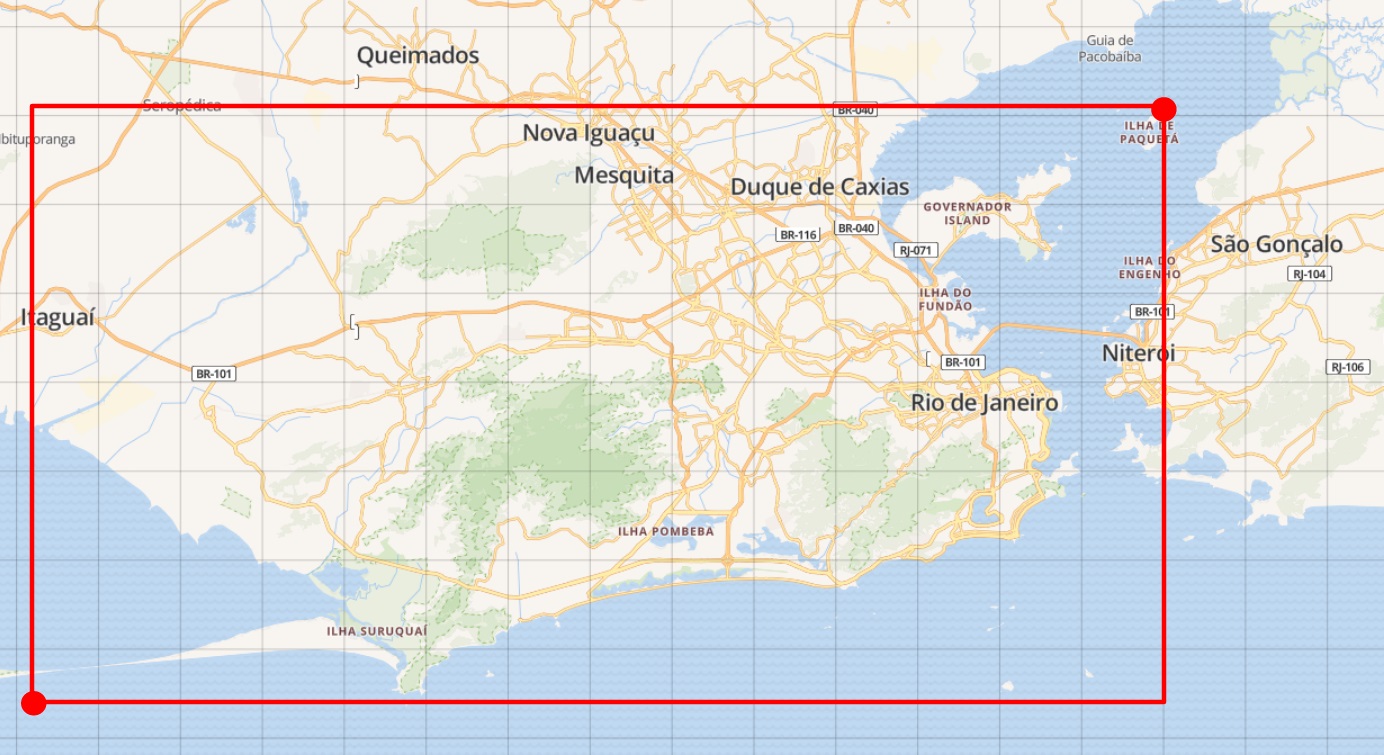}
		\caption{Rio de Janeiro}
		\label{fig:riodejaneiro_bounding_box}
	\end{subfigure}
	\quad
	\begin{subfigure}[t]{0.38\textwidth}
		\centering
		\includegraphics[width=1\linewidth]{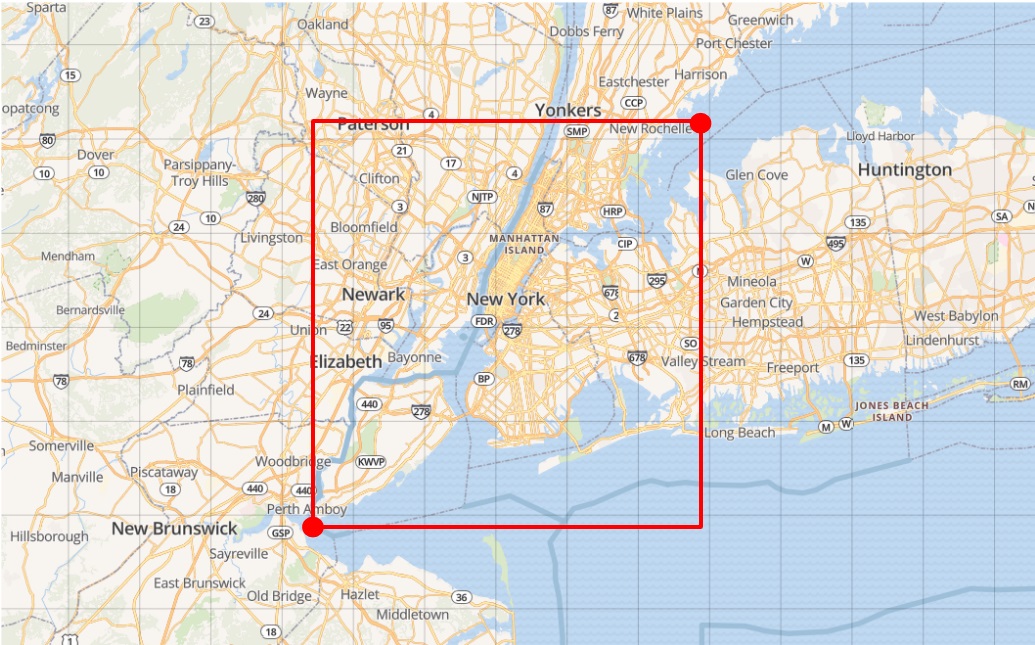}
		\caption{New York}
		\label{fig:newyork_bounding_box}
	\end{subfigure}
	\caption{Search Bounding-boxes for the data collection}
	\label{fig:bounding_boxes}
\end{figure}

As previously mentioned, in Section~\ref{sec:data_collection}, we exploit an auxiliary \textit{online} tool to generate the coordinates for the bounding-boxes used in the collection process. The visual representation of the each city bounding-box is illustrated in Figure~\ref{fig:bounding_boxes}, as well as its the corresponding coordinates which are presented in Table~\ref{tab:bbs_points}.

\begin{table}[b]
	\centering
	\setlength\extrarowheight{3pt}
	\caption{Collecting Bounding-boxes Coordinates (South-West and North-East)}
	\label{tab:bbs_points}
	\begin{tabular}{l|c|c}
		\hline
		\multicolumn{1}{c|}{\textbf{City}} & \textbf{South-West} & \textbf{North-East} \\ \hline
		\textbf{Rio de Janeiro} & (-43.7950599, -23.0822288) & (-43.0969042, -22.7460327) \\
		\textbf{São Paulo} & (-46.825514, -24.0082209) & (-46.3650844, -23.3566039) \\
		\textbf{New York City} & (-74.2590899, 40.4773991) & (-73.7002721, 40.9175771) \\
		\textbf{London} & (-0.3514683, 51.3849401) & (0.148271, 51.6723432) \\
		\textbf{Melbourne} & (144.5937418, -38.4338593) & (145.5125288, -37.5112737) \\ \hline
	\end{tabular}
\end{table}

Taking a careful observation into to coordinates used within each bounding-box, we can affirm that Rio de Janeiro present the broadest bounding-box comparatively to the others cities.

\begin{figure}[!htbp]
	\centering
	\begin{subfigure}[htbp]{0.4\textwidth}
		\centering
		\includegraphics[width=1\linewidth]{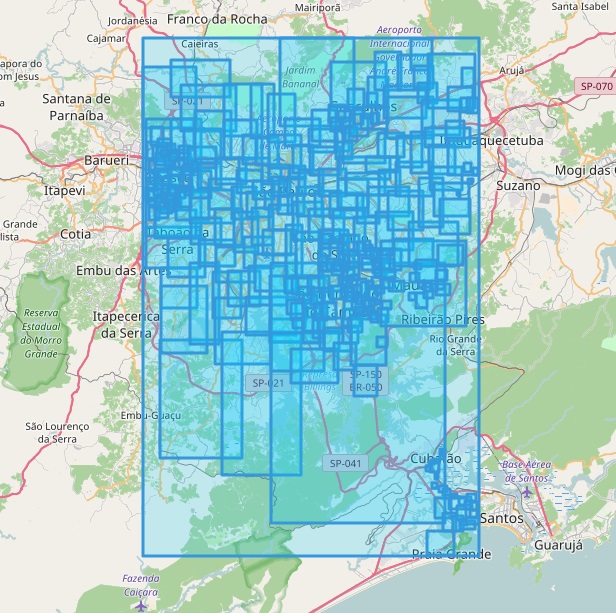}
		\caption{}
		\label{subfig:saopaulo_bounding_boxes}
	\end{subfigure}
	\quad
	\begin{subfigure}[htbp]{0.5\textwidth}
		\centering
		\includegraphics[width=1\linewidth]{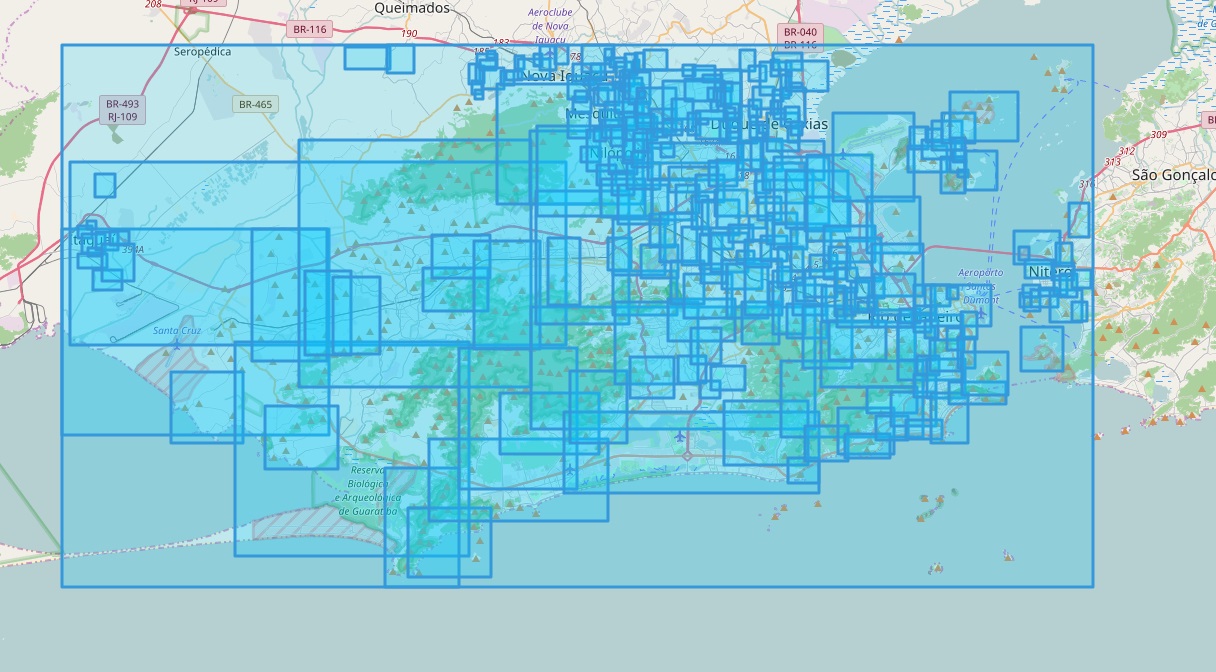}
		\caption{}
		\label{subfig:riodejaneiro_bounding_boxes}
	\end{subfigure}
	
	\medskip
	
	\begin{subfigure}[htbp]{0.4\textwidth}
		\centering
		\includegraphics[width=1\linewidth]{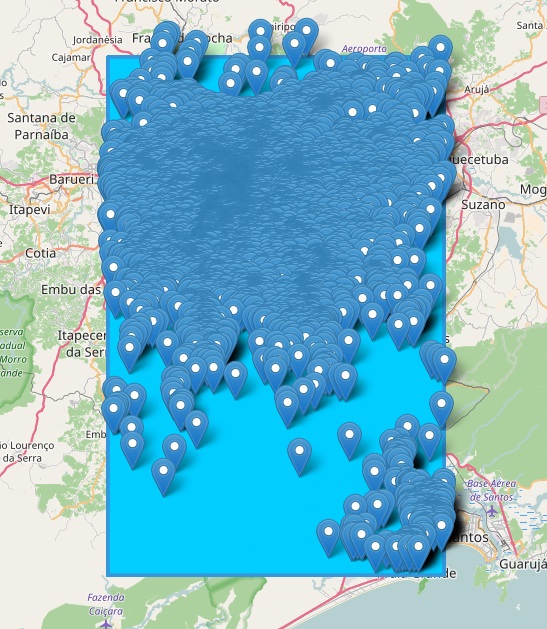}
		\caption{}
		\label{subfig:saopaulo_markers}
	\end{subfigure}
	\quad
	\begin{subfigure}[htbp]{0.5\textwidth}
		\centering
		\includegraphics[width=1\linewidth]{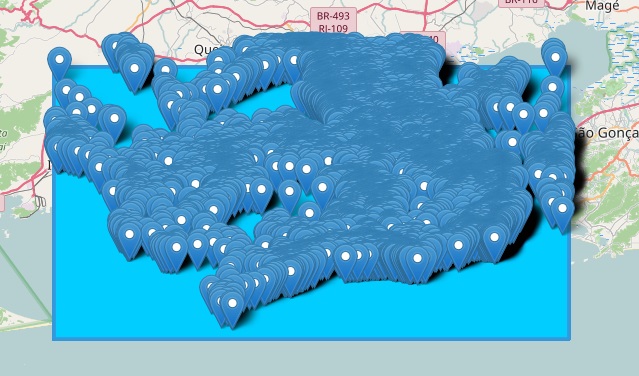}
		\caption{}
		\label{subfig:riodejaneiro_markers}
	\end{subfigure}
	
	\medskip
	
	\begin{subfigure}[htbp]{0.4\textwidth}
		\centering
		\includegraphics[width=1\linewidth]{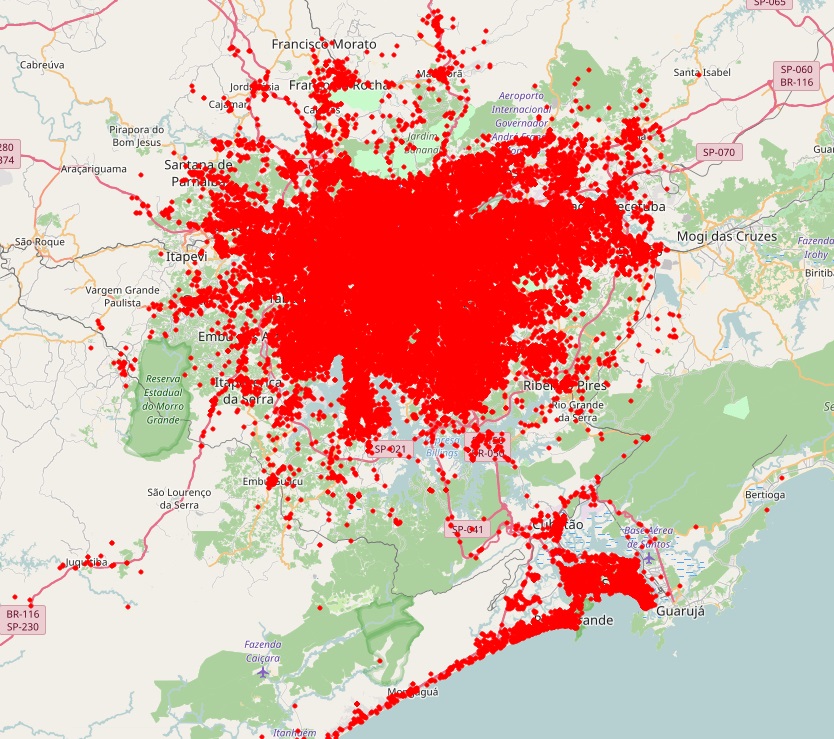}
		\caption{}
		\label{subfig:saopaulo_points}
	\end{subfigure}
	\quad
	\begin{subfigure}[htbp]{0.5\textwidth}
		\centering
		\includegraphics[width=1\linewidth]{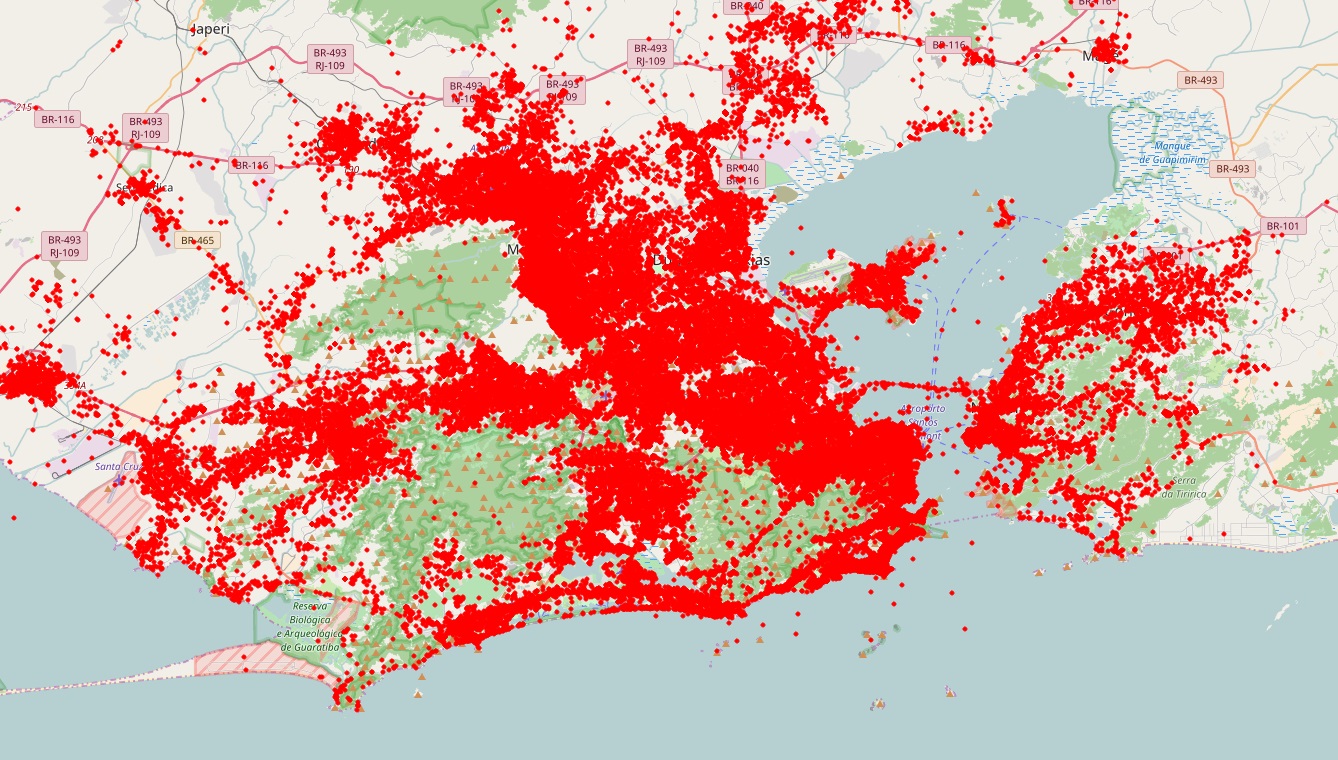}
		\caption{}
		\label{subfig:riodejaneiro_points}
	\end{subfigure}
	
	\caption[Exploratory analysis in Brazilian cities]{São Paulo (a, c, e) and Rio de Janeiro (b, d, f) Geographical Distributions: (a, b) Bounding-boxes of places (c, d) Specific places (e, f) Geo-tagged tweets}
	\label{fig:rio_sp_geographical_distribution}
\end{figure}

\begin{figure}[!htbp]
	\centering
	\begin{subfigure}[htbp]{0.3\textwidth}
		\centering
		\includegraphics[width=1\linewidth]{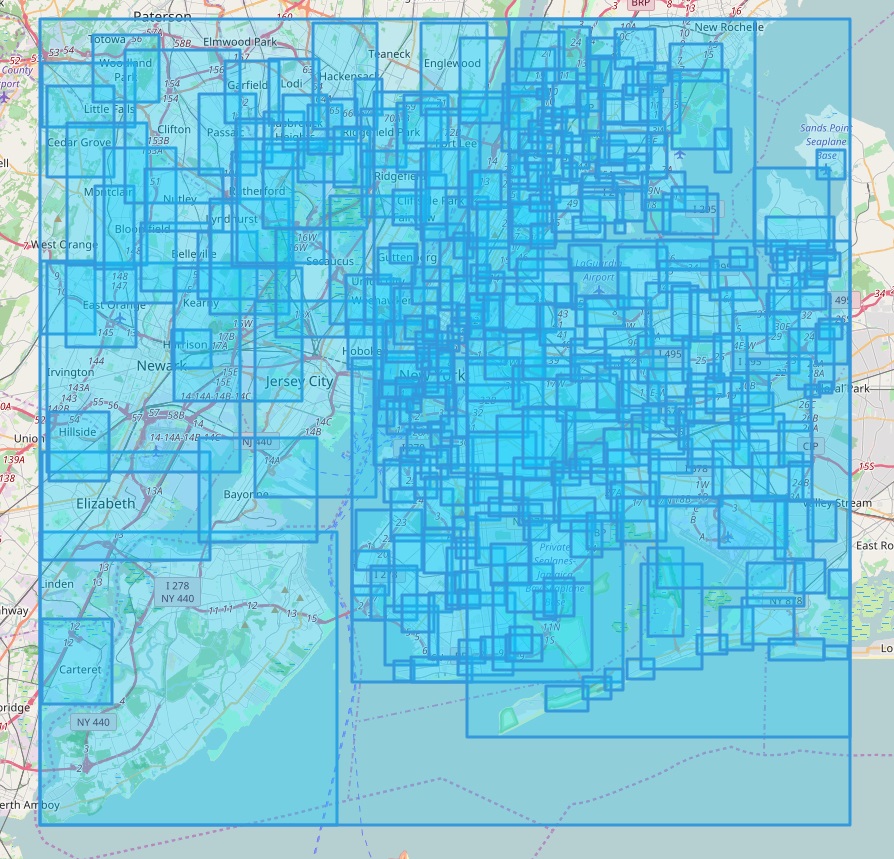}
		\caption{}
		\label{subfig:nyc_bounding_boxes}
	\end{subfigure}
	\quad
	\begin{subfigure}[htbp]{0.3\textwidth}
		\centering
		\includegraphics[width=1\linewidth]{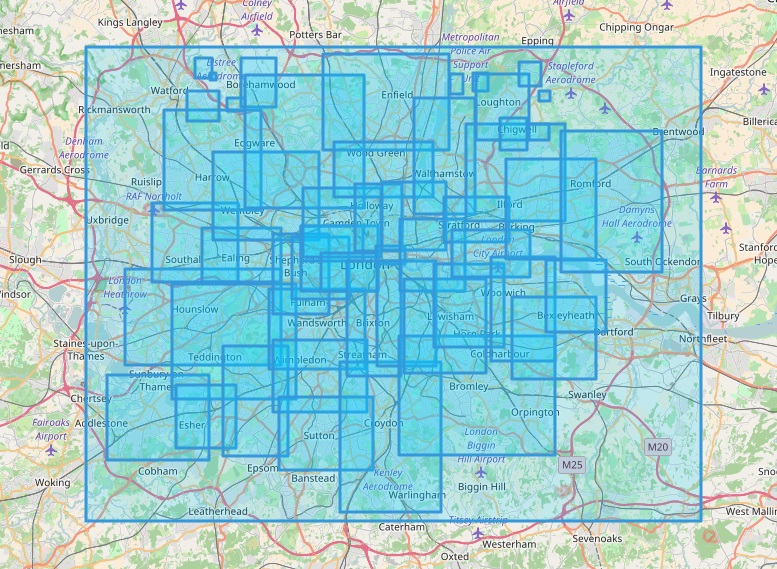}
		\caption{}
		\label{subfig:london_bounding_boxes}
	\end{subfigure}
	\quad
	\begin{subfigure}[htbp]{0.3\textwidth}
		\centering
		\includegraphics[width=1\linewidth]{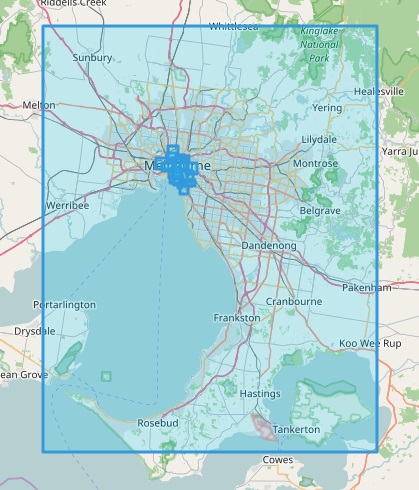}
		\caption{}
		\label{subfig:melbourne_bounding_boxes}
	\end{subfigure}
	
	\medskip
	
	\centering
	\begin{subfigure}[htbp]{0.3\textwidth}
		\centering
		\includegraphics[width=1\linewidth]{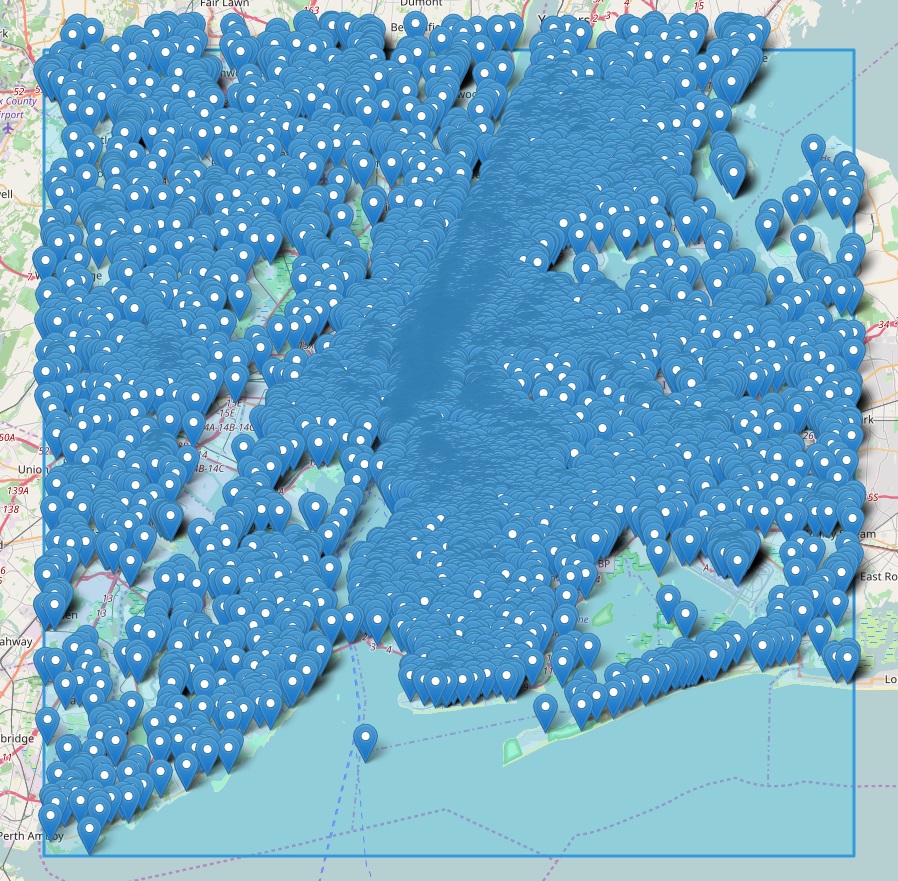}
		\caption{}
		\label{subfig:nyc_markers}
	\end{subfigure}
	\quad
	\begin{subfigure}[htbp]{0.3\textwidth}
		\centering
		\includegraphics[width=1\linewidth]{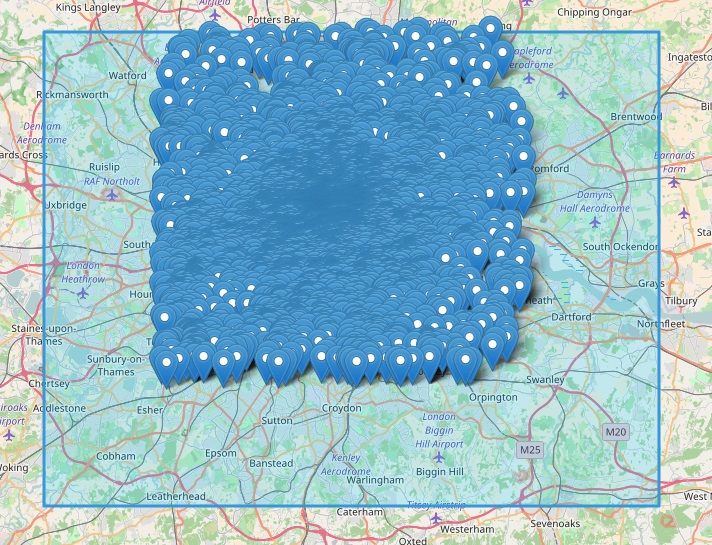}
		\caption{}
		\label{subfig:london_markers}
	\end{subfigure}
	\quad
	\begin{subfigure}[htbp]{0.3\textwidth}
		\centering
		\includegraphics[width=1\linewidth]{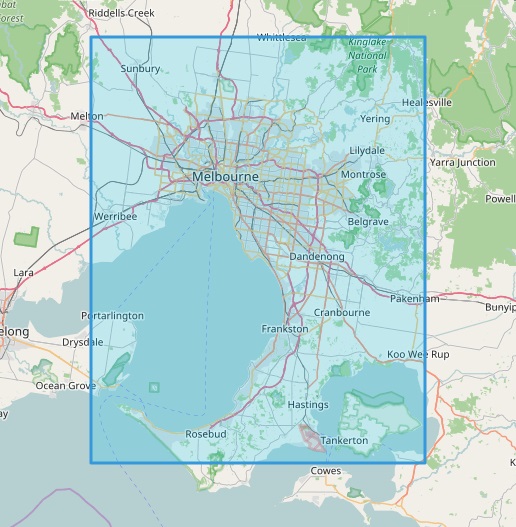}
		\caption{}
		\label{subfig:melbourne_markers}
	\end{subfigure}
	
	\medskip
	\centering
	\begin{subfigure}[htbp]{0.3\textwidth}
		\centering
		\includegraphics[width=1\linewidth]{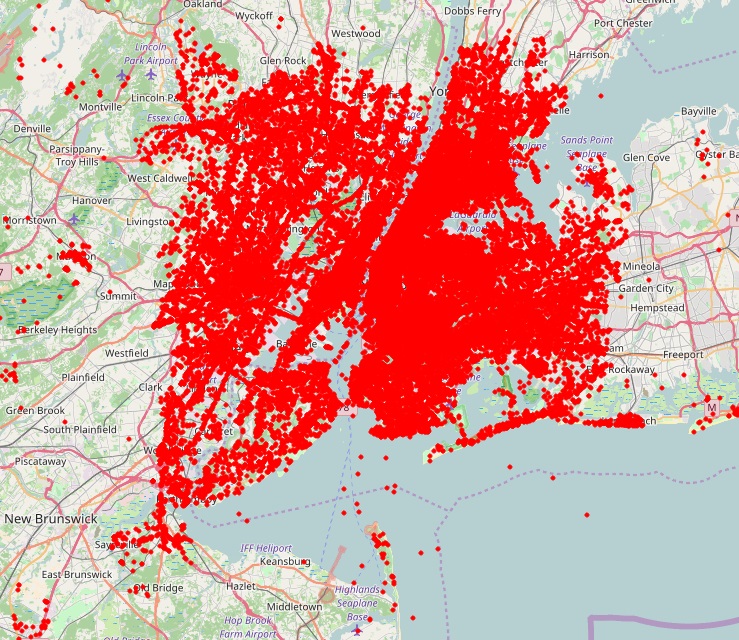}
		\caption{}
		\label{subfig:nyc_points}
	\end{subfigure}
	\quad
	\begin{subfigure}[htbp]{0.3\textwidth}
		\centering
		\includegraphics[width=1\linewidth]{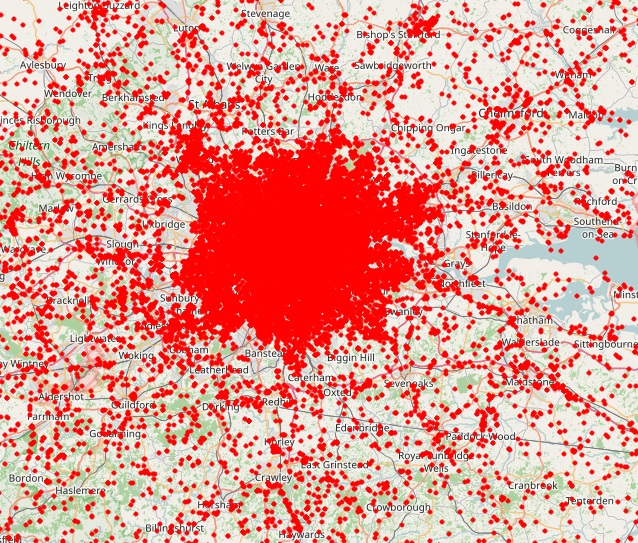}
		\caption{}
		\label{subfig:london_points}
	\end{subfigure}
	\quad
	\begin{subfigure}[htbp]{0.3\textwidth}
		\centering
		\includegraphics[width=1\linewidth]{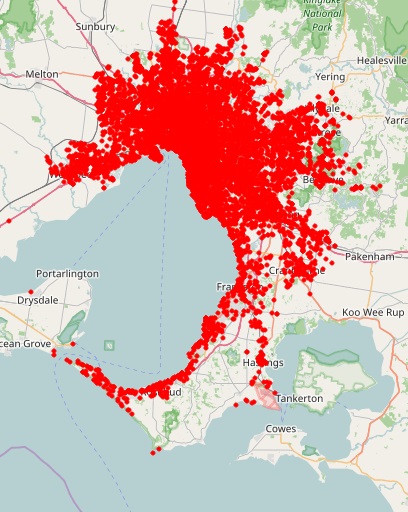}
		\caption{}
		\label{subfig:melbourne_points}
	\end{subfigure}
	
	\caption[Exploratory analysis in English-speaking cities]{New York City (a, d, g), London (b, e, h) Geographical Distributions: (a, b) Bounding-boxes of places (c, d) Specific places (e, f) Geo-tagged tweets}
	\label{fig:nyc_london_melbourne_geographical_distribution}
\end{figure}

To conduct the data filtering process, we extracted from data the Twitter default bounding-boxes for each city in study, being possible to observe their corresponding South-West and North-East coordinates in Table~\ref{tab:bbs_filter}. The map visualization of these bounding-boxes is demonstrated in Figures~\ref{fig:rio_sp_geographical_distribution} (subfigures~\ref{subfig:riodejaneiro_bounding_boxes} and~\ref{subfig:saopaulo_bounding_boxes}) and~\ref{fig:nyc_london_melbourne_geographical_distribution} (subfigures~\ref{subfig:nyc_bounding_boxes},~\ref{subfig:london_bounding_boxes} and~\ref{subfig:melbourne_bounding_boxes}), where the biggest rectangle represents the Twitter default bounding-boxes for each city.

\begin{table}[htbp]
	\centering
	\setlength\extrarowheight{3pt}
	\caption{Twitter Default Bounding-boxes Coordinates (South-West and North-East)}
	\label{tab:bbs_filter}
	\begin{tabular}{l|c|c}
		\hline
		\multicolumn{1}{c|}{\textbf{City}} & \textbf{South-West} & \textbf{North-East} \\ \hline
		\textbf{Rio de Janeiro} & (-43.795449, -23.08302) & (-43.087707, -22.739823) \\
		\textbf{São Paulo} & (-46.826039, -24.008814) & (-46.365052, -23.356792) \\
		\textbf{New York City} & (-74.255641, 40.495865) & (-73.699793, 40.91533) \\
		\textbf{London} & (-0.510365, 51.286702) & (0.334043, 51.691824) \\
		\textbf{Melbourne} & (144.593742, -38.433859) & (145.512529, -37.511274) \\ \hline
	\end{tabular}
\end{table}

The final volume of tweets located inside and outside the cities correspondent bounding-boxes are presented in Table~\ref{tab:geographic_counts_bb}. Alongside with the location analysis, the language count was also performed since future experiments only took into consideration tweets with the native language of the city in study and not foreign ones. In the abovementioned table (\ref{tab:geographic_counts_bb}) it is possible to verify a vast difference regarding the activity on Twitter in Rio de Janeiro. Numbers tell that such activity, with respect to geo-located tweets, is almost two times more than São Paulo and New York City, four times London and twenty five times Melbourne. A possible justification for this noticeable difference may be associated to the area of the bounding-box used in the collection process, but, on the other hand, according to some sources related to the demographic measures, for the case Rio De Janeiro \textit{versus} São Paulo, the population volume has an opposite behavior, where São Paulo~\footnote{\url{https://cidades.ibge.gov.br/v4/brasil/sp/sao-paulo/panorama} (Accessed on 17/06/2017)} has almost 12 millions habitants while Rio de Janeiro~\footnote{\url{https://cidades.ibge.gov.br/v4/brasil/rj/rio-de-janeiro/panorama} (Accessed on 17/06/2017)} has 6 million. Having only this amount of information it is impossible, at the moment, formulate a explanation to this phenomenon.

\begin{table}[htbp]
	\centering
	\caption[Datasets composition according bounding-box analysis]{Datasets composition after verification of the tweets inside the corresponding bounding-box}
	\label{tab:geographic_counts_bb}
	\resizebox{\textwidth}{!}{\begin{tabular}{l|c|cl|cl|cl|cl|cl}
			\hline
			\multicolumn{1}{c|}{\multirow{2}{*}{\textbf{City}}} & \multirow{2}{*}{\textbf{All}} & \multicolumn{2}{c|}{\textbf{PT/EN}} & \multicolumn{2}{c|}{\textbf{Non-PT/EN}} & \multicolumn{2}{c|}{\textbf{\begin{tabular}[c]{@{}c@{}}In\\ Bounding-Box\end{tabular}}} & \multicolumn{2}{c|}{\textbf{\begin{tabular}[c]{@{}c@{}}Out\\ Bounding-Box\end{tabular}}} & \multicolumn{2}{c|}{\textbf{\begin{tabular}[c]{@{}c@{}}PT/EN and In\\ Bounding-Box\end{tabular}}} \\ \cline{3-12} 
			\multicolumn{1}{c|}{} &  & \textbf{No. tweets} & \multicolumn{1}{c|}{\textbf{\%}} & \textbf{No. tweets} & \multicolumn{1}{c|}{\textbf{\%}} & \textbf{No. tweets} & \multicolumn{1}{c|}{\textbf{\%}} & \textbf{No. tweets} & \multicolumn{1}{c|}{\textbf{\%}} & \textbf{No. tweets} & \multicolumn{1}{c}{\textbf{\%}} \\ \hline
			\textbf{Rio de Janeiro} & 18,803,774 & 15,906,680 & 84,59\% & 2,897,094 & 15,41\% & 12,976,048 & 69,01\% & 5,827,726 & 30,99\% & 11,060,136 & 58,82\% \\
			\textbf{São Paulo} & 9,319,624 & 7,203,115 & 77,29\% & 2,116,509 & 22,71\% & 6,237,427 & 66,93\% & 3,082,197 & 33,07\% & 4,886,626 & 52,43\% \\
			\textbf{New York City} & 8,507,145 & 7,260,829 & 85,35\% & 1,246,316 & 14,65\% & 6,972,312 & 81,96\% & 1,534,833 & 18,04\% & 5,956,355 & 70,02\% \\
			\textbf{London} & 5,596,551 & 4,774,310 & 85,31\% & 822,241 & 14,69\% & 4,752,918 & 84,93\% & 843,633 & 15,07\% & 4,040,092 & 72,19\% \\
			\textbf{Melbourne} & 789,927 & 669,435 & 84,75\% & 120,492 & 15,25\% & 742,946 & 94,05\% & 46,981 & 5,95\% & 629,424 & 79,68\% \\ \hline
		\end{tabular}}
	\end{table}

Later, after the filtering process, we tried to understand the volume, as well as the location of each tweet. Through this kind of analysis it was possible to find out that a tweet which \textit{coordinates }field was empty and is, actually, represented with a bounding-box, can also be a specific place, i.e. a place that has a precise coordinate. Not all places were represented by a bounding-box in which each point that composed it are different. An example to that is \texttt{Estádio do Maracanã} which although its location field being represented by a bounding-box format, all four points are equal. A division was made considering this three types of location - (1) bounding-box with four different points; (2) bounding-box with four equal points; (3) precise coordinate - in order to have a perception of how different specific places and bounding-boxes as so which is the volume of tweets that are related to it.

\begin{table}[htbp]
	\centering
	\caption{Volume of tweets for each type of geo-location}
	\label{tab:volume_geolocation}
	\resizebox{\textwidth}{!}{\begin{tabular}{l|c|ccc|ccc|ccc}
			\hline
			\multicolumn{1}{c|}{\multirow{2}{*}{\textbf{City}}} & \multirow{2}{*}{\textbf{Total}} & \multicolumn{3}{c|}{\textbf{Bounding-boxes}} & \multicolumn{3}{c|}{\textbf{Specific Places}} & \multicolumn{3}{c|}{\textbf{Precisely}} \\ \cline{3-11} 
			\multicolumn{1}{c|}{} &  & \multicolumn{1}{c|}{\textbf{Distinct}} & \multicolumn{1}{c|}{\textbf{No. Tweets}} & \textbf{Percentage (\%)} & \multicolumn{1}{c|}{\textbf{Distinct}} & \multicolumn{1}{c|}{\textbf{No. Tweets}} & \textbf{Percentage (\%)} & \multicolumn{1}{c|}{\textbf{Distinct}} & \multicolumn{1}{c|}{\textbf{No. Tweets}} & \textbf{Percentage (\%)} \\ \hline
			\textbf{Rio de Janeiro} & 11060136 & 297 & 10237280 & 92,56\% & 11159 & 49440 & 0,45\% & 163748 & 773416 & 6,99\% \\
			\textbf{São Paulo} & 4886626 & 325 & 4284795 & 87,68\% & 7189 & 21022 & 0,43\% & 100028 & 580809 & 11,89\% \\
			\textbf{New York City} & 5956355 & 328 & 4210854 & 70,70\% & 16078 & 85204 & 1,43\% & 138123 & 1660297 & 27,87\% \\
			\textbf{London} & 4040092 & 53 & 3196043 & 79,11\% & 8123 & 53412 & 1,32\% & 95317 & 790637 & 19,57\% \\
			\textbf{Melbourne} & 629424 & 22 & 523870 & 83,23\% & 0 & 0 & 0,00\% & 21826 & 105554 & 16,77\% \\ \hline
		\end{tabular}}
	\end{table}

The final counts of the analysis for each identified type of geo-location are presented in Table~\ref{tab:volume_geolocation}. Looking at the numbers it is possible to conclude some facts applicable to all cities. Citizens tend to geo-locate themselves with a location which has variable bounding-box size since more than 70\% of the tweets are of this type. Furthermore, only a few percentage of tweets, between 0\% and 1.43\%, are located in specific places, although the existence of a higher number of distinct specific places comparatively to the bounding-boxes with variable size, with exception of Melbourne that has zero specific places in our dataset.

Other interesting point to enhance is the considerable percentage of tweets with precise location (i.e. tweets that people tagged himself using the GPS). The Brazilian cities proved to be less supportive of precisely located tweets, while the English cities were more contributive. The distribution of each type of geo-located tweet is illustrated in Figures~\ref{fig:rio_sp_geographical_distribution} and~\ref{fig:nyc_london_melbourne_geographical_distribution}. The variable bounding-boxes are showed in ~\ref{subfig:saopaulo_bounding_boxes},~\ref{subfig:riodejaneiro_bounding_boxes},~\ref{subfig:nyc_bounding_boxes},~\ref{subfig:london_bounding_boxes} and~\ref{subfig:melbourne_bounding_boxes} proving that our filter method was able to correctly agglomerate places that were, indeed, inside of the Twitter default bounding-boxes. In~\ref{subfig:saopaulo_markers},~\ref{subfig:riodejaneiro_markers},~\ref{subfig:nyc_markers},~\ref{subfig:london_markers} and~\ref{subfig:melbourne_markers} is illustrated the distribution of the specific places found out in our datasets for each city. A particular point identified was the absence of specific places in Melbourne and the limited places in a certain area of London. With a first look at the image of London, there may be doubts about the results concerning the filter method, however the bounding-box used to that process was the same in both cases, and so the only viable explanation for such result is the absence of specific locations for that area in the predefined list of places provided by the Twitter applications. Lastly, in~\ref{subfig:saopaulo_points},~\ref{subfig:riodejaneiro_points},~\ref{subfig:nyc_points},~\ref{subfig:london_points} and~\ref{subfig:melbourne_points} is illustrated the distribution of precisely located tweets. Through a careful observation in this distribution it was possible the arising of another doubt relatively to the first aforementioned heuristic of the Twitter Streaming API. There were tweets retrieved that not matched the bounding-box used in the collection process and this fact conducts to uncertainty and mistrust regarding the performance of this type of collection available on Twitter. 

\section{Temporal Frequencies}

Another interesting analysis in our datasets concerns the temporal distribution of the data. The volume of tweets posted per hour, per day, as well as the activity by day-of-the-week or hour-of-the-day are statistics that enable the possibility of finding out patterns or variations which can be correlated to some events or incidents happening in a city.

\begin{figure}[htbp]
	\centering
	\begin{subfigure}[htbp]{0.8\textwidth}
		\includegraphics[width=\linewidth]{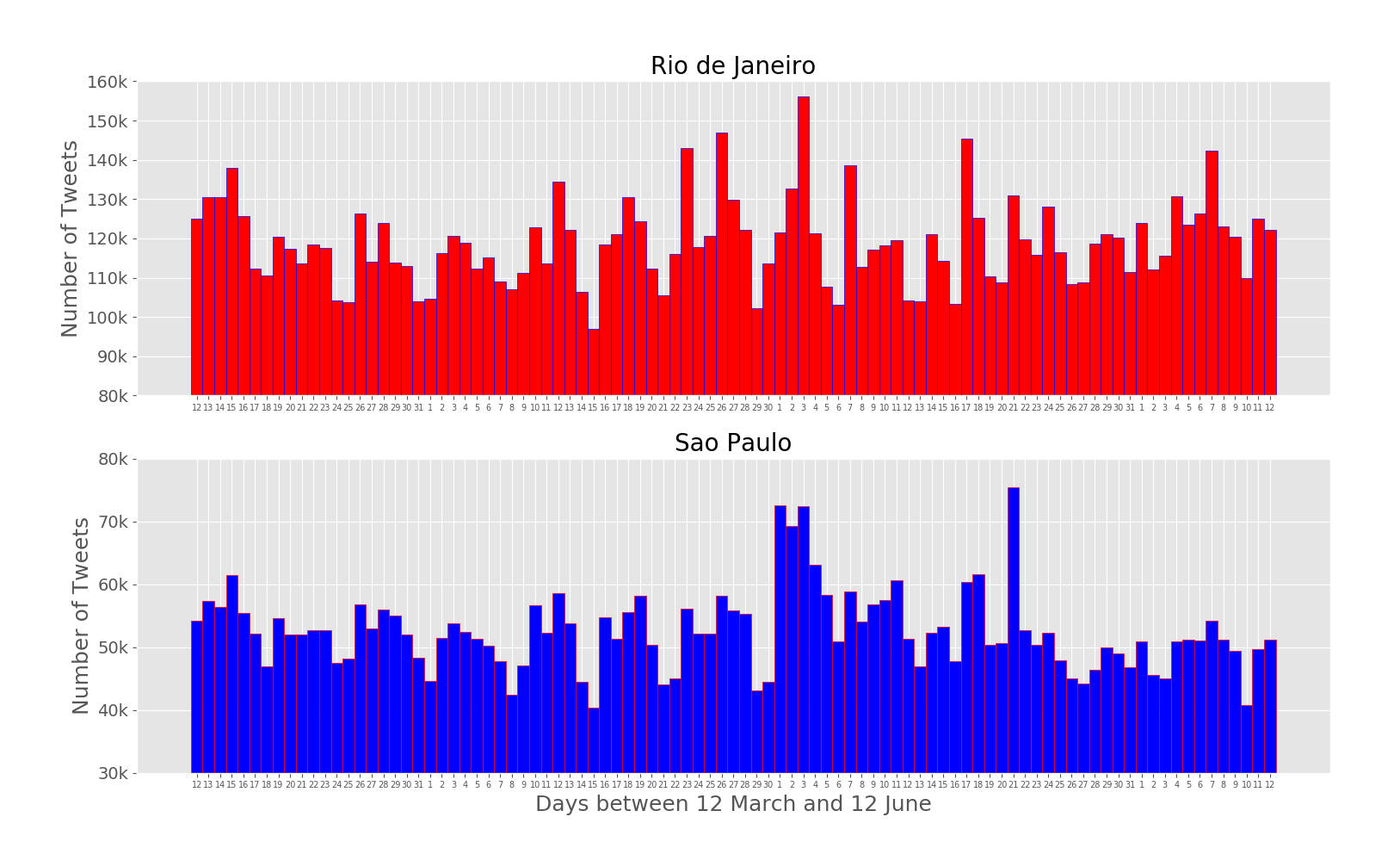}
		\caption{}
		\label{subfig:portuguese_cities_whole_months} 
	\end{subfigure}
	
	\begin{subfigure}[htbp]{0.8\textwidth}
		\includegraphics[width=\linewidth]{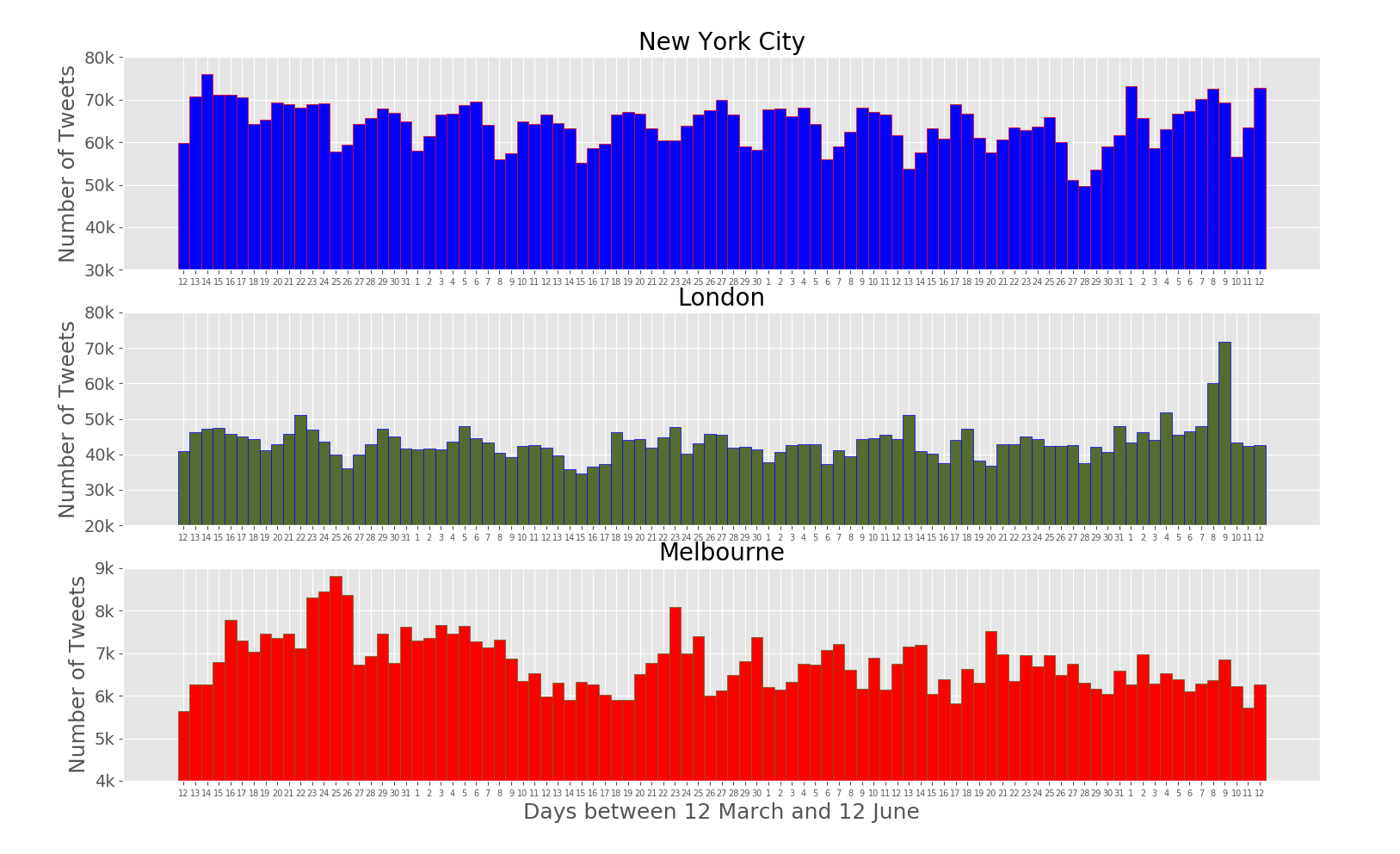}
		\caption{}
		\label{subfig:english_cities_whole_months}
	\end{subfigure}
	
	\caption[Daily volume of tweets]{Daily volume of tweets (a) Rio de Janeiro and São Paulo - Portuguese Cities (b) New York City, London and Melbourne - English Cities}
	\label{fig:daily_distribution}
\end{figure}

During and after remarkable events, citizens are impelled to share their feelings, opinions or even report their safety and well-being conditions (e.g. in cases of terrorist attack) through mobile applications. This share of information increases the activity of social media platforms, which can be potentially used for the identification of uncommon events. Figure~\ref{fig:daily_distribution} illustrates the daily distribution of all cities for the period of collection, three whole months, between 12 March and 12 June, 2017. The Brazilian cities present high level of variation between consecutive days (with the volume varying in a tens of thousands of tweets) and so the task of identifying remarkable events turns out to be much harder. On the other hand, the English speaking cities in our study are very similar, with exception of Melbourne whose activity is very low comparatively to the other cities (New York City and London). In the particular case of London, we can identify an abrupt increase of volume during days 8 and 9 of June. With the support of external sources such as news websites, we learnt about the United Kingdom General Elections 2017~\footnote{\url{https://www.theguardian.com/politics/general-election-2017} (Accessed on 17/06/2017)} occurred on that period which suggests that an increase of the Twitter activity might be associated with that event. 

\begin{figure}[htbp]
    \centering
    \begin{subfigure}[htbp]{0.45\textwidth}
        \centering
        \includegraphics[width=1\linewidth]{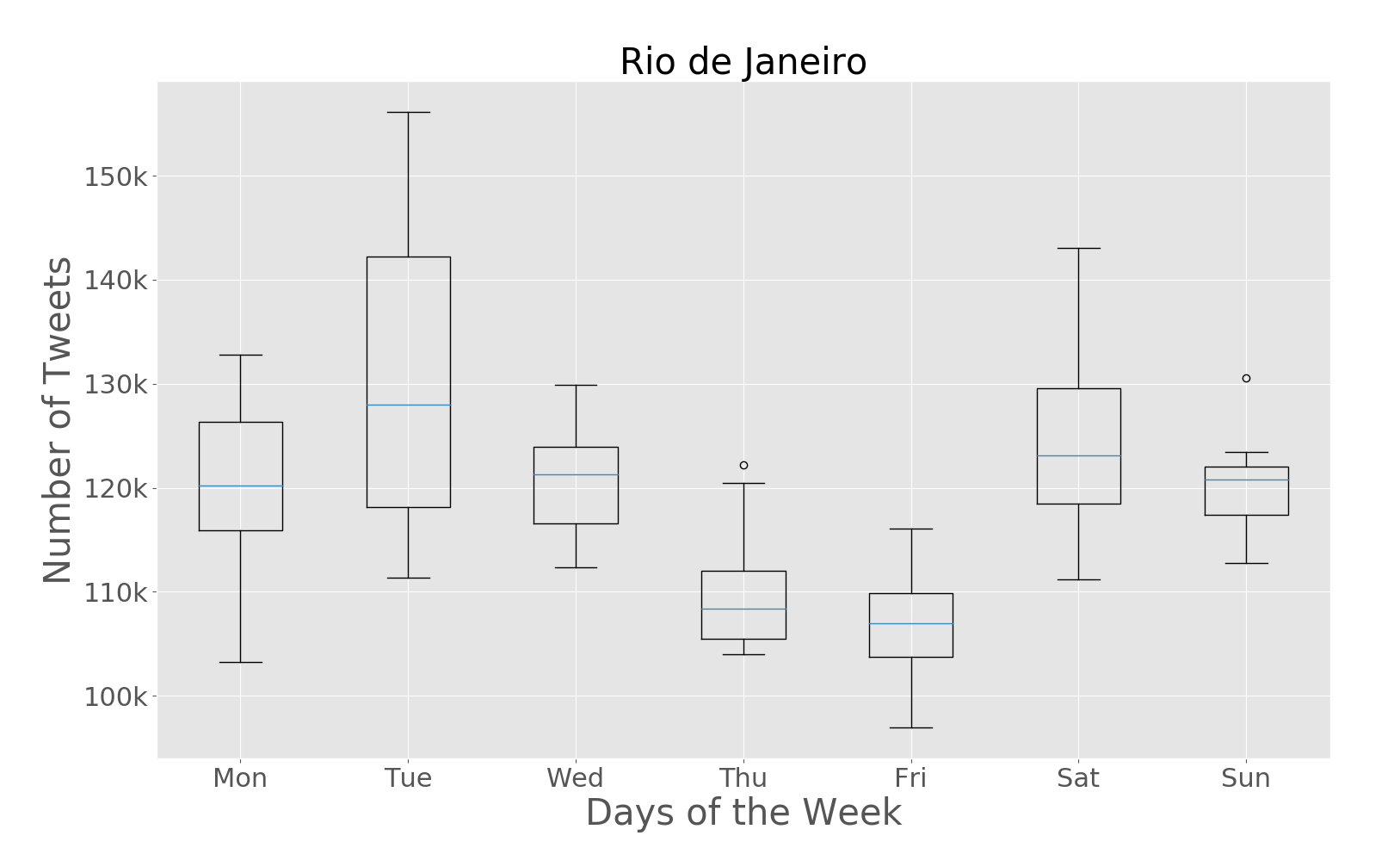}
        \caption{}
        \label{subfig:riodejaneiro_box_plot_day_of_week}
    \end{subfigure}%
    \quad
    \begin{subfigure}[htbp]{0.45\textwidth}
        \centering
        \includegraphics[width=1\linewidth]{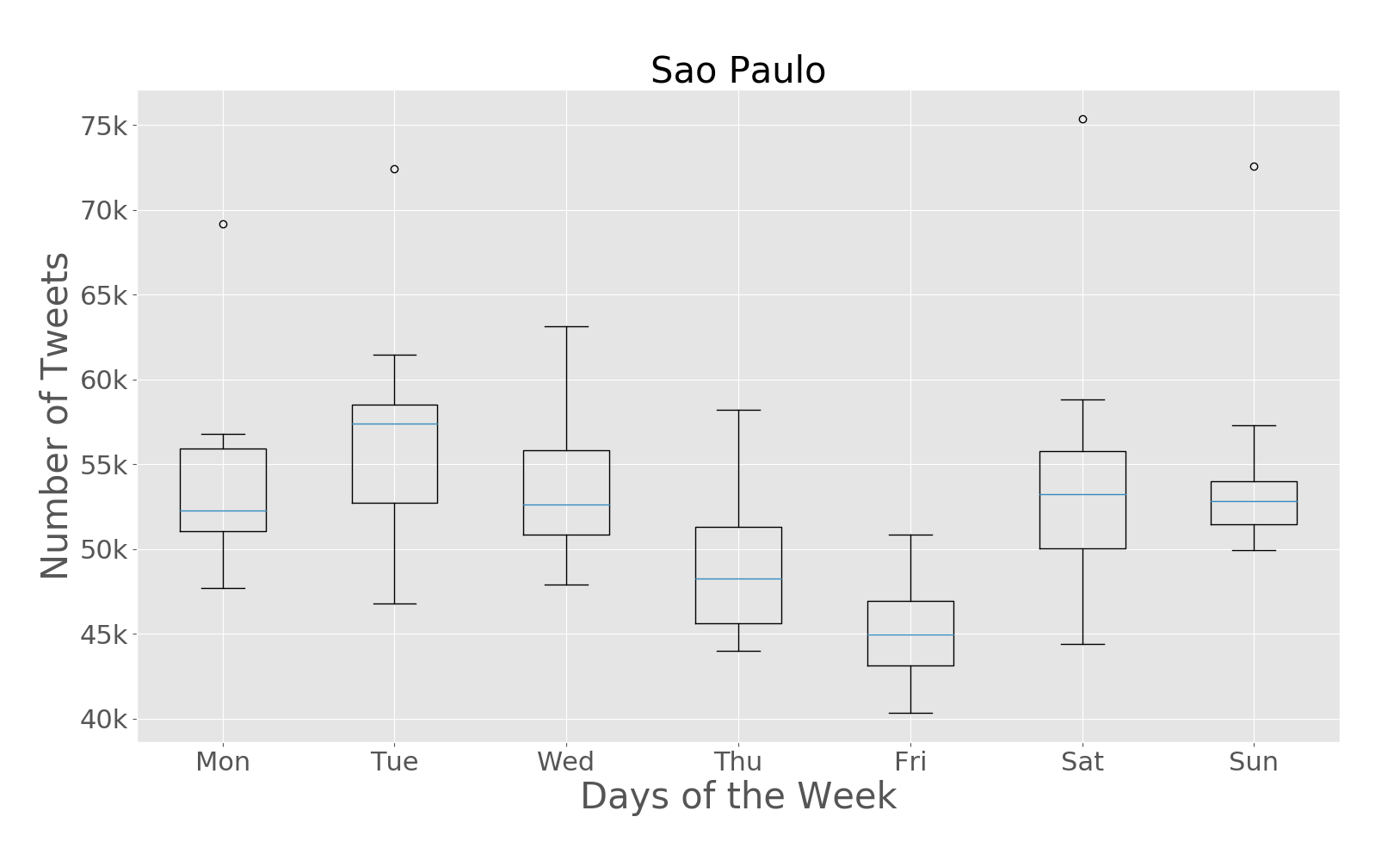}
        \caption{}
        \label{subfig:saopaulo_box_plot_day_of_week}
    \end{subfigure}

    \medskip

    \begin{subfigure}[htbp]{0.45\textwidth}
        \centering
        \includegraphics[width=1\linewidth]{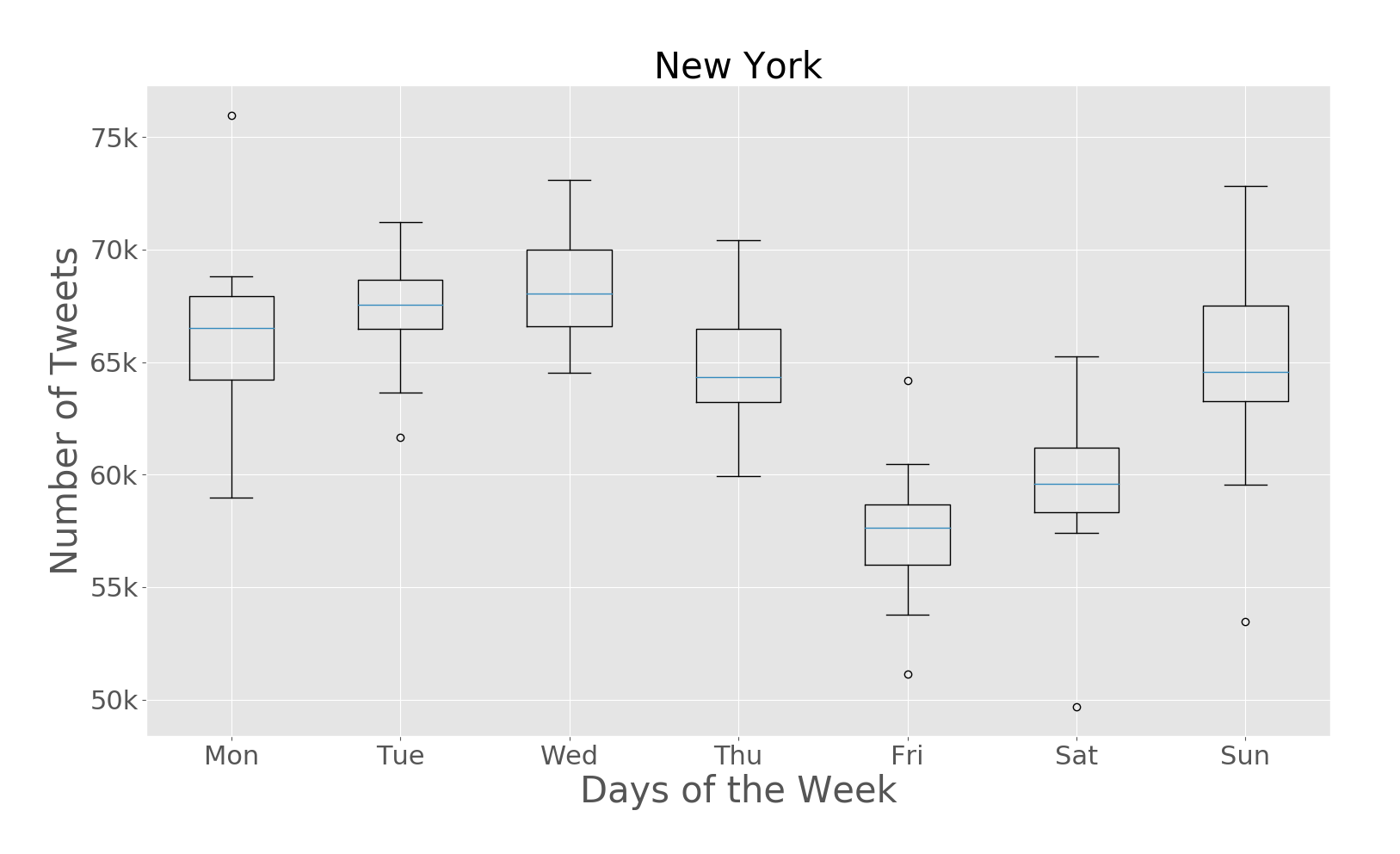}
        \caption{}
        \label{subfig:newyork_box_plot_day_of_week}
    \end{subfigure}
    \quad
    \begin{subfigure}[htbp]{0.45\textwidth}
        \centering
        \includegraphics[width=1\linewidth]{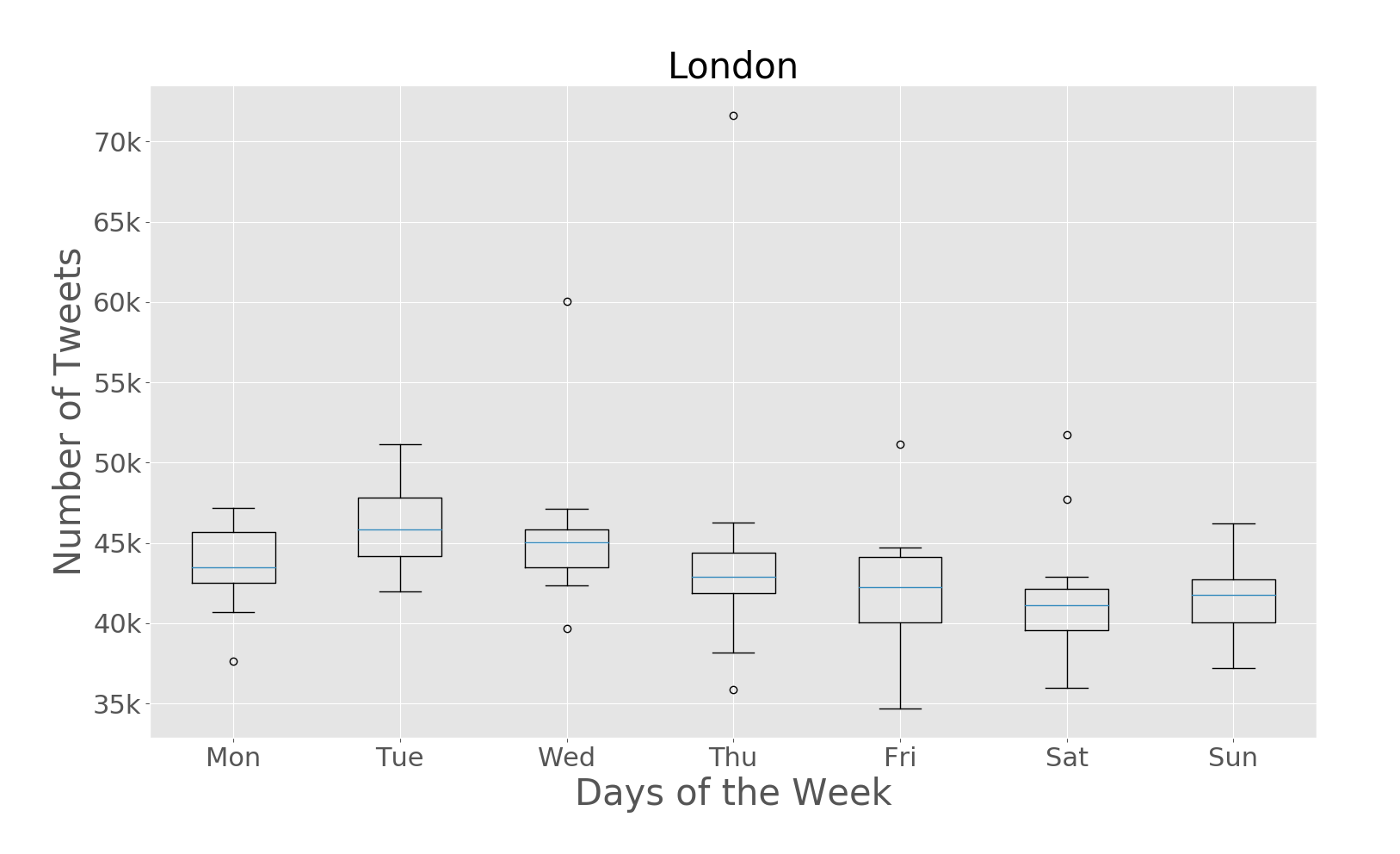}
        \caption{}
        \label{subfig:london_box_plot_day_of_week}
    \end{subfigure}

	\medskip
    
     \begin{subfigure}[htbp]{0.45\textwidth}
        \centering
        \includegraphics[width=1\linewidth]{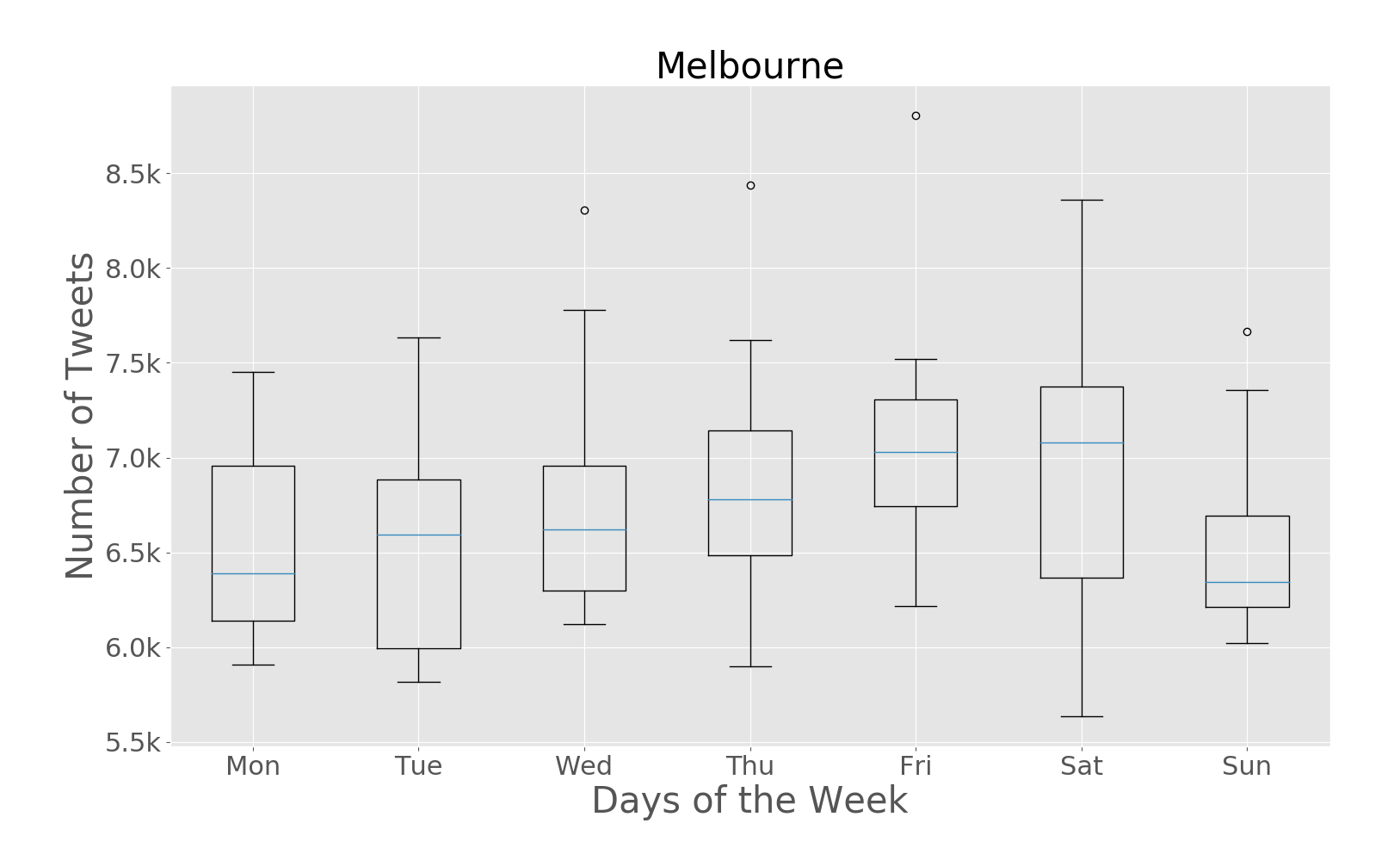}
        \caption{}
        \label{subfig:melbourne_box_plot_day_of_week}
    \end{subfigure}
    
\caption[Days-of-the-week box-plots for the volume of tweets]{Days-of-the-week box-plots for the volume of tweets (a) Rio de Janeiro (b) São Paulo (c) New York City (d) London (e) Melbourne}
\label{fig:box_plots_day_of_week}
\end{figure}

\begin{figure}[!htbp]
	\centering
	\begin{subfigure}[htbp]{0.45\textwidth}
		\centering
		\includegraphics[width=1\linewidth]{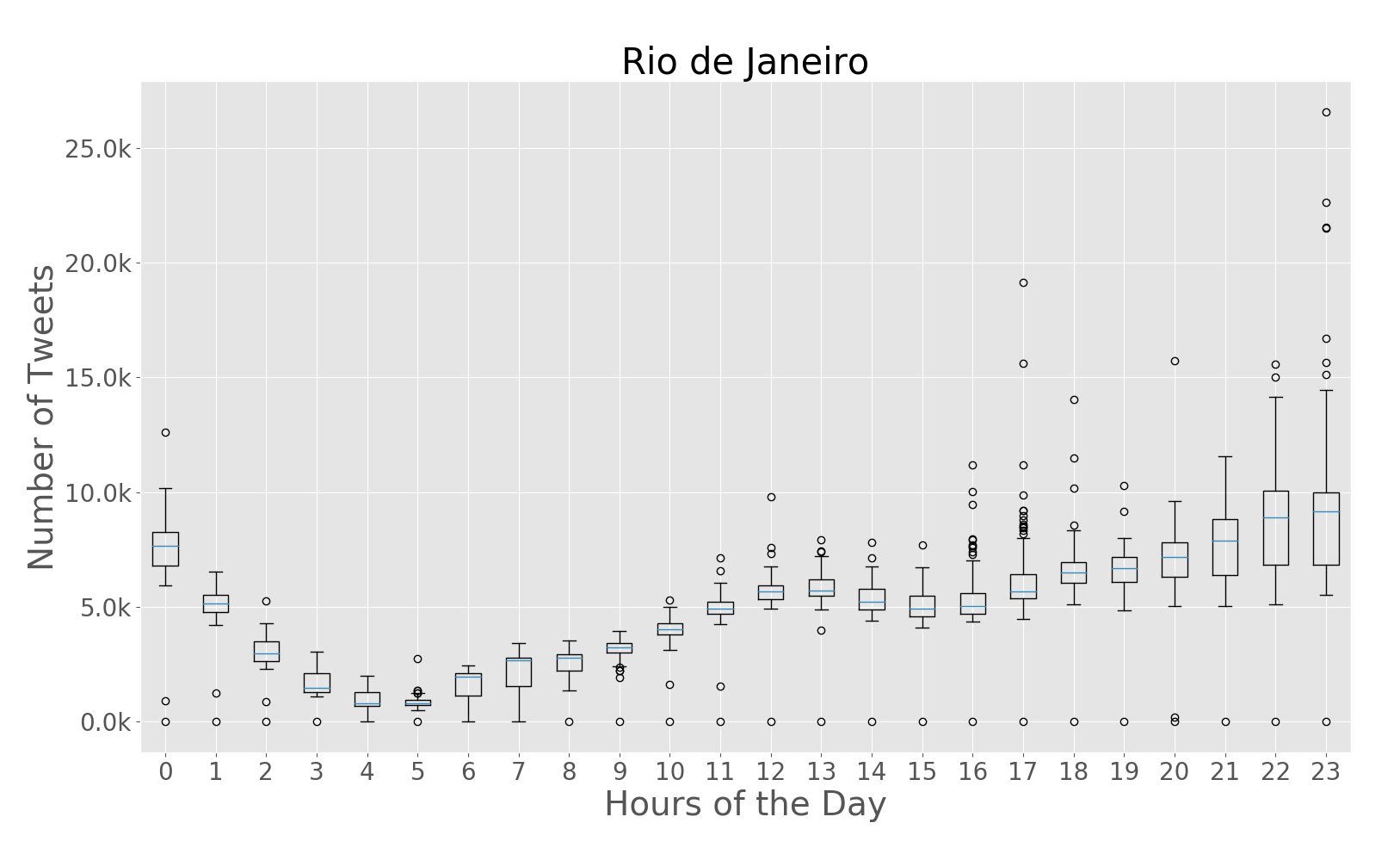}
		\caption{}
		\label{subfig:riodejaneiro_box_plot_hour_of_day}
	\end{subfigure}%
	\quad
	\begin{subfigure}[htbp]{0.45\textwidth}
		\centering
		\includegraphics[width=1\linewidth]{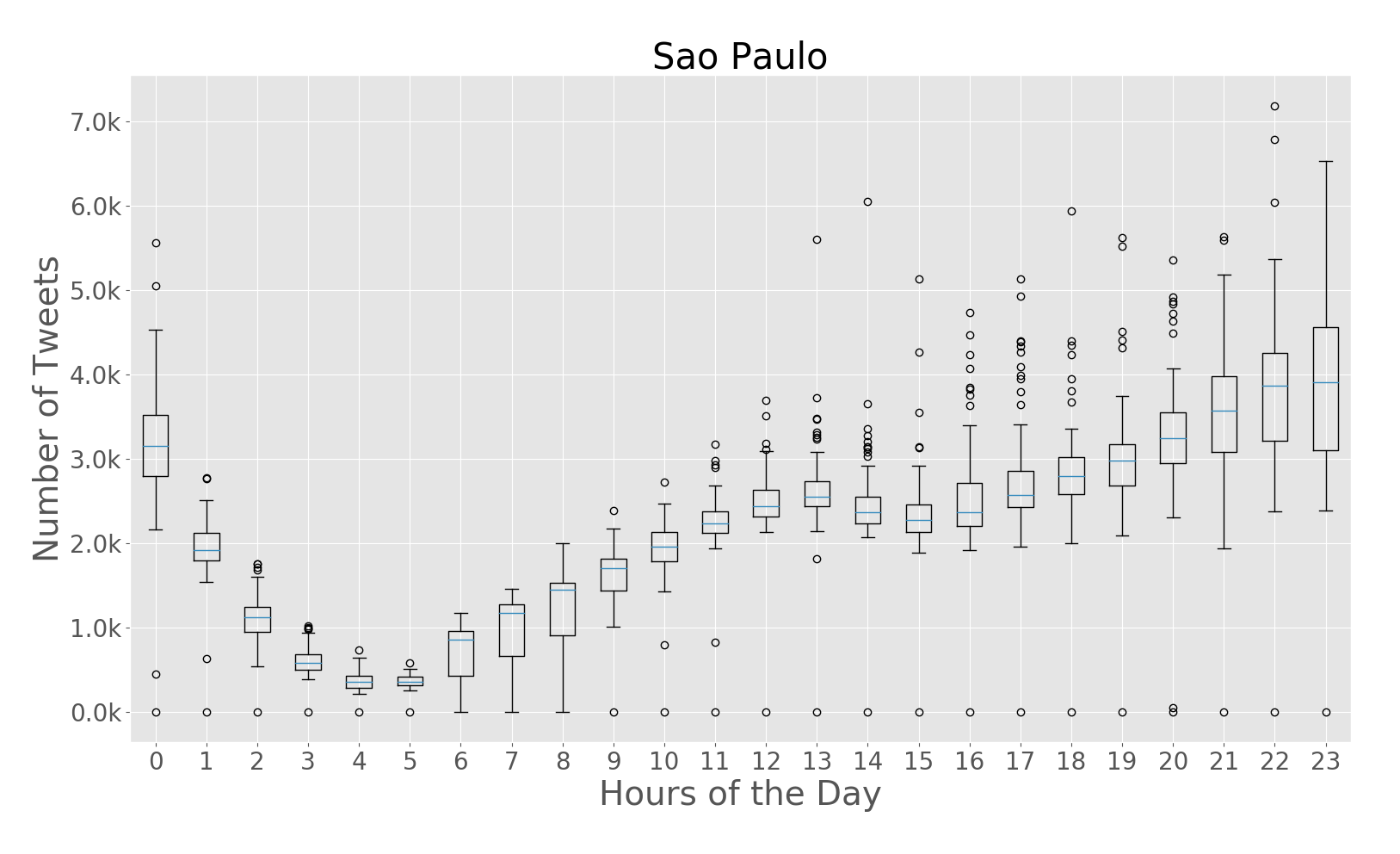}
		\caption{}
		\label{subfig:saopaulo_box_plot_hour_of_day}
	\end{subfigure}
	
	\medskip
	
	\begin{subfigure}[htbp]{0.45\textwidth}
		\centering
		\includegraphics[width=1\linewidth]{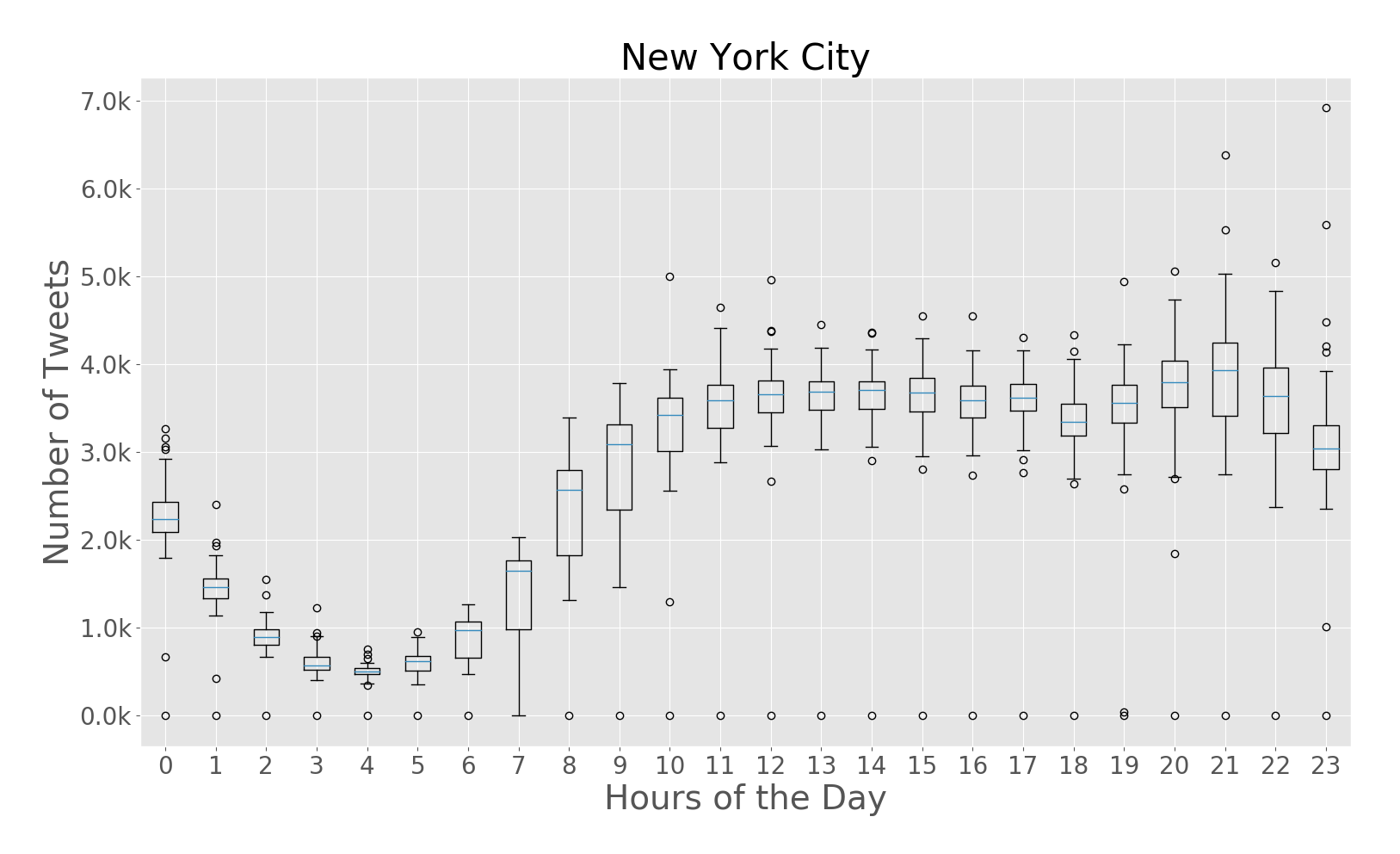}
		\caption{}
		\label{subfig:newyork_box_plot_hour_of_day}
	\end{subfigure}
	\quad
	\begin{subfigure}[htbp]{0.45\textwidth}
		\centering
		\includegraphics[width=1\linewidth]{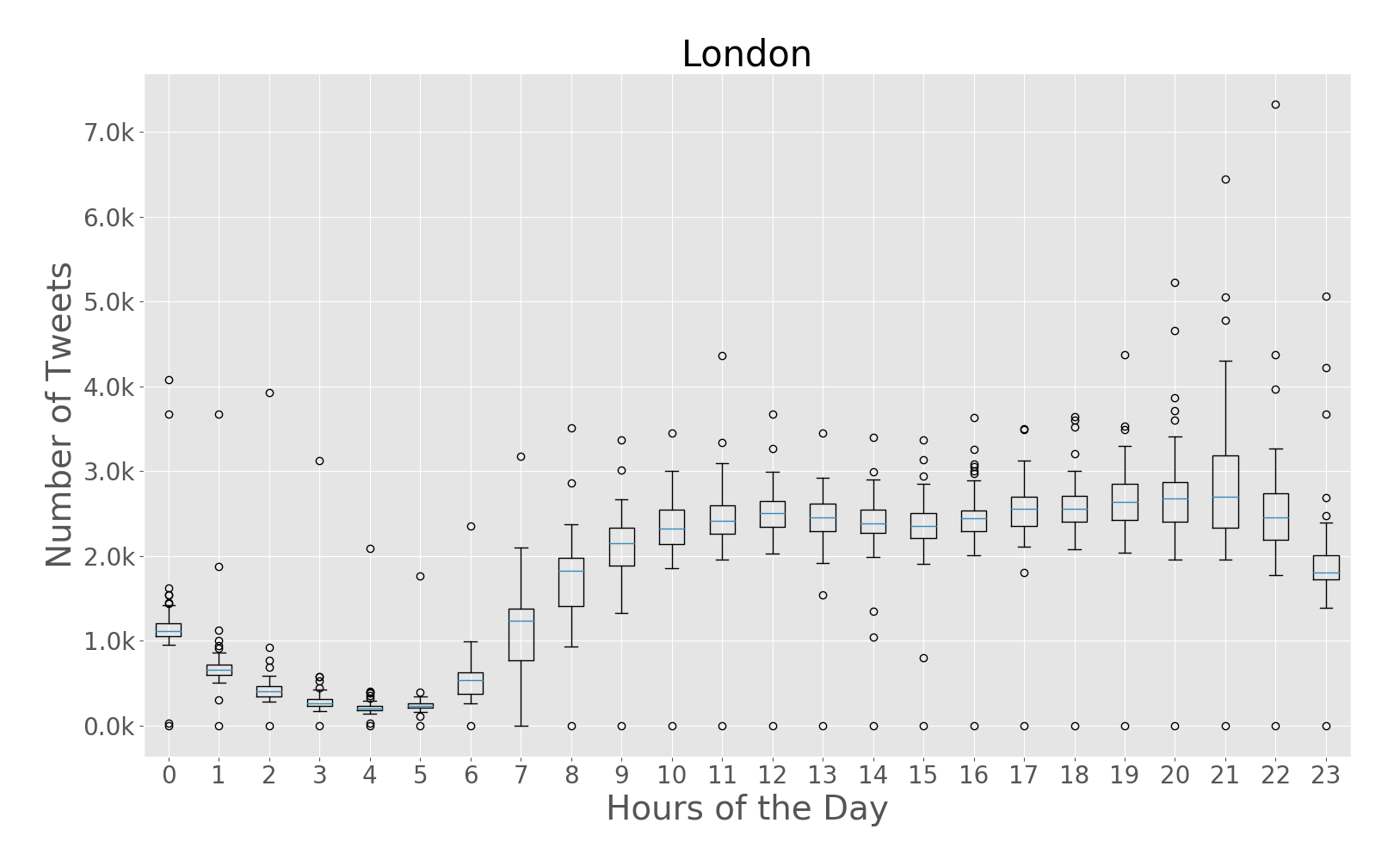}
		\caption{}
		\label{subfig:london_box_plot_hour_of_day}
	\end{subfigure}
	
	\medskip
	
	\begin{subfigure}[htbp]{0.45\textwidth}
		\centering
		\includegraphics[width=1\linewidth]{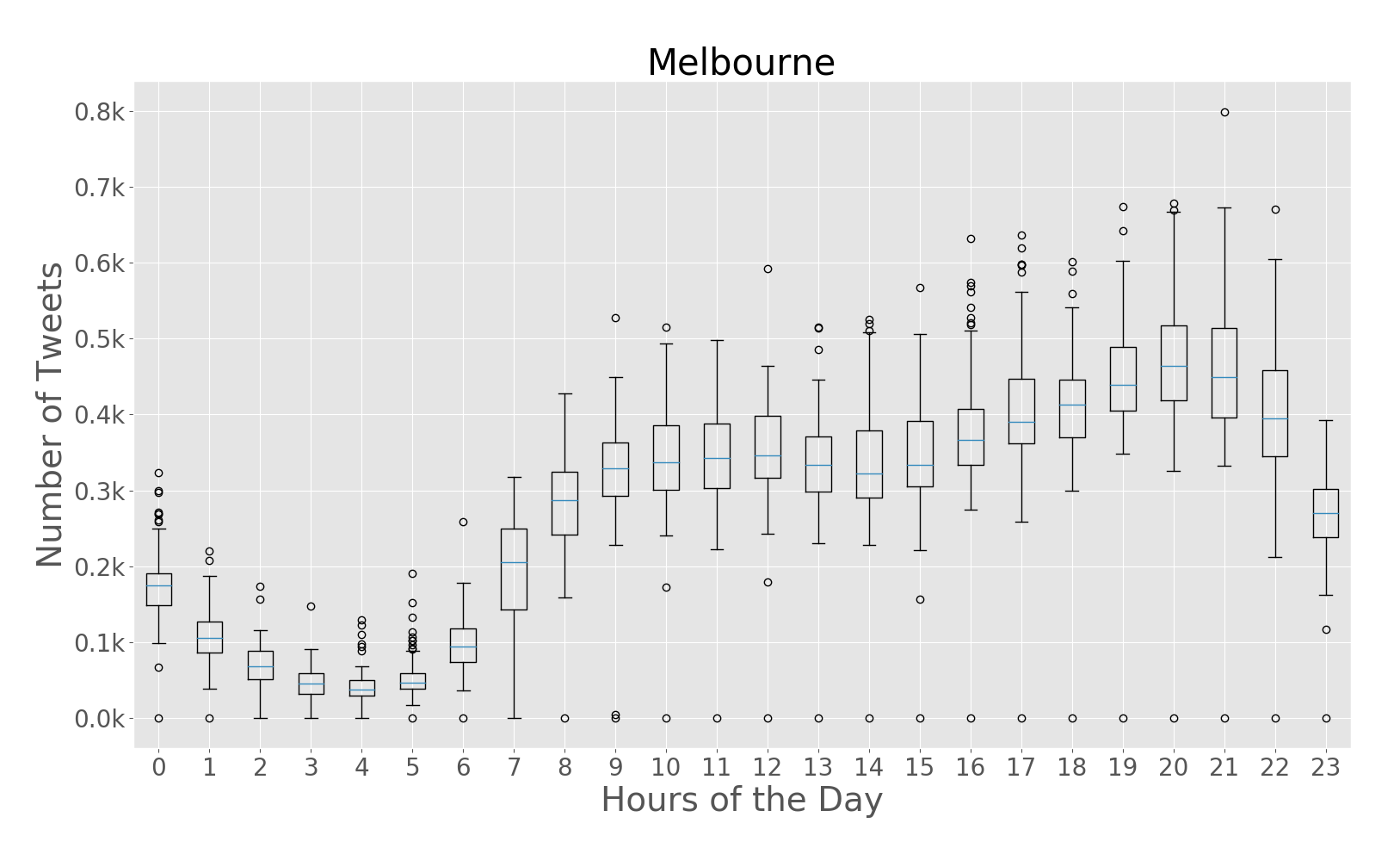}
		\caption{}
		\label{subfig:melbourne_box_plot_hour_of_day}
	\end{subfigure}
	
	\caption[Hour-of-the-day box-plots for the volume of tweets]{Hour-of-the-day box-plots for the volume of tweets (a) Rio de Janeiro (b) São Paulo (c) New York City (d) London (e) Melbourne}
	\label{fig:box_plots_hour_of_day}
\end{figure}

In order to understand the most active days and hours in Twitter, for all cities under this study, we aggregate the datasets by these attributes and represented the final results in a box-plot representation. This type of data visualization allows, in a standardized way, the displaying of distributions of data based on the six different values: (1) minimum and (2) maximum values for each day/hour regarding the activity on Twitter; (3) median value for the each day/hour, (4) first and (5) third quartiles as well as (6) the \gls{IQR}. Figures~\ref{fig:box_plots_day_of_week} and~\ref{fig:box_plots_hour_of_day} illustrated this type of data visualization for the whole three months of data collected. Taking into analysis the city of Rio de Janeiro, it was possible to observe and enhance Tuesdays as the day of the week where there is more activity on Twitter. Moreover, Fridays revealed to be the day less active, not only for the city of Rio de Janeiro, but for all remaining cities with exception of Melbourne. Particularly, the activity on Twitter in Melbourne is centered in the weekend days while the other cities the highest levels of activity is spread between week and weekend days. The interquartile range in the plots can tell us the amount of days whose activity was above and behold the median value, and through that we identify Rio de Janeiro and Melbourne as the cities where this phenomenon happen more times. São Paulo, New York City and London present an almost regular \gls{IQR} which means that the days of weeks are similarly regarding the activity on Twitter.

Looking at the hour-of-the-day box-plot (\ref{fig:box_plots_hour_of_day}), it is possible to verify an decrease in terms of activity on Twitter during the night period to all cities. More specifically, there were cases in which the volume of tweets was inexistent and based on this fact, two possible reason are suggested: (1) the absence of tweets during this period is explained through the zero activity of users in the city, regarding geo-located tweets; (2) the service on Twitter was in maintenance and due to that, any tweet was retrieved by the API. Although the observable increase of activity during day-time, the peak of it is similiar to all cities and it is established between the 19 and 23 hours.

\section{Content Composition}

Tweets although its classification as text messages, also contain other kind of \textit{metadata} which exploration of it can sometimes be transformed in added-value information. The \textit{metadata} present in a tweet is represented by the \textit{hashtags}, \textit{user mentions}, \textit{URLs} and \textit{media} attached to it. Other point to explore is the number of distinct users that contributed to the datasets composition. Users which number of posts are unnatural may sometimes be \textit{bots}. If there is a time pattern associated to the post of tweets by a user, for example, the user posts a tweet in a period of 5 minutes over the whole day, then this user is a potential \textit{bot}. The existence of \textit{bots} is not considered in this dissertation because the information provide by such automatic system can also be valuable. In this subsection, we demonstrated the distribution of users over the number of posts made by themselves, as well as the counts of the different type of \textit{metadata }contained in the data. 

\begin{figure}[h]
	\centering
	\begin{subfigure}[t]{0.45\textwidth}
		\centering
		\includegraphics[width=1\linewidth]{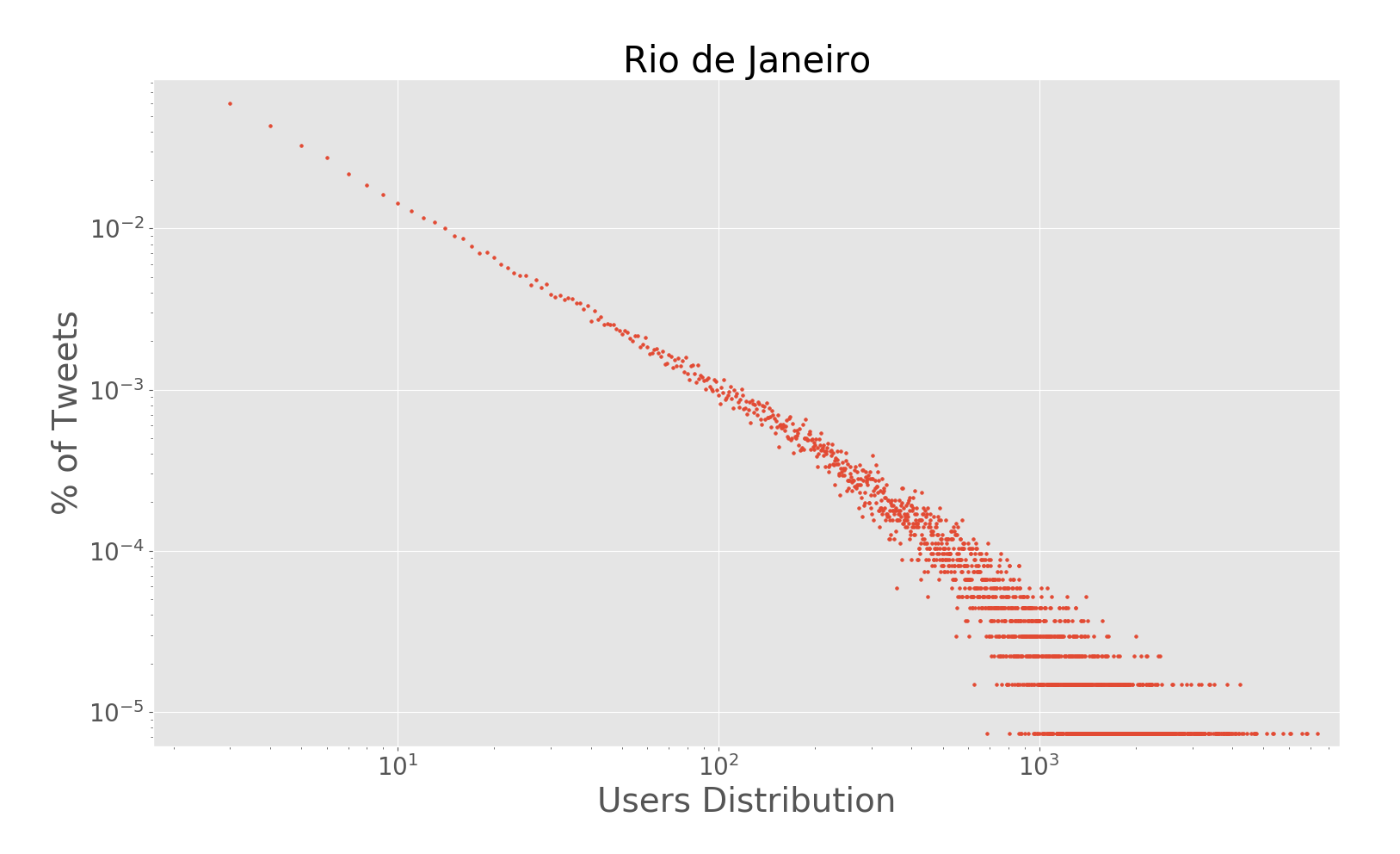}
		\caption{}
		\label{subfig:riodejaneiro_loglog_users}
	\end{subfigure}%
	\quad
	\begin{subfigure}[t]{0.45\textwidth}
		\centering
		\includegraphics[width=1\linewidth]{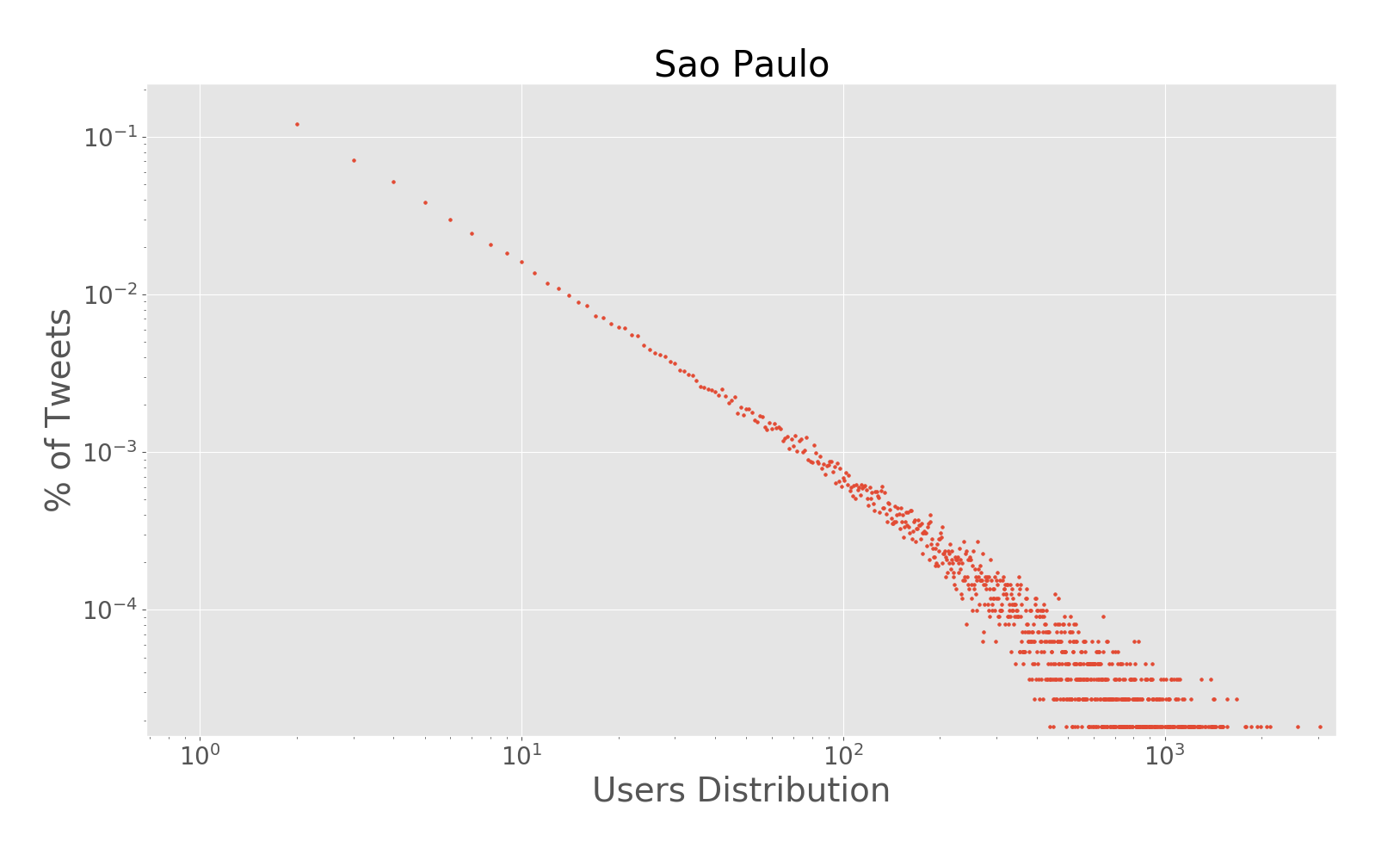}
		\caption{}
		\label{subfig:saopaulo_loglog_users}
	\end{subfigure}
	
	\medskip
	
	\begin{subfigure}[t]{0.45\textwidth}
		\centering
		\includegraphics[width=1\linewidth]{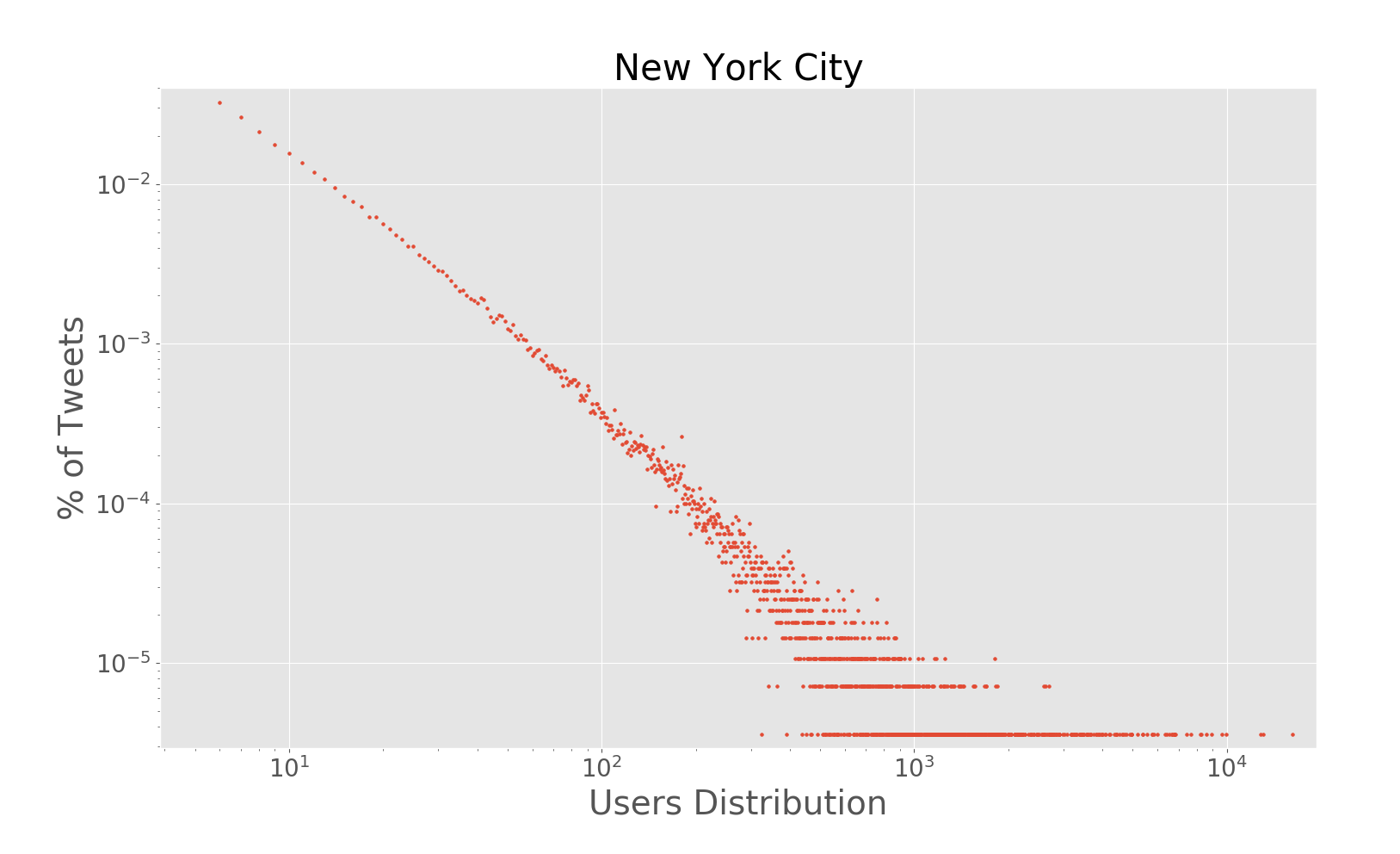}
		\caption{}
		\label{subfig:newyork_loglog_users}
	\end{subfigure}
	\quad
	\begin{subfigure}[t]{0.45\textwidth}
		\centering
		\includegraphics[width=1\linewidth]{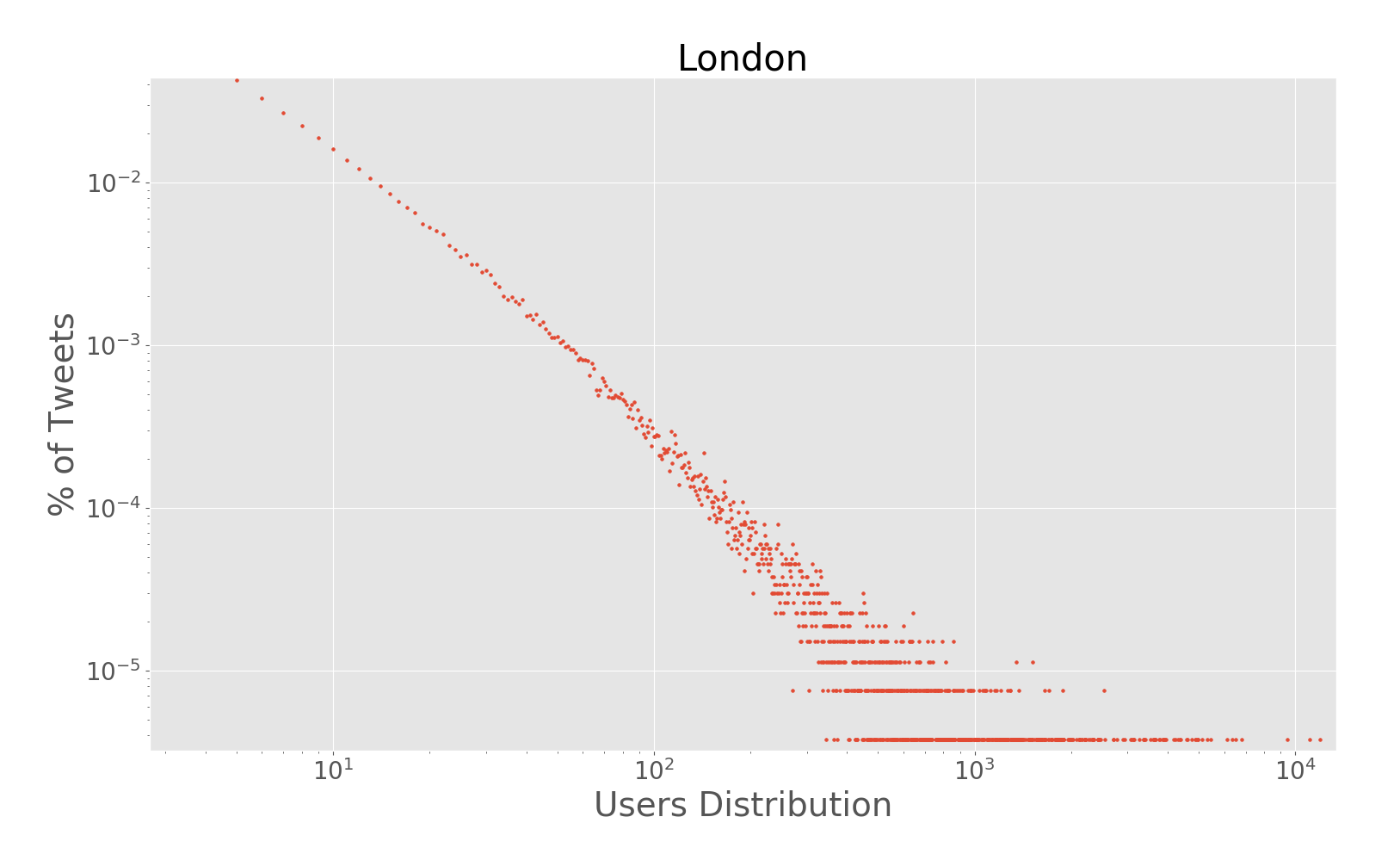}
		\caption{}
		\label{subfig:london_loglog_users}
	\end{subfigure}
	
	\medskip
	
	\begin{subfigure}[t]{0.45\textwidth}
		\centering
		\includegraphics[width=1\linewidth]{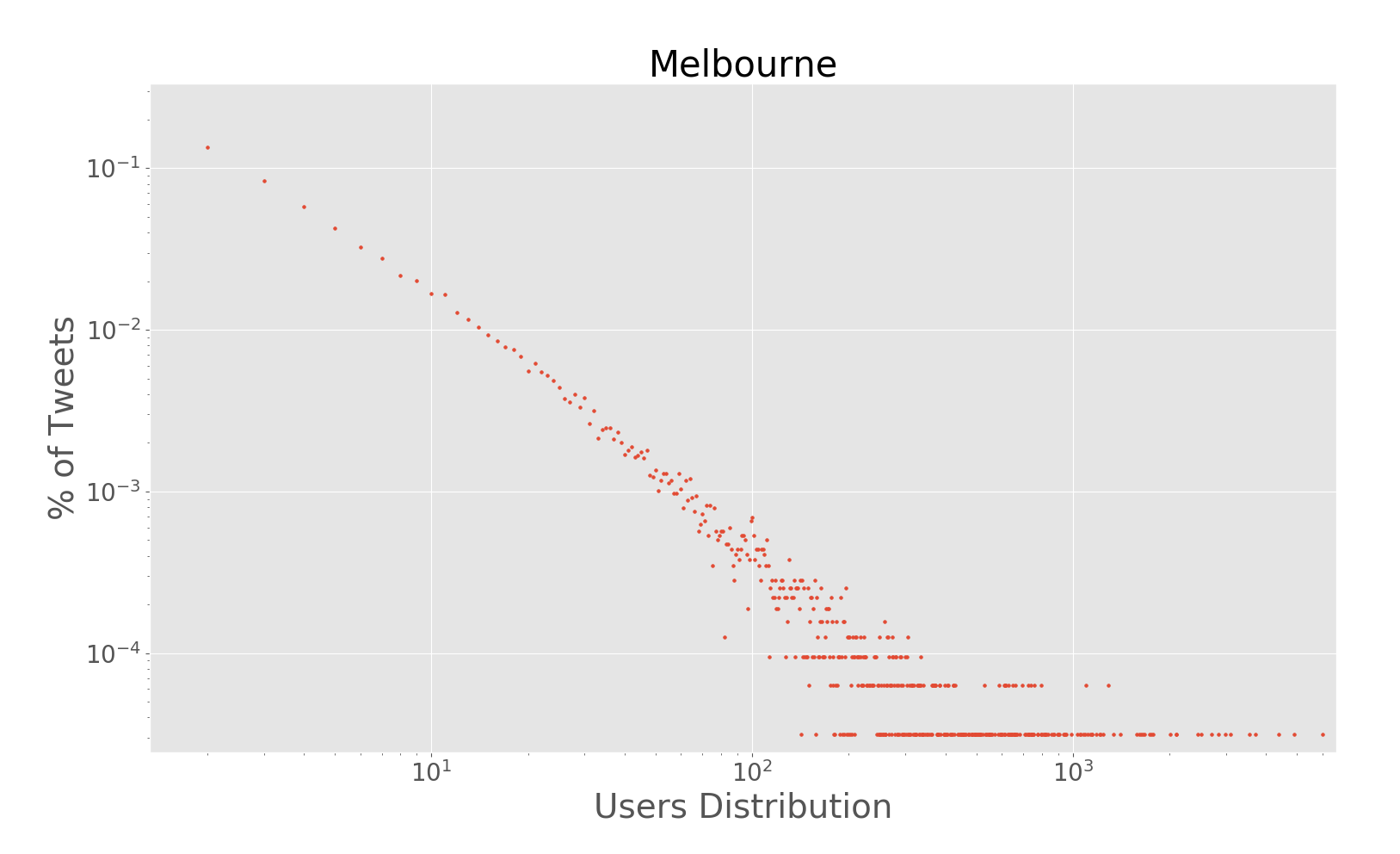}
		\caption{}
		\label{subfig:melbourne_loglog_users}
	\end{subfigure}
	
	\caption[Log-log plots of users distribution]{Log-log plots for the users distribution over the number of tweets posted (a) Rio de Janeiro (b) São Paulo (c) New York City (d) London (e) Melbourne}
	\label{fig:loglog-plots-users}
\end{figure}

Social media platforms present similar characteristics between themselves. One of the most studied ones is the behviour of the its users activity in its services (social media services). The visualization of users activity usually is similar to the power-law distribution long tail~\cite{muchnik2013origins}. Here, we tried to reproduce such visualization in order to establish this kind of correlation as so to prove this behaviour over social media services. The results are present in Figure~\ref{fig:loglog-plots-users}. Each city proved to have a high number of users with few posts and that is observable in the long-tail showed in the cities corresponding sub-figures (~\ref{subfig:riodejaneiro_loglog_users},~\ref{subfig:saopaulo_loglog_users},~\ref{subfig:newyork_loglog_users},~\ref{subfig:london_loglog_users},~\ref{subfig:melbourne_loglog_users}).

The counts and percentages of users that have posted a certain number of tweets was calculated in order to assure the trustiness of the aforementioned distribution. Rio de Janeiro although the highest number of tweets in the datasets only was composed by 135,449 distinct users followed by São Paulo with a lower number 110,352 individuals. The English speaking cities revealed to be very different comparatively to the Portuguese speaking cities in this factor. New York City dataset was composed by 279,554 distinct users, London presented 266,128 users and Melbourne only was composed by 31,733 individuals. Looking at these numbers, we may conclude that Rio de Janeiro has a high percentage of users with more than a certain number of tweets and following this assumption, the log-log distribution made to correlate the behaviour of a power-law distribution must be different from the other cities, at least the English speaking ones.

For example, the percentage of users that posted 20 tweets in a period of three months was almost 63\% for the city of Rio de Janeiro, São Paulo registered 75\%, New York City presented 84\%, London showed 87\% while Melbourne had 87\% of his users with that number of tweets shared. Only taking this example in consideration we proved the assumption mentioned before. The distributions also presented differences if the x-axis is considered. The scale at such axis is one magnitude higher for the English speaking cities, and this means that the number of users with lower number of tweets posted in a three months period is much higher than the users with the same number for the city of Rio de Janeiro.

The last analysis presented in this subsection is related to the \textit{metadata} contained in the tweets. Here, we want to characterize the different cities with respect to the amount of extra content used by the users in the posts and what kind of information such results suggests for each city.

Having this considered, we counted the volume of each element constituting the previously mentioned \textit{metadata} and calculate the percentage of tweets containing it. In Table~\ref{tab:metadata} are listed the counts and the corresponding percentage of it relatively to the datasets. The resulting analysis and results were performed over the tweets with the city's native language and located inside the bounding-box area used in the filtering process.
The most observable evidence in the results is the greater use of this elements in the English speaking cities. User mentions, as well as \textit{URLs} are the most used \textit{metadata}. This elements may suggest that citizens tend to tag other people in their messages when posting and also share information about certain topic through urls. Regarding the Brazilian cities, the \textit{metadata} usage is not so noticeable. This fact may me related to the number of users composing each dataset because, as it was previously mentioned, the English speaking cities possesses almost two times more users than the Brazilian cities and this characteristic contributes to the increase of this type of \textit{metadata} usage since when someone tag another one in a message, usually a re-post is sent tagging the person responsible by the starting of the conversation. To prove this so, an intensive study about social media tracking and mapping of the flow of each Twitter conversation is needed.

\begin{table}[htbp]
	\centering
	\caption{Percentage of Metadata composing the datasets}
	\label{tab:metadata}
	\resizebox{\textwidth}{!}{\begin{tabular}{l|c|cc|cc|cc|cc}
		\hline
		\multicolumn{1}{c|}{\multirow{2}{*}{\textbf{City}}} & \multicolumn{1}{l|}{\multirow{2}{*}{\textbf{Total}}} & \multicolumn{2}{c|}{\textbf{Hashtags (\#)}} & \multicolumn{2}{c|}{\textbf{User Mentions (@)}} & \multicolumn{2}{c|}{\textbf{URLs}} & \multicolumn{2}{c}{\textbf{Media}} \\ \cline{3-10} 
		\multicolumn{1}{c|}{} & \multicolumn{1}{l|}{} & \multicolumn{1}{c|}{\textbf{Total (tweets)}} & \textbf{\%} & \multicolumn{1}{c|}{\textbf{Total (tweets)}} & \textbf{\%} & \multicolumn{1}{c|}{\textbf{Total(tweets)}} & \textbf{\%} & \multicolumn{1}{c|}{\textbf{Total (tweets)}} & \textbf{\%} \\ \hline
		\textbf{Rio de Janeiro} & 11,060,136 & 504,835 & 4,56\% & 1,336,329 & 12,08\% & 1,783,060 & 16,12\% & 409,500 & 3,70\% \\
		\textbf{São Paulo} & 4,886,626 & 593,952 & 12,15\% & 1,030,341 & 21,08\% & 1,111,749 & 22,75\% & 325,385 & 6,66\% \\
		\textbf{New York City} & 5,956,355 & 1,697,416 & 28,50\% & 1,752,839 & 29,43\% & 2,839,794 & 47,68\% & 535,945 & 9,00\% \\
		\textbf{London} & 4,040,092 & 1,163,981 & 28,81\% & 1,744,051 & 43,17\% & 1,812,152 & 44,85\% & 465,610 & 11,52\% \\
		\textbf{Melbourne} & 629,424 & 195,967 & 31,13\% & 271,970 & 43,21\% & 258,278 & 41,03\% & 65,941 & 10,48\% \\ \hline
	\end{tabular}}
\end{table}

\section{Summary}
In this chapter we tried to identify interesting patterns and valuable information recurring only to the simple characteristics provided by a tweet: location, date of creation and \textit{metadata} content. First, it was possible to find out existing problems regarding the collection of geo-located tweets. During the last few years, the research community is proposing lots of studies using geo-located tweets, however, for the best of our knowledge none of them report the problem we identified regarding the geographic analysis. Our datasets represent only three months of data, however supporting in the analysis made, we conclude that the majority (> 70\%) of tweets are tagged with variable sized bounding-boxes instead of precisely geo-coordinates. This problem turns on difficult challenges when proposals of studies about human activity patterns our even human mobility need to be conducted using social media content, with respect to geo-located tweets.

Furthermore, we tried to instigate temporal patterns using the tweets already filtered and proved that it is possible to learn about remarkable events only seeing abrupt activity on Twitter for some days.

By studying the Twitter users distribution it was possible correlate the behaviour of it with the well-known power-law distribution. Nonetheless, we were able to identify high differences of the amount of tweets posted by some users. It is worth noting that some of them may be robot users, and due to the limited time in the implementation of our framework we did not perform any filtering regarding this problem.

Last but not least, a brief analysis of the \textit{metadata} was performed in order to see the amount of possible topics identified on it (hashtags), the volume of tweets mentioning another user and how many information can be share through the use of urls in this microblog. \textit{Hashtags} combined with temporal analysis can provide quickly and easier identification of remarkable events while \textit{user mentions}, can prove that people are using microblogs services as a communication service and for this reason human behaviour studies can be explored through this platform - Twitter. 
\chapter{Text Analytics Experiments}
\label{chap:experiments}

\minitoc \mtcskip \noindent

In this section we present two different text analytics experiments regarding topic modelling and travel-related tweets classification. First, we manually labelled topics to characterize two Brazilian cities, namely Rio de Janeiro and São Paulo. Later on, we built two travel-related classification models to discriminate tweets for two different speaking languages. For both experiments related to travel classification we manually annotated a English-speaking and a Portuguese-speaking training and test datasets. The remainder sections described each experiment as well as discussion and analysis of the obtained results.



\section{Topic Modelling}\label{sec:topic_modeling}
In this section we describe the experiment of automatic characterization of tweets in two different Brazilian cities, Rio de Janeiro and São Paulo. Here, as previous mentioned in Section~\ref{sec:topic_modelling_framework}, we use \gls{LDA} model to find out which are the latent topics in both cities.
We conduct this experiment using data collected from a period of two months, between March 12 and May 12, 2017. After the data filtering and text pre-processing steps, we obtain a total of 6.6M tweets for Rio de Janeiro and 2.7M tweets for São Paulo.

 We tried training LDA model with 5, 10, 20, 25 and 50 latent topics and manually inspected the top most probable words for each topic. Models with 5, 10, 20 and 25 presented a high number number of overlap terms between topics. Therefore, we opted to proceed with the experiment using 50 latent topics. The number of iterations to train the model was set to 20, in line with the work of Lansley and Longley~\cite{lansley2016geography}.
	
\subsection{Results and Analysis}
\label{subsec:lda_results}
\begin{table}[!ht]
	\centering
	\caption{Example of the topics labels}
	\label{tab:topics_classification}
	\resizebox{\textwidth}{!}{
		\begin{tabular}{c|c}
			\hline
			\textbf{\begin{tabular}[c]{@{}c@{}}Words\\ (20 words most frequent words)\end{tabular}} & \textbf{\begin{tabular}[c]{@{}c@{}}Topic\\ Labels\end{tabular}} \\ \hline
			\begin{tabular}[c]{@{}c@{}}paulo, vai, hoje, dia, jogo, ser, melhor, time, vamo, brazil, \\ todo, santo, brasil, gol, cara, aqui, agora, corinthiam, ano, palmeiro, vem, ...\end{tabular} & \begin{tabular}[c]{@{}c@{}}Sports and\\ Games\end{tabular} \\ \hline
			\begin{tabular}[c]{@{}c@{}}vou, dia, dormir, queria, hoje, ficar, casa, semano, quero, ter, \\ ainda, hora, agora, sono, aula, acordar, acordei, cedo, fazer, prova, ...\end{tabular} & \begin{tabular}[c]{@{}c@{}}Wake-up\\ Messages\end{tabular} \\ \hline
			\begin{tabular}[c]{@{}c@{}}top, social, artist, vote, the, award, army, bom, voting, doi, \\ bogo, oitenta, sipda, today, vinte, prepara, cypher, oito, quatro, man, ...\end{tabular} & \begin{tabular}[c]{@{}c@{}}Voting and\\ Numbers\end{tabular} \\ \hline
			\begin{tabular}[c]{@{}c@{}}marco, nada, falar, emilly, gente, quer, nao, pessoa, nunca, fala, \\ vai, falando, sobre, chama, agora, manda, vem, mensagem, vivian, bbb, ...\end{tabular} & \begin{tabular}[c]{@{}c@{}}Big Brother\\ Brazil 2017\end{tabular} \\ \hline
			\begin{tabular}[c]{@{}c@{}}paulo, brazil, sao, santo, vila, just, parque, posted, photo, shopping, \\ paulista, centro, bernardo, jardim, cidade, avenido, praia, santa, campo, academia\end{tabular} & \begin{tabular}[c]{@{}c@{}}Tourism and\\ Places\end{tabular} \\ \hline
		\end{tabular}
	}
\end{table}




The LDA model does not provide a semantic label for each topic, such as "topic x is about Sports". We inspected the most frequent words of each topic and manually assigned a semantic label. Table~\ref{tab:topics_classification} presents the most frequent words for 5 random topics and the corresponding labels manually assigned. For instance, the topic containing frequent words such as "gol", "jogo", "time" is labelled as "Sports and Games". We follow the semantic taxonomy proposed by Lansley and Longley~\cite{lansley2016geography} to manually labelling each topic.

We observed that many different latent topics were about the same semantic subject but at different levels of granularity. For instance, "European Football \emph{vs} Brazilian Football". We decided to manually aggregate these overlapping topics to create a simpler and easier to analyse list of topics, resulting in a total of 29 aggregated topics. We performed this process to both cities independently.

\begin{table}[!ht]
	\centering
	\caption[LDA model final results]{Final results of the LDA topics aggregation. (*) topic labels different from Lansley and Longley~\cite{lansley2016geography}}
	\label{tab:topic_labels}
	\resizebox{\textwidth}{!}{
		\begin{tabular}{l|S[table-format=7.0]S[table-format=2.2]|S[table-format=7.0]S[table-format=2.2]|rS[table-format=2.2]}
			\hline
			\multicolumn{1}{c|}{\multirow{2}{*}{\textbf{Topic Label}}} & \multicolumn{2}{c|}{\textbf{Rio de Janeiro}} & \multicolumn{2}{c|}{\textbf{S\~ao Paulo}} & \multicolumn{1}{c}{\multirow{2}{*}{\textbf{Diff (\%)}}} \\ \cline{2-5}
			\multicolumn{1}{c|}{} & \textbf{No. Tweets} & \textbf{Percentage (\%)} & \textbf{No. Tweets} & \textbf{Percentage (\%)} & \multicolumn{1}{c}{} \\ \hline
			Academic Activities (*) & 101,590 & 1.54 & 90,616 & 3.30 & -1.76 \\
			Actions or Intentions & 600,030 & 9.12 & 128,710 & 4.69 & \textbf{+4.43} \\
			Antecipation and Socialising & 132,606 & 2.01 & 0 & 0.00 & \textbf{+2.01} \\
			BBB17 (*) & 122,054 & 1.85 & 68,385 & 2.49 & -0.64 \\
			Body, Appearances and Clothes & 160,342 & 2.44 & 71,447 & 2.60 & -0.17 \\
			Food and Drink & 167,204 & 2.54 & 58,407 & 2.13 & +0.41 \\
			Health & 119,013 & 1.81 & 0 & 0.00 & \textbf{+1.81} \\
			Holidays and Weekends & 104,695 & 1.59 & 79,610 & 2.90 & -1.31 \\
			Informal Conversations & 272,502 & 4.14 & 138,848 & 5.06 & -0.92 \\
			Live Shows, Social Events and Nightlife & 359,342 & 5.46 & 140,240 & 5.11 & +0.35 \\
			Mood & 139,287 & 2.12 & 138,399 & 5.04 & \textbf{-2.92} \\
			Movies and TV & 285,198 & 4.33 & 39,778 & 1.45 & \textbf{+2.89} \\
			Music and Artists & 84,407 & 1.28 & 78,142 & 2.85 & 1.56 \\
			Negativism, Pessimism and Anger(*) & 229,104 & 3.48 & 183,050 & 6.67 & \textbf{-3.18} \\
			Numbers, Quantities and Classification & 86,897 & 1.32 & 78,160 & 2.85 & -1.53 \\
			Optimism and Positivism & 106,714 & 1.62 & 39,725 & 1.45 & +0.18 \\
			Personal Fellings & 375,735 & 5.71 & 532,331 & 19.38 & \textbf{-13.67} \\
			Politics & 81,254 & 1.23 & 46,758 & 1.70 & 0.47 \\
			Relationships and Friendship (*) & 1,524,804 & 23.17 & 187,541 & 6.83 & \textbf{+16.34} \\
			Religion & 183,174 & 2.78 & 66,788 & 2.43 & +0.35 \\
			Routine Activities & 334,216 & 5.08 & 82,421 & 3.00 & +2.08 \\
			Slang and Profinities & 241,676 & 3.67 & 44,620 & 1.62 & +2.05 \\
			Social Media Applications & 105,809 & 1.61 & 44,073 & 1.60 & +0.01 \\
			Sport and Games & 382,479 & 5.81 & 133,047 & 4.84 & +0.97 \\
			Tourism and Places & 59,288 & 0.90 & 86,519 & 3.15 & -2.25 \\
			Transportation and Travel & 130,261 & 1.98 & 63,923 & 2.33 & -0.35 \\
			Weather (*) & 91,302 & 1.39 & 42,588 & 1.55 & -0.16 \\
			Shopping (*) & 0 & 0.00 & 44,470 & 1.62 & \textbf{-1.62} \\
			Voting (*) & 0 & 0.00 & 37,687 & 1.37 & \textbf{-1.37} \\ \hline
			\textbf{Total} & 6,580,983 & 100.00 & 2,746,283 & 100.00 &  \\			\hline
		\end{tabular}
	}
\end{table}

The resulting list of topics and their distribution (number of tweets) for both cities is depicted in Table~\ref{tab:topic_labels}. We first observe that the majority of topics is common in both cities, with exception of 4 topics:  Weather and Shopping are not discussed in Rio de Janeiro; Antecipation \& Socializing and Health are not discussed in São Paulo. We were not able to assign labels using the Lansley and Longley~\cite{lansley2016geography} taxonomy to 7 different topics. In such cases, we created our own labels, e.g., "BBB17" is about a highly popular reality TV show in Brazil named Big Brother Brazil. 

There is a wide range of topics covered by both cities, from "Food and Drink", "Politics" and "Religion" to "Sports and Games" or "Transportation and Travel". The most talked topics in Rio de Janeiro are "Relationships and Friendship", "Actions and Intentions" and "Sports and Games", while in São Paulo, the most talked about topics are "Personal Feelings", "Relationships and Friendship" and "Negativism, Pessimism and Anger". Comparing both cities, the topics with higher relative difference are "Relationships and Friendships" (+16\% in Rio de Janeiro) and "Personal Feelings" (+13\% in São Paulo).

\begin{figure}[!t]
	\centering
	\begin{subfigure}{0.49\textwidth}
		\centering
		\includegraphics[width=1.0\linewidth]{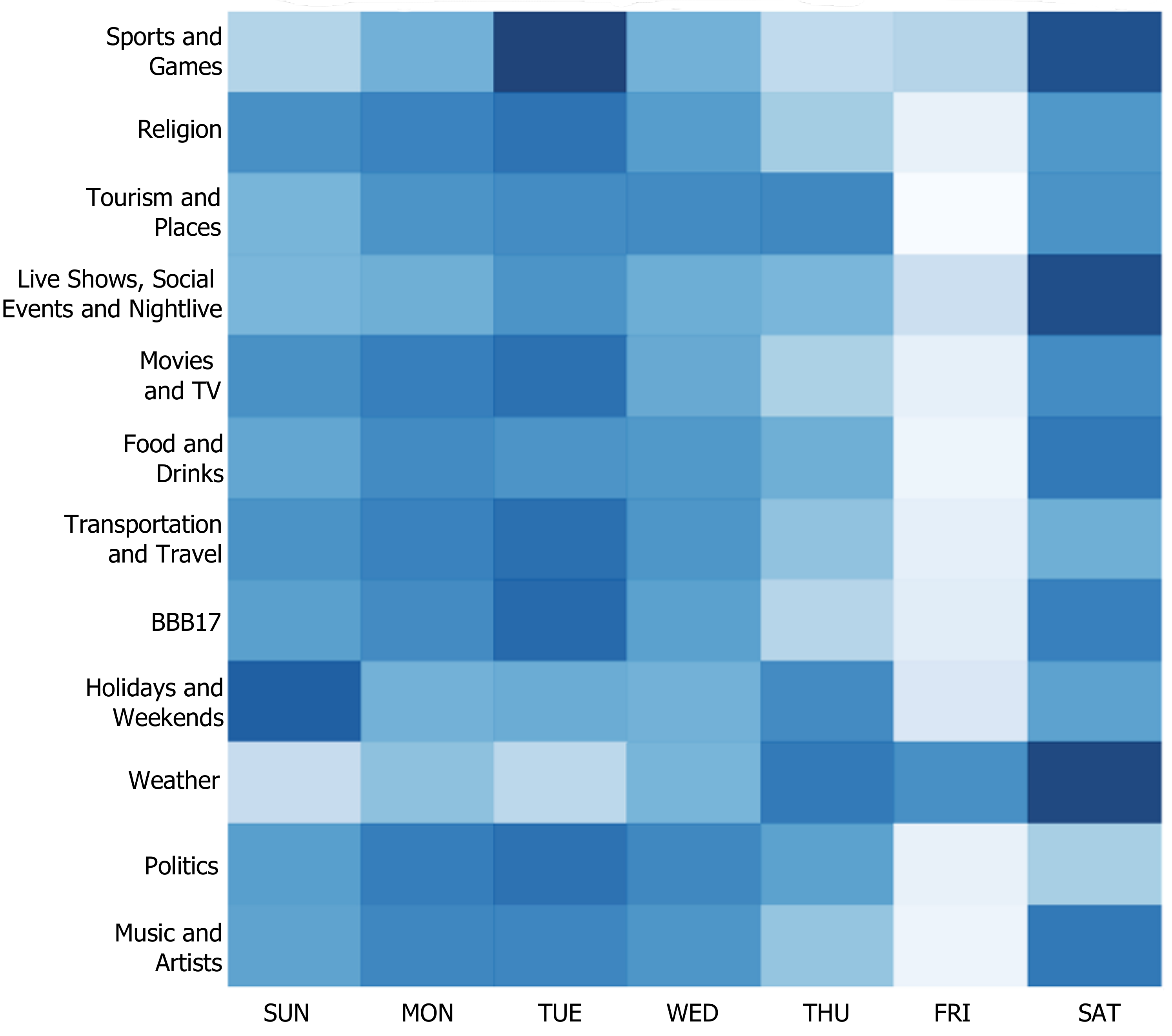}
		\caption{Rio de Janeiro}
		\label{fig:rio}
	\end{subfigure}
	\begin{subfigure}{0.49\textwidth}
		\centering
		\includegraphics[width=1.0\linewidth]{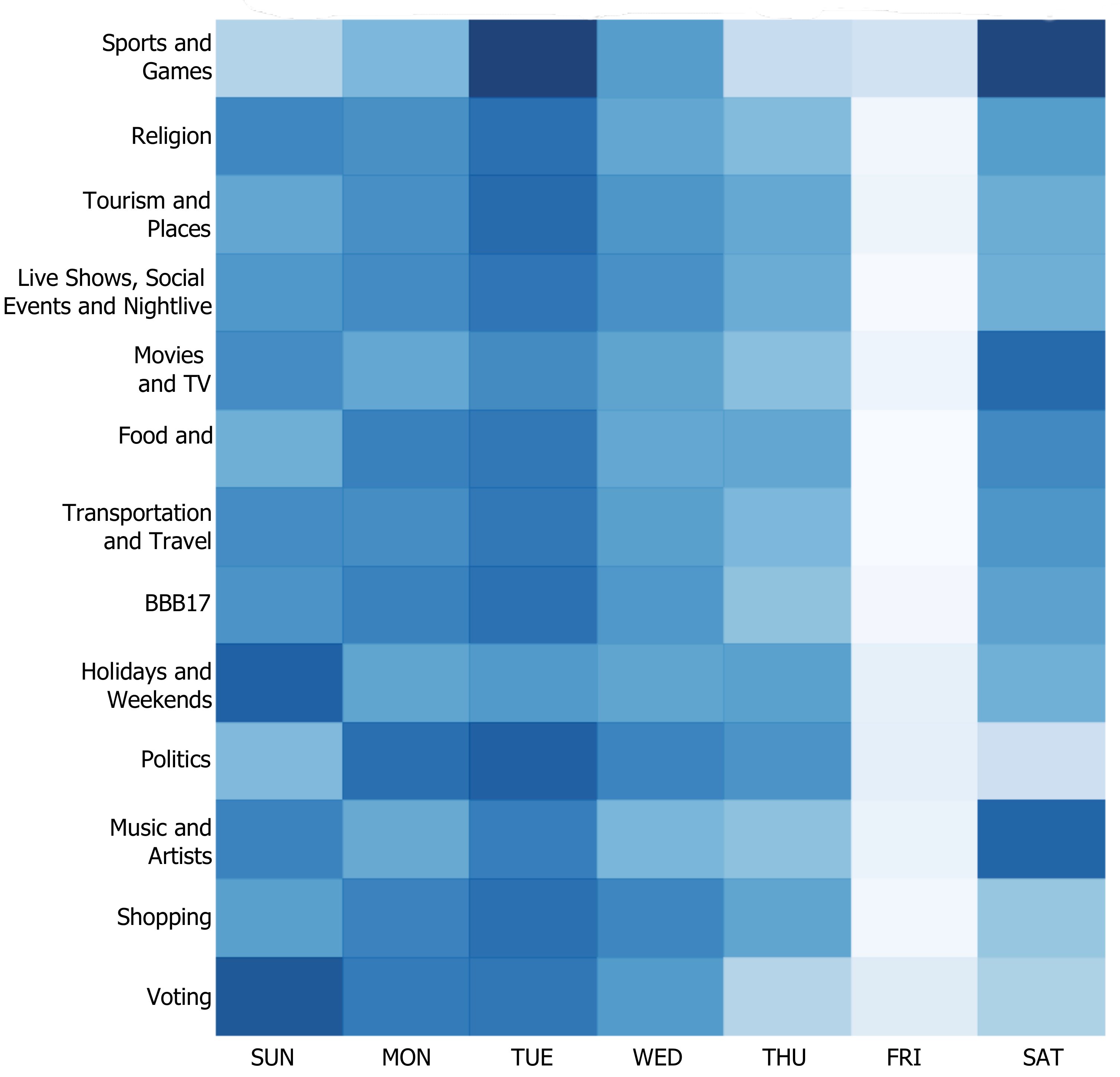}
		\caption{São Paulo}
		\label{fig:sp}
	\end{subfigure}
	\caption[Day-of-the-week Twitter activity]{Day-of-the-week activity per each topic in both cities}
	\label{fig:topics_heat_maps}
\end{figure}

We also produced a day-of-the-week temporal distribution of topics in both cities as depicted in Figure~\ref{fig:topics_heat_maps}. We selected 12 topics for Rio de Janeiro and 13 for São Paulo that are more prone to temporal shift of popularity, such as "Religion" and "Sports and Games" which are presumably more popular on the weekends. For both cities the topic "Sports and Games" is more mentioned on Tuesdays and Saturdays. Indeed, this observation correlates with Tuesdays where \textit{UEFA Champions League} competition happens and Saturdays when occur \textit{Brazilian Football League} matches. Also in both cities, "Holidays and Weekends" presented Sundays as the day where more people talk about it, while "Religion" and "Tourism and Places" are less prone to be talked about in Fridays, which is similar to all the remaining topics. "Live Shows, Social Events and Nightlife" are more talked about on Saturdays in Rio de Janeiro, while on São Paulo we identify Tuesdays as the day more popular. Sundays and Tuesdays are the days when the well-famous reality show TV program is emitted, and due to that exists more popularity in these days for the "BBB17" topic. "Weather" topic presents more activity in Saturday probably due to people going out to take a walk.

\subsection{Final Remarks}
To the best of our knowledge, this is the first large scale analysis of topic modelling of geo-located tweets from Rio de Janeiro and São Paulo. Most of the topics are common to both cities. Is is interesting to notice that people in Brazil post geo-located tweets about general purpose topics, such as reality TV show, health issues and relationship and friendship. Transportation and travel are marginal topics with less 2\% of relative frequency in Rio de Janeiro and 2.3\% in São Paulo.



This experiment demonstrates the capability of our framework to handle different topic modelling analysis under unregulated and non-conventional data such as the content found in most social media. The application of topic modeling technique to tweets from two different cities enables interesting comparisons between them since the whole analytics process accounts for what inhabitants talk about in their social networks. Through these analysis, cities' services are capable of monitoring human behaviour, activity patterns as well as of identifying regions where there may be some levels of intolerance on certain topics, making it possible to trigger preventive measures to solve problems in those specific areas.

Future direction for this research will include application of spatio-temporal aggregation methods over both datasets in order to create meta-documents (tweets group by day/hour/location) and verify whether results can be different taking into consideration temporal and spatial factors. To pursue this, it is required that a large dataset for both cities is available, which is expectable only in mid- to long-term.

\section{Travel-related Classification}
\label{sec:travel_related_classification}
The main goal of this section is to describe the experiments conducted to discriminate travel-related tweets in Rio de Janeiro, São Paulo and New York City. Considering the volume of the collected data for each scenario, it is necessary to automatically identify tweets whose content somehow suggests to be related to the transportation domain. Conventional approaches would require us to specify travel-related keywords to classify such tweets. On the contrary, our approach consisted in training a classification model to automatically discriminate travel-related tweets from non-related ones. 

One big challenge always present in text analysis is the sparse nature of data, which is especially the case in Twitter messages. Conventional techniques such as bag-of-words tend to produce sparse representations, which become even worse when data is composed by informal and noisy content.

Word embeddings, on the other hand, is a text representation technique that tries to capture syntactic and semantic relations from words. The result is a more cohesive representation where similar words are represented by similar vectors. For instance, \emph{"taxi"/"uber"}, \emph{"bus/busão/ônibus"}, \emph{"go to work"/"go to school"/"ir para a escola"} would yield similar vectors respectively.
We are particularly interested in exploring the characteristics of word embeddings techniques to understand which extent it is possible to improve the performance of our classifier to capture such travel-related expressions. In the reminder subsections, we describe two different text classification experiments following distinct approaches across two speaking languages - Portuguese and English. We trained the word embeddings using all tweets from the two different use cases with support of the Python's library, \texttt{Gensim}, and window size of 2.

Support Vector Machines (SVM), Logistic Regression (LR) and Random Forests (RF) were the classifiers used in these experiments. The SVM classifier was tested under three different kernels, namely \textit{rbf}, \textit{sigmoid} and \textit{linear}; the latter proved to obtain the best results for both experiments. 

The LR classifier was used with the standard parameters, whereas the RF classifier used 100 trees in the forest. The gini criterion and the maximum number of features were limited to those as aforementioned in Section~\ref{sec:travel_features}, in the case of the RF classifier.

To evaluate the performance of classifiers in our experiences we used five different metrics: precision, recall, F1-score, ROC and AUC.

We established the use of different groups of features to train our classification model, namely bag-of-words, bag-of-embeddings - word embeddings dependent technique - and both combined (horizontally combination of bag-of-words and bag-of-embeddings matrices into a single one).

\subsection{Rio de Janeiro and São Paulo}
\label{subsec:rio_de_janeiro_sao_paulo_experiment}

Messages were collected for a period of a whole month, between days March 12 and April 12, 2017, and the resulting datasets sum up a total of 6.1M and 2.9M tweets for Rio de Janeiro and São Paulo, respectively. Due to the problem detected in Section~\ref{sec:geographical_distribution}, we filtered the data in order to use only tweets that were actually inside the cities' areas. The data considered in this experiment sum up a total of 7.7M tweets -  5.3M and 2.4M tweets for Rio de Janeiro and São Paulo, respectively.

\subsubsection{Training and Test Datasets}
\label{subsec:training_test_datasets_portuguese}
The construction of the training and test sets followed a semi-automatic labeling approach. We tried to build a balanced training set, consisting of 2,000 travel-related tweets and 2,000 non-related. We start by searching tweets using specific travel-terms as described in study of Maghrebi et al.~\cite{maghrebi2016transportation}. Table~\ref{tab:terms} shows the terms used for querying each travel-mode. We found 30,000 tweets matching those terms. From this subset, we created a stratified random sample of 3,000 tweets and manually annotated them. We ended up with 2,000 tweets annotated as positive (travel-related). To select negative examples we randomly sampled 2,000 tweets and manually verified if they were negative (non-travel-related).

To create a test set, we randomly selected 1,000 tweets different from the training set and then manually labelled them as travel-related or non-travel-related. We forced the positive test examples to contain travel-modes terms that were not used in the training set construction. For instance, the word "uber" is relted to the taxi travel-mode but is just available in the tests set and not in the training set. The same happens with the word "busão" which is an informal word meaning "bus". The idea of behind this approach is to try to access the robustness and generalization of the travel-related classifiers. In the end, 71 tweets were found to be travel-related and whereas 929 were not.

\begin{table}[htbp]
	\centering
	\small
	\caption{Travel terms used to build the training set}
	\label{tab:terms}
	\begin{tabular}{c|c|c}
		\hline
		\multirow{2}{*}{\textbf{Mode of Transport}} & \multicolumn{2}{c}{\textbf{Terms}} \\ \cline{2-3} 
		& \multicolumn{1}{l|}{\textbf{Portuguese Language}} & \textbf{English Language} \\ \hline
		\textbf{Bike} & bicicleta, moto & bicycle, bike \\
		\textbf{Bus} & onibus, ônibus & bus \\
		\textbf{Car} & carro & car \\
		\textbf{Taxi} & taxi, táxi & taxi, cab \\
		\textbf{Train} & metro, metrô, trem & metro, train, subway \\
		\textbf{Walk} & caminhar & walk \\ \hline
	\end{tabular}
\end{table}

\subsubsection{Results and Analysis}
\label{subsubsec:results_rio_de_janeiro_sao_paulo}
Table~\ref{classifiers} presents the results obtained using the different features combination for our test set composed by 1,000 tweets manually annotated. According to the evaluation metrics we conclude that the bag-of-word and bag-of-embeddings combined produced better classification models. The model produced by the Linear SVM performed slightly better than the LR and the RF. Interesting to note is that \gls{BoW} features have influence on the precision scores obtained from our results, producing more conservative classifiers. Regarding the recall results, we can see that the Logistic Regression using only bag-of-embeddings features was the model with best results; perhaps if the precision is taken into consideration, the same conclusions will not be possible. Analysing the scores provided in Table~\ref{classifiers}, the best model under the F1-score was the Linear SVM, with a score of 0.85. It is worth noting that combining Bag-of-words and Bag-of-embedding with size 100 was the group of features with best performance taking into consideration the evaluation metrics used in this experiment.

\begin{table}[!bp]
	\small
	\centering
	\caption{Performance results with 100 sized vectors for BoE}
	\label{classifiers}
	\begin{tabular}{c|c|c|c|c}
		\hline
		\textbf{Classifier}                  & \textbf{Features} & \textbf{Precision} & \textbf{Recall} & \textbf{F1-score} \\ \hline
		\multirow{3}{*}{Linear SVM}          & BoW               & 1.0                & 0.6761          & 0.8067            \\
		& BoE               & 0.4338             & 0.8309          & 0.5700            \\
		& \textbf{BoW + BoE}         & \textbf{1.0}       & \textbf{0.7465} & \textbf{0.8548}   \\ \hline
		\multirow{3}{*}{Logistic Regression} & BoW               & 1.0                & 0.6338          & 0.7759            \\
		& BoE               & 0.4444             & 0.8451          & 0.5825            \\
		& BoW + BoE         & 1.0                & 0.6761          & 0.8067            \\ \hline
		\multirow{3}{*}{Random Forest}       & BoW               & 1.0                & 0.6338          & 0.7759            \\
		& BoE               & 0.2298             & 0.8028          & 0.3574            \\
		& BoW + BoE         & 1.0                & 0.6338          & 0.7759            \\ \hline
	\end{tabular}
\end{table}

\begin{figure}[!htp]
	\centering
	\includegraphics[width=0.7\textwidth]{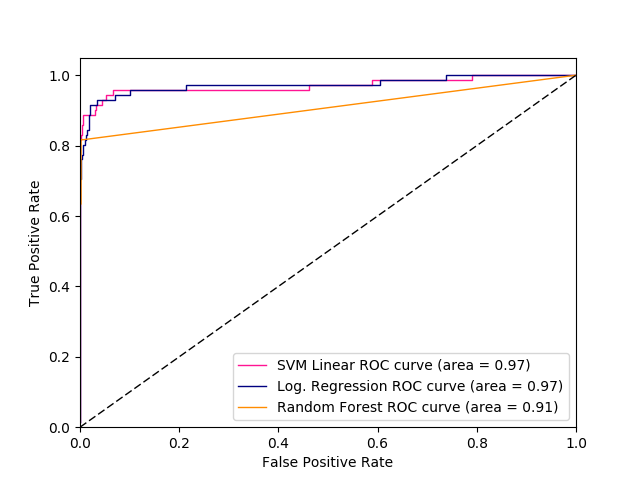}
	\caption{ROC Curve of SVM, LR and RF experiences}
	\label{fig:roc_curve}
\end{figure}

\begin{figure}[!hbp]
	\centering
	\includegraphics[width=0.7\textwidth]{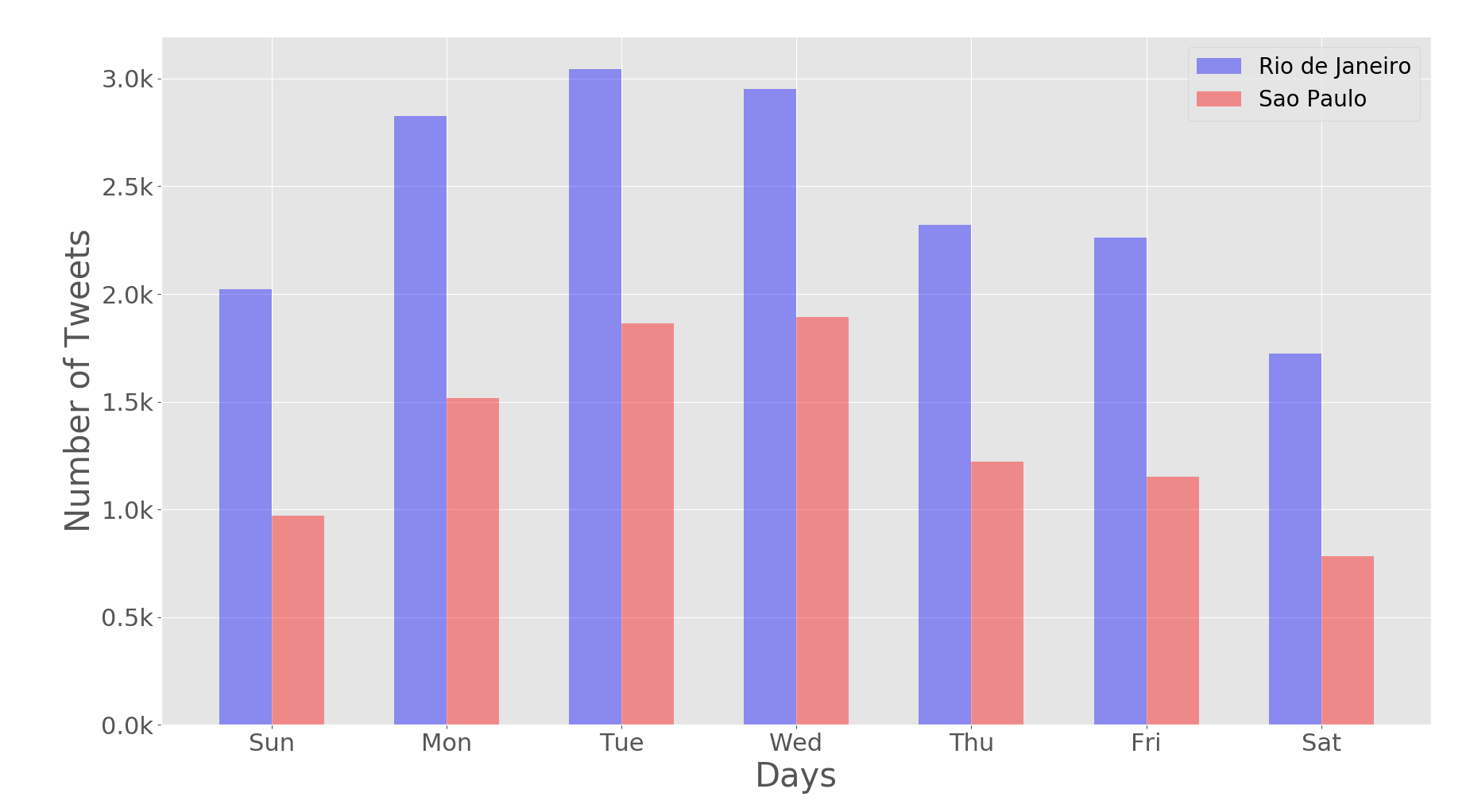}
	\caption{Positive Predicted Tweets per Day of Week}
	\label{fig:predicted}
\end{figure}

The performance of all three classifiers is illustrated using the ROC Curve in Figure~\ref{fig:roc_curve}.

The area under the curve of the Receiver Operating Characteristic (AUROC) was very similar for both the Logistic Regression and the Linear SVM models. The results obtained from the Random Forest model were not so promising as expected since this classifier is pretty tough to beat due to its versatility.

After the selection of our classification model, we decided to classify all the Portuguese dataset and draw some statistics from the results. The trained Linear SVM classifier was used to predict whether tweets were travel-related or not, since it was the model presenting the best score under the F1-score metric (as shown in Table~\ref{classifiers}). From a total of 7.8M tweets, our classifier was able identified 37,300 travel-related entries.

\begin{figure}[!bp]
	\centering
	\includegraphics[width=0.7\textwidth]{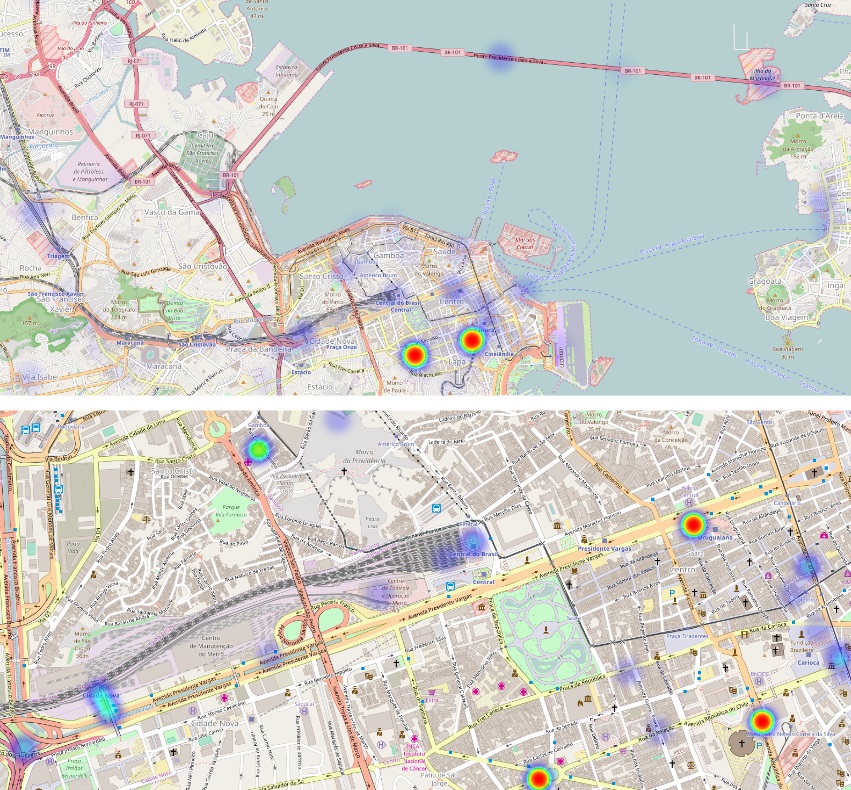}
	\caption{Rio de Janeiro heat map to the positive tweets}
	\label{subfig:rio_heatmap}
\end{figure}

\begin{figure}[!htp]
	\centering
	\includegraphics[width=0.7\textwidth]{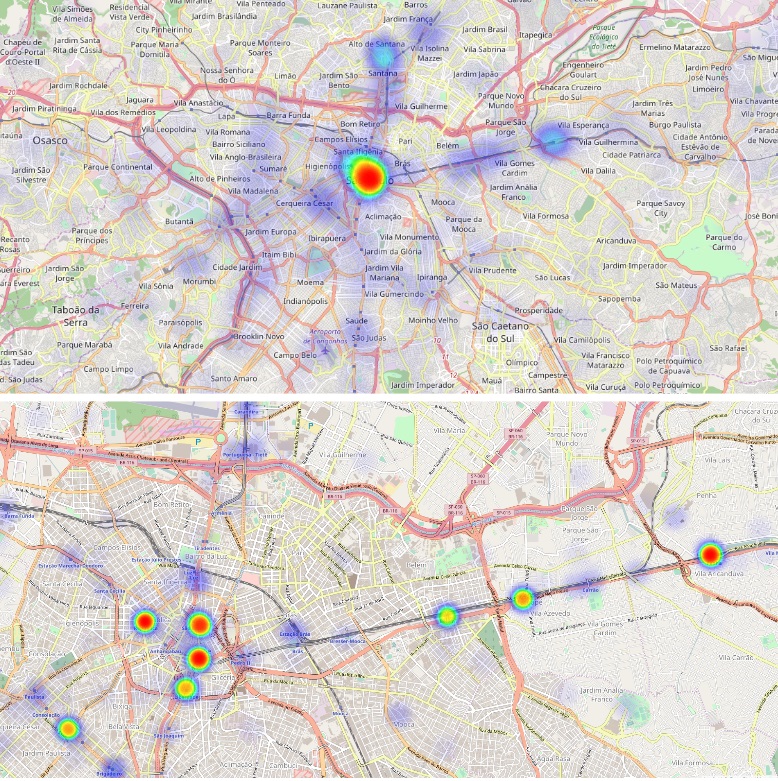}
	\caption{São Paulo heat map to the positive tweets}
	\label{subfig:sp_heatmap}
\end{figure}

Figure~\ref{fig:predicted} depicts the distribution of travel-related tweets over the days of the week. We can see that the first three business days (Monday, Tuesday and Wednesday) are the ones on which the Twitter activity is higher for both cities in our study.

In order to understand the spatial distribution of travel-related tweets we generated a heatmap for both cities. From the heatmap of Rio de Janeiro, illustrated in Figure~\ref{subfig:rio_heatmap}, it is possible to identify that some agglomerations of tweets are located at Central do Brasil, Cidade Nova and Triagem train stations, as well as at Uruguaiana, Maracanã and Carioca metro stations. The Rio-Niterói bridge, connecting Rio de Janeiro to Niterói, as well as the piers on both sides also presented considerable clouds of tweets classified as travel-related.

The heatmap for the city of SP, illustrated in Figure~\ref{subfig:sp_heatmap}, was also an interesting case to observe. Almost every agglomeration matched some metro or train station. Estação Brás, Tatuapé, Belém, Estação Paulista, Sé, Liberdade were some of the stations highlighted in the heatmap. We could also identify a little agglomeration of travel-related tweets at Congonhas airport, even though no tweets seemed to mention the word \textit{plane} explicitly in the training of our classification model.

\subsubsection{Final Remarks}
The previous described experiment explores an approach of supervised learning using as training examples a set of manually annotated tweets extracted from the whole datasets with the support of a term-based regular expression. The overall methodology is concerned with the problem of construct a fine-grained Twitter training set for the travel domain and also the automatic identification of travel-related tweets from a large scale corpus. We combined different word representations to verify whether our classification model could learn relations between words at both syntactic and semantic levels. After using standard techniques such as bag-of-words and bag-of-embeddings, we have used them combined yielding results that showed that these different groups of features can complement each other, with respect to Portuguese-speaking tweets. Modes of transport are always evolving and new services emerges making the identification of tweets related to it difficult. Overall, our experiment proved that word-embeddings features are actually an advantage regarding its applicability into instable real-world scenarios such as the transportation domain. 

\subsection{New York City}
\label{subsec:new_york_city_experiment}

Similar to the experiment of Portuguese-speaking travel-related classification of tweets, we built a model to discriminate English-speaking travel-related tweets. However, the construction of the training and test sets in this experiment follows a different approach. While in the Rio de Janeiro and São Paulo experiment we explore an semi-automatic approach and tweets were almost instantaneous formed as a group, here we were obligate to follow a two-phase approach due to the polysemy level of English travel terms.

Differently from the Brazilian cities experiment, tweets were collected from New York City during a period of two months, between days March 12 and May 12, 2017. Ignoring all non-English, as well as tweets located outside the bounding-box of New York City, the resulting dataset comprehends 4M tweets.

Regarding the preparation of data, we used the same preprocessing operations in both experiments, Brazilian and North-American. The operations were lowercasing, transformation of repeated characters and cleaning of \emph{entities} (user mentions and URLs) from the message content.

\subsubsection{Training and Test Datasets}

In the Portuguese dictionary, travel-related terms do not have more than one meaning. For instance \emph{"caminhar"} or even \emph{"comboio"} possesses only one meaning. Regarding the English dictionary, travel-related tweets may have more than one meaning since some of them present high level of polysemy. Terms such as \emph{"walk"} may be used to describe the action of walk or, for example, the action of \emph{walk into}. On the other hand, the term \emph{"train"} can be used to describe the mode of transport train or a type of behaviour through practice and instruction.

The polysemy level of such terms was took into consideration while the construction process of our training set of tweets for the English-language travel-related classification model. In the first stage of the construction process, we used the same strategy of the Portuguese training set. By take support on a semi-automatic labeling technique using a regular expression, we find out almost 16,000 tweets. The next step in the construction process was a manually verification followed by a manually annotation. Overall, 1,686 tweets were selected for each of both binary classes, travel-related and non-related. The travel-related set was strictly balanced in order to have almost the same amount of examples for each of the travel-modes involved in this study. The non-related training set is composed of several subjects that are not related to travel, e.g. football, leisure, politician, personal tweets, among others.

Nonetheless, we include into the training set tweets which polysemy level may induce doubts regarding the context of the message in order to make possible higher levels of discrimination in our model. This inclusion may help the learning process of our model making it capable of correctly identify which are the tweets that are actually related to the travel and transportation domain.
The final composition of the training datasets is presented in Table~\ref{tab:new_york_first_dataset}.

\begin{table}[!tp]
	\centering
	\caption[English-speaking tweets datasets]{Composition of the training and test datasets for the English travel-related tweets classification}
	\label{tab:new_york_first_dataset}
	\begin{tabular}{c|c|c|}
		\cline{2-3}
		\textbf{}                                       & \multicolumn{2}{c|}{\textbf{Training Set}}                          \\ \hline
		\multicolumn{1}{|c|}{\textbf{Mode of Tranport}} & \textbf{Travel-related} & \multicolumn{1}{l|}{\textbf{Non-related}} \\ \hline
		\multicolumn{1}{|c|}{Bike}                      & 300                     & \multirow{6}{*}{1686}                     \\
		\multicolumn{1}{|c|}{Bus}                       & 311                     &                                           \\
		\multicolumn{1}{|c|}{Car}                       & 317                     &                                           \\
		\multicolumn{1}{|c|}{Taxi}                      & 314                     &                                           \\
		\multicolumn{1}{|c|}{Train}                     & 317                     &                                           \\
		\multicolumn{1}{|c|}{Walk}                      & 217                     &                                           \\ \hline
		\multicolumn{1}{|c|}{\textbf{Total}}            & \multicolumn{2}{c|}{3372}                                           \\ \hline
	\end{tabular}
\end{table}

\subsubsection{Preliminary Results}\label{subsec:preliminar_results}
Due to the laborious and time-consuming effort made in the construction of the training set, we opt to apply a different approach in the training phase of our model classification model. In order to enhance the differences between tweets whose terms present high levels of polysemy, the model was trained using a \gls{k-fold-cross-validation} technique with 10 iterations for all groups of features: bag-of-words and bag-of-embeddings and both combined. Results showed good performance for all models regarding the selected evaluation metrics. The best model in this experiment was the Logistic Regression classifier trained with bag-of-words and bag-of-embeddings features, presenting a F1-score of 0,98324.

\begin{table}[!bp]
	\centering
	\caption[New York City First Experiment Results]{Preliminary results (it is only demonstrated the best result for the bag-of-embeddings group)}
	\label{tab:first_experiment}
	\begin{tabular}{|c|c|c|c|c|}
		\hline
		\textbf{Classifier} & \textbf{Features} & \textbf{Precision} & \textbf{Recall} & \textbf{F1-score} \\ \hline
		\multirow{3}{*}{\textbf{Linear SVM}} & BoE(200) & 0,90883 & 0,83634 & 0,87089 \\
		& BoW & 0,96298 & 0,97652 & 0,96962 \\
		& \textbf{BoE(200) + BoW} & \textbf{0,97251} & \textbf{0,99114} & \textbf{0,98170} \\ \hline
		\multirow{3}{*}{\textbf{Logistic Regression}} & BoE(100) & 0,90172 & 0,84948 & 0,87447 \\
		& BoW & 0,96431 & 0,98042 & 0,97222 \\
		& \textbf{BoE(200) + BoW} & \textbf{0,97391} & \textbf{0,99285} & \textbf{0,98324} \\ \hline
		\multirow{3}{*}{\textbf{Random Forests}} & BoE(100) & 0,81283 & 0,83600 & 0,82394 \\
		& \textbf{BoW} & \textbf{0,96569} & \textbf{0,98997} & \textbf{0,97764} \\
		& BoE(50) + BoW & 0,93688 & 0,99939 & 0,96701 \\ \hline
	\end{tabular}
\end{table}

The fact that all models performed incredibly well, in particular models using the features group of \gls{BoW} and \gls{BoW}+\gls{BoE} raise to us some questions and doubts about the robustness of the features used in the training process. First, in the Brazilian cities experiment, by following the same approach over the training set construction process we did not obtain results of this kind. Second, the selected tweets are very specific and our model may be overfitted due to training data. In order to pursue and have answers to our questions, we designed another experiment using the same dataset.

\subsubsection{\emph{Leave-one-group-out}}\label{subsec:leave_one_group_out}

It is worth noting that in our first experiment all travel-mode classes were known by the model before the classification of the test set (the remaining sub-dataset in the 10-fold cross-validation). Comparing with real-world scenarios, this may not be true since new modes of transport and companies, such as Uber, Lyft and Cabify, arise from unpredictable moments. This second experiment follows a \emph{leave-one-group-out} strategy, meaning that one travel-mode class if left out of the training set and moved into the test set. Hence, the behaviour of the learned model when facing a completely unknown travel-mode class can be evaluated. A model for each hidden transport-mode class was built and evaluated using the same training conditions and metrics. The datasets composition of each experiment led in this strategy can be observed in Table~\ref{tab:leave}.

\begin{table}[htbp]
	\small
	\centering
	\caption[\emph{Leave-one-group-out expirement datasets}]{Datasets composition used in the \emph{leave-one-group-out} strategy}
	\label{tab:leave}
	\begin{tabular}{|c|c|c|c|c|}
		\hline
		\multirow{2}{*}{\textbf{\begin{tabular}[c]{@{}c@{}}Travel-Mode \\ Class\end{tabular}}} & \multicolumn{2}{c|}{\textbf{Training Set}} & \multicolumn{2}{c|}{\textbf{Test Set}} \\ \cline{2-5} & \textbf{Pos.} & \textbf{Neg.} & \textbf{Pos.}  & \textbf{Neg.}  \\ \hline
		Taxi & 1,372 & \multirow{6}{*}{1,686}   & 314 & \multirow{6}{*}{300}  \\
		Train & 1,369 & & 317 & \\
		Car  & 1,369 & & 317 & \\
		Bike & 1,386 & & 300 & \\
		Walk & 1,469 & & 217 & \\
		Bus  & 1,375 & & 311 & \\ \hline
	\end{tabular}
\end{table}

For each experiment of the learning models, we maintain a 10-fold cross-validation approach, however it was built a test set with a hidden travel-mode class and 300 non-related tweets (negative class). Here, only bag-of-words and bag-of-embeddings features were fed into our models classification routine since the main goal of this experiment is to check the features robustness. Table~\ref{tab:results} presents the best results for each model, as so the group of features feeding it. To achieve the final results of this experiment, we calculated the mean between all models' results to each of the hidden transport-mode classes.

\begin{table}[!bp]
	\small
	\centering
	\caption{\emph{Leave one group out} experiments results for SVM, LR and RF classifiers}
	\label{tab:results}
	\begin{tabular}{|c|c|c|c|c|}
		\hline
		\textbf{Classifier} & \textbf{Features}  & \textbf{Precision} & \textbf{Recall}  & \textbf{F1-score} \\ \hline
		\multirow{2}{*}{\textbf{Random Forests}} & BoW & 0,40774 & 0,07474 & 0,12629  \\
		& \textbf{BoE (50)}  & \textbf{0,80278} & \textbf{0,76194} & \textbf{0,78447}  \\ \hline
		\multirow{2}{*}{\textbf{Logistic Regression}} & BoW & 0,40774 & 0,07474 & 0,12629  \\
		& \textbf{BoE (50)}  & \textbf{0,84882} & \textbf{0,75702} & \textbf{0,80219}  \\ \hline
		\multirow{2}{*}{\textbf{Linear SVM}} & BoW & 0,41527 & 0,07153 & 0,12203  \\
		& \textbf{BoE (200)} & \textbf{0,86374} & \textbf{0,75715} & \textbf{0,81289}  \\ \hline
	\end{tabular}
\end{table}

According to results, all classification models have performed reasonably well under the bag-of-embeddings features group, although the dimensionality used being different for the Linear SVM classifier.

After testing each model with a hidden travel-mode class, the models trained with bag-of-words features demonstrated poor performance when facing unknown travel-modes, revealing higher sensitivity and lower generalization capabilities in comparison to the bag-of-embeddings version. The generalization power is an important and crucial characteristic for our desired solution since in a real world scenario is very likely that we will face a higher variety of categories that were not taken into consideration in the training phase of our model. Having this considered, the bag-of-words features group presents lack of robustness as we doubt in our first experiment (Section~\ref{subsec:preliminar_results}).

\begin{figure}[htbp]
	\centering
	\includegraphics[scale=0.7]{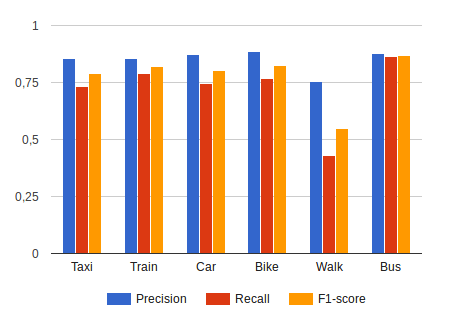}
	\caption{SVM model with BoE(200) for each travel mode}
	\label{fig:svm_leave}
\end{figure}

The best result of the \emph{leave-one-group-out} was the Linear SVM model, with the dimensionality of 200 in the size of the feature vectors. Figure~\ref{fig:svm_leave} presents the results of each experiment led for the different hidden travel-mode classes. An interesting point to observe is the low performance obtained to the experiment with the travel-mode class "Walk" hidden. This is due to the different semantic and syntactic contexts that the word \emph{walk} is used. Although all other classes can be used in the same context, for example, \emph{car}, \emph{train}, or \emph{bus}, usually the word \emph{walk} is not applied in the same way.

\begin{figure}[htbp]
	\centering
	\begin{subfigure}[htbp]{0.7\textwidth}
		\centering
		\includegraphics[width=0.9\columnwidth]{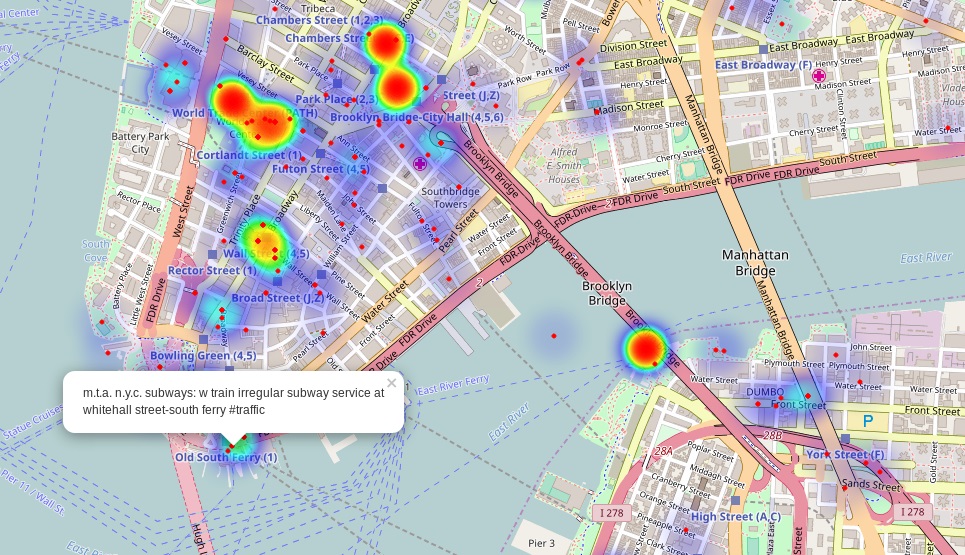}
		\caption{}
		\label{fig:brooklyn}
	\end{subfigure}
	
	\medskip
	
	\centering
	\begin{subfigure}[htbp]{0.7\textwidth}
		\centering
		\includegraphics[width=0.9\columnwidth]{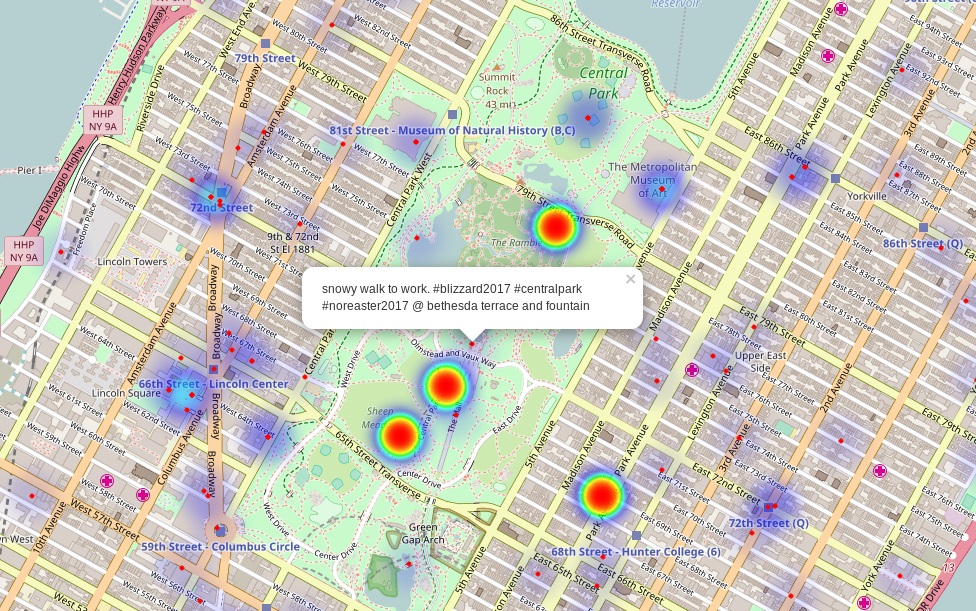}
		\caption{}
		\label{fig:central_park}
	\end{subfigure}
	
	\caption[Spatial density of the predicted tweets]{Spatial density of the travel-related predicted tweets in New York City: (a) South of Manhattan and over the Brooklyn Bridge, (b) Central Park}
	\label{fig:nyc__geographical_distribution}
\end{figure}

Having the experiments concluded, we used the best model, in this case, Linear SVM for the dimensionality of 200, to predict the 4M tweets that composed the NYC dataset. Almost 300,000 tweets were classified as travel-related. After the classification step, a sample of 10,000 tweets was taken from all the travel-related classified tweets and it was produced a heat-map distribution in order to verify which are the most concentrated zones. Such distribution enables the identification of associations with metro, train, bus stations. In Figure~\ref{fig:brooklyn}, that shows the south of the Manhattan island and also the Brooklyn bridge, it is possible no note some agglomerations over the bridge and also in the port and closed to the Wall Street(4.5) where there are some metro stations. The Central Park is one place that also took our attention since presented several agglomerations of tweets. In this particular place, tweets related to the walk class were correctly identified.

\begin{table}[htbp]
	\centering
	\small
	\caption{Sample of tweet messages correctly classified}
	\label{tab:tweets_examples}
	\resizebox{\textwidth}{!}{\begin{tabular}{|c|}
			\hline
			when you get into your uber and he has a pipe in the back \\
			a ground stop for \#ewr is no longer in effect \#flightdelay \\
			snowy walk to work. \#blizzard2017 \#centralpark \#noreaster2017 \@ bethesda terrace fountain -  \textbf{Figure~\ref{fig:central_park}} \\
			m.t.a. n.y.c subways: w train irregular subway service at whitehall street-south ferry \#traffic - \textbf{Figure~\ref{fig:brooklyn}} \\ \hline
		\end{tabular}}
	\end{table}

\subsection{Concluding Remarks}
The main objective of this experiment was to devise a travel-related tweet classifier using word embeddings trained with geo-located English-speaking tweets. Similar to the Portuguese travel-related classification, we tried to build our model using a combined approach relying on bag-of-words and bag-of-embeddings features; however, results presented signs of dependency in the bag-of-words features which is not desired when facing real-world scenarios and lots of changes happen in short periods of time. On the other hand, by looking at the results, the almost perfect performance lead us to doubt about the existence of overfitting, and so, a \emph{leave-one-group-out} strategy was applied to validate the robustness of features. There, we excluded one of the travel-modes classes, which resulted in the fact that models using bag-of-words features could not maintain the performance previously demonstrated. Comparatively to the approach based on bag-of-words, the models using bag-of-embeddings features revealed consistency and robustness in the classification task. The Linear SVM model proved to be the best option with respect to the performance metrics considered in this work. We thus used that model trained with bag-of-embeddings to predict all the travel-related English tweets from our NYC dataset, whose results showed significant improvement over a standard bag-of-words baseline. Finally, we applied the resulting classifier to a stream of geo-located tweets in New York City, which was able to depict important spatio-temporal patterns.

\section{Summary}
This chapter has the purpose of report the experiments conducted over the period of this dissertation in order to help and validate the implementation of the different modules designed in our framework architecture.

Firstly, topic modelling techniques were applied under Portuguese-speaking tweets for two different \textit{megacities}, Rio de Janeiro and São Paulo, in order to extract information that may enabling interesting characterizations in different regions/zones of the cities regarding temporal and geographical distributions. Although huge restrictions regardind the labeling of each topic, results show promising contributions and informations to the \textit{smart cities} entities, allowing until this point possible identifications of the most \textit{hot} topics through time. The location of these topics is hard since, as it was mentioned in Section~\ref{sec:geographical_distribution}, the design of a geographical distribution is difficult because the majority of tweets do not have precisely location. In the future, this problem will need to be tackled in order to allow the cities' services possible geographical recognition of the topics.

Moreover, two different classification models for travel-related tweets were developed taking into consideration two possible languages in texts, Portuguese and English. Under the implementation of the Portuguese classification, we were able to prove that the combination of conventional techniques (bag-of-words) and recent ones (word embeddings) performed very well. However, for the English-speaking messages, where polysemy levels are higher than Portuguese-speaking messages, the same group of features did not perform well. Furthermore, by following a \textit{leave-one-group-out} strategy, we study and proved robustness regarding word-ebeddings features. The omission of a transport-mode class cause the model fed with bag-of-words features to perform worst than the one using only bag-of-embeddings. Such results need to be seen as a positive point since through this experiment we were able to capable of produce two classification models with higher levels of generalization. The resulting models were used in the development of our frameworks' travel-related classification module. We can conclude that is now able of discriminate travel-related tweets for two distinct speaking languages based on two important factors: consistency and robustness.
\chapter{Conclusions and Future Work} \label{chap:conclusions}

\minitoc \mtcskip \noindent

\section{Final Remarks}

This work tackles the problem of extracting meaningful and actionable knowledge from user generated content in the scope of smart cities and intelligent transportation systems. We designed and developed a framework for collection, processing and mining of geo-located Tweets. More specifically, it provides functionalities for parallel collection of geo-located tweets from multiple pre-defined bounding boxes (cities or regions), including filtering of non complying tweets, text pre-processing for Portuguese and English language, topic modeling, and transportation-specific text classifiers, as well as, aggregation and data visualization.

The Twitter Streaming API has three different heuristics to collect tweets: terms-based retrieval, following the activity of users and collect all tweets inside a specific bounding-box. The final stakeholders of our framework are cities and to provide content and geographical analysis, we opt for the locations heuristic. However, we found that several retrieved tweets do not respect the pre-defined bounding-box, i.e., the limits the tweets coordinates are outside the pre-defined bounding-box. Most of previous work using geo-located tweets do not take into account this phenomenon. We designed a filtering model capable to cope with noisy geo-located tweets. This module also filters out tweets written in any language besides the pre-defined English and Portuguese languages.

To analyse the text message of tweets, we design a text analytics module which is composed by two sub-modules. These sub-modules are in charge of performing topic modelling and travel-related classification tasks. In the module responsible for characterize tweets over latent topics, we needed to submit each tweet to a pre-defined group of text pre-processing operations. Lower casing, transformation of repeated characters, removal of \textit{metadata}, punctuation. short tokens and Portuguese stop words are the pre-processing operations used to clean the text message and facilitate the identification of latent topics. With respect to the travel-related classification of tweets, we tried to combine different text representation to fed as features group in the training routine of our models. To clean and make easier the task of classification, we perform lower casing, transformation of repeated characters and removal of \textit{metadata} and Portuguese and English stop words. 

To aggregate the final results provided by the text analytics module, we used the MongoDB aggregation framework since it present high performance and scalability, dealing well with large volumes of data. Finally, the visualization of results explores a library capable of saving locally the graphical representations of results. By using this saving method, the framework only needs to execute its text and statistical analysis in specific periods of time since when a user requests for a visualization, there is no need to consult the data warehouse. Lower response time is one of the most important factors in systems dealing with high volumes of data.

To illustrate our approach we conduct an exploratory data analysis and performed experiments for each text analytics module in real-world scenarios. We performed empirical studies and implemented illustrative examples for 5 cities: Rio de Janeiro, São Paulo, New York City, London and Melbourne, comprising a total of more than 43 millions tweets in a period of 3 months. Through the exploratory data analysis we identify that more than 70\% of tweets do not have a fine-grained geographic coordinate but they refer to bounding-boxes representing large areas on each city, such as "Ipanema". This fact reduces the ability to perform thorough spatial analysis of geo-located tweets.

Our framework focuses on content analysis of geo-located tweets. We performed a topic modeling experiment for a large volume of tweets from the two most populous and active Brazilian cities, Rio de Janeiro and São Paulo. The latent topics discovered might serve as actionable information to cities helping in the monitoring of what is being talked about in its urban regions. Both Rio de Janeiro and São Paulo presented similar latent topics in which 25 of them were equal and only 2 latent topics were specific of each city, summing up a total of 29 distinct topics. It is worth to notice that our latent topic model was capable of recognize general purpose topics such as "Relationship ad Friendship" and "Personal Feelings", making us to question why Brazilian people talk about such personal subjects into social media networks.

In the experiment of travel-related tweets classification, we constructed different gold standard datasets for two distinct speaking-languages, Portuguese and English. In the features construction process, we take support of recent advanced text mining techniques such as word-embeddings. This technique enhances the semantic and syntactic similarities between text messages allowing the identification that "busão" (a Brazilian informal term for the bus transport-mode) is used in the same context that is formal term "ônibus". We have further used the embedding matrix (bag-of-embeddings) of tweets in combination with bag-of-words features to train our text classifiers. The Portuguese classification model performed very well and was able to discriminate tweets which travel terms were omitted from the training process. The classification of tweets having "Uber" and "Busão" as travel-related ones is a proof of the robustness and generalization of our model.

On the other hand, the English text classifier revealed high levels of dependency with the bag-of-words features. The training set of this model was conducted using a \gls{k-fold-cross-validation} technique since English travel-terms, such as "Walk" and "Train", have high levels of polysemy. The preliminary scores of our model were almost perfect and beacuse of that, we designed a new strategy to conduct the remainder of the experiment. By following a leave-one-group-out strategy, we verified that models trained with word embeddings features maintain their performance while models trained with bag-of-words have drastically decrease its performance. Such strategy was enough to conclude the lack of robustness in the bag-of-words features making us to decide to use of a model trained with word-embeddings into the framework implementation for English speaking cities. 

Regarding the Portuguese cities, we chose to use the model trained with both type of features: \gls{BoE} and \gls{BoW}.


\section{Contributions}
\label{seCc:contributions}
At the end of this dissertation, we summarise the contributions achieve in three main dimensions:

\begin{itemize}
	\item \textbf{Technical Contributions}
	We designed and developed an open-source framework implemented in Python programming language using the Tweepy library for collecting geo-located tweets. Our implementation allows the collection of multiple and parallel bounding-boxes (cities or regions) and it is complying with the Twitter Streaming API usage limits. We opted to use a no-SQL database (MongoDB) as data storage software which provides flexibility, scalability and adaptability to the framework. We rely on Python's LDA library fro topic modelling, Gensim library to train paragraph2vec embeddings from geo-located tweets and Sckit-learn to train the test classifiers. The framework also provide flexible aggregations and on-time visualization using the Plotly library. The framework can be used in multiple application scenarios, from different content languages to different levels of geographic granularity (streets vs cities vs regions vs countries).

	\item \textbf{Applicational Contributions}
	To the best of our knowledge this work is the first large scale comparative topic modelling study of geo-located tweets in Brazilian \textit{megacities}. We are also the first to explore the recent advances in word-embeddings with application to text classification in the scope of smart cities and intelligent transportation cities.
	
	\item \textbf{Scientific Contributions}
	We performed empirical evaluation on the applicability and robustness of word-embeddings representation as features to train a travel-related classifier of geo-located tweets. To perform these studies, we had to create new gold standard data that can be used by the community for further experimentation.
	
	The main finding of the analysis carried out were documented in papers submitted to conferences as follows. 
\end{itemize}

\section{Publications}
\label{sec:publications}
During the period of this dissertation, we published three different scientific papers in order to share our experiments' methodologies and results.

\begin{itemize}
	\item
	João Pereira, Arian Pasquali, Pedro Saleiro and Rosaldo J. F. Rossetti. {\color{blue}Transportation in Social Media: an automatic classifier for travel-related tweets}. In \emph{Portuguese Conference on Artificial Intelligence} (EPIA), 2017. In Press~\cite{pereira2017transport}.
	
	\item
	João Pereira, Arian Pasquali, Pedro Saleiro, Rosaldo J. F. Rossetti and Javier Sanchez-Medina. {\color{blue}Classifying Travel-related Tweets Using Word Embeddings}. In \emph{IEEE 20th International Conference on Intelligent Transportation Systems} (IEEE ITSC), 2017. Under review.
	
	\item
	João Pereira, Arian Pasquali, Pedro Saleiro, Rosaldo J. F. Rossetti and Nélio Cacho. {\color{blue}Characterizing Geo-located Tweets in Brazilian Megacities}. In \emph{The Third International Smart Cities Conference} (ISC2), 2017. Under review.
\end{itemize}

\section{Future Work}

The dissertation purpose had as it main focus the conception of an automatic system capable of analyse real-time data streams from social media platforms in order to produce valuable information for users of services or even its responsible entities. For achieve the proposed goals, we tried to explore already consistent state-of-the-art methodologies as well as unexplored ones regarding specific domains. Since this framework can be seen as a prototype of a future complex system, several improvements can be invested here. Although already existent modules and text analysis devised, it worth noting the conjecture of a additional sentiment analysis module in order to infer the sentiment polarity value regarding specific zones where the travel-related tweets were located in, as so the overall sentiment in an identified topic.

In this dissertation we trained word embeddings using geo-located tweets to further use such embeddings matrices as features to train our transportation text classifiers. Later we will perform an intrinsic evaluation of these word embeddings in order to publicly share benchmarks for travel-related terms as Saleiro et al.~\cite{saleiro2017embeddings} have made using general tweets to study practical aspects of this text representation.

The extension of our training and test sets is other future work to take into consideration, as well as the application of deep learning classifiers into our classification tasks.

Nonetheless, we can explore a way of predicting future events in a city or even the impact that certain transportation entities will have in specific places~\cite{saleiro2016learning} in order to monitor correctly the the agglomeration of people in these places and the roads traffic.

Another important work to pursue in the future is to correlate the results of this study with official sources of transportation agencies relatively to traffic congestions and other events on the transportation network, including all modes of transports and their integration interfaces and modules. This kind of association will be useful both to validate the proposed approach as well as to improve the inference process and knowledge extraction. The automatic classifier herein presented will then be integrated into data fusion routines to enhance transportation supply and demand prediction processes alongside other sensors and sources of information. Additionally, such correlations and inference results can be very useful as input to traffic simulation tools~\cite{passos2011towards, azevedo2015state}, and to support multi-resolution and multi-purpose transportation analysis~\cite{timoteo2010trasmapi, ferreira2008cooperative}.

A possible future direction to improve the topic modelling approach is the application of spatio-temporal aggregation methods under a sample of data to create more complex documents, retrain the model and verify if the results can be different taking into consideration some of the factors that distinguish both cities: demographics, culture and location. An attempt to pursue good performances using supervised LDA models also needs to be enhanced here.

Lastly, there is a need of creation of other specific models to other fields of a \textit{smart city} in order to assure equally performances for any of its fields. Nonetheless, this work will also be integrated into the MAS-Ter Lab framework~\cite{rossetti2007towards}, whose main objective is to support the design, evaluation, and implementation of transportation engineering solutions and smart mobility relying on the concept of Artificial Transportation Systems~\cite{rossetti2014advances}, combining both data- and model-driven methodologies.




\glsaddall
\glstoctrue
\glsnogroupskiptrue
\printglossary[type=\acronymtype]
\printglossary

\PrintBib{myrefs}

\end{document}